\documentclass[12pt,preprint]{aastex}

\usepackage{natbib}
\usepackage{longtable}
\slugcomment{Draft version from \today}

\shorttitle{Spectral line data toward IRDCs}
\shortauthors{Beuther \& Sridharan}

\begin{document}

\title{Outflow and dense gas emission from massive Infrared Dark Clouds} 


\author{H.~Beuther$^1$ \& T.K. Sridharan$^2$}
\altaffiltext{1}{Max-Planck-Institute for Astronomy, K\"onigstuhl 17, 69117 Heidelberg, Germany}
\altaffiltext{2}{Harvard-Smithsonian Center for Astrophysics, 60 Garden Street, Cambridge, MA 02138, USA}
\email{beuther@mpia.de, tksridha@cfa.harvard.edu}

\begin{abstract}
  Infrared Dark Clouds are expected to harbor sources in different,
  very young evolutionary stages. To better characterize these
  differences, we observed a sample of 43 massive Infrared Dark
  Clouds, originally selected as candidate high-mass starless cores,
  with the IRAM\,30\,m telescope covering spectral line tracers of
  low-density gas, high-density gas, molecular outflows/jets and
  temperature effects. The SiO(2--1) observations reveal detections
  toward 18 sources.  Assuming that SiO is exclusively produced by
  sputtering from dust grains, this implies that at least in 40\% of
  this sample star formation is on-going. A broad range of SiO
  line-widths is observed (between 2.2 and 65\,km\,s$^{-1}$), and we
  discuss potential origins for this velocity spread.  While the
  low-density tracers $^{12}$CO(2--1) and $^{13}$CO(1--0) are detected
  in several velocity components, the high-density tracer
  H$^{13}$CO$^+$(1--0) generally shows only a single velocity
  component and is hence well suited for kinematic distance estimates
  of IRDCs.  Furthermore, the H$^{13}$CO$^+$ line-width is on average
  1.5 times larger than that of previously observed NH$_3$(1,1). This
  is indicative of more motion at the denser core centers, either due
  to turbulence or beginning star formation activity. In addition, we
  detect CH$_3$CN toward only six sources whereas CH$_3$OH is observed
  toward approximately 40\% of the sample.  Estimates of the CH$_3$CN
  and CH$_3$OH abundances are low with average values of $1.2\times
  10^{-10}$ and $4.3\times 10^{-10}$, respectively. These results are
  consistent with chemical models at the earliest evolutionary stages
  of high-mass star formation. Furthermore, the CH$_3$OH abundances
  compare well to recently reported values for low-mass starless
  cores.
\end{abstract}


\section{Introduction}
\label{intro}

While observational research in high-mas star formation over the last
few decades has focused on relatively evolved regions like
Ultracompact H{\sc ii} regions (UCH{\sc ii}s) and High-Mass
Protostellar Objects (HMPOs) (e.g., see reviews by
\citealt{kurtz2000,churchwell2002,beuther2006b}), the earliest stages
of massive star formation have become regularly accessible only
through the advent of large-scale mid-infrared surveys of the Galactic
plane by satellite missions like the Infrared Space Observatory (ISO),
the Midcourse Space Experiment (MSX) and most recently Spitzer. The
basic observational characteristics of the earliest stages of
(massive) star formation, prior to the formation of any embedded
heating source, should be that they are strong cold dust emitters at
(sub)mm wavelengths, and weak or non-detections in the mid-infrared
because they have not yet heated a warm dust cocoon. In the low-mass
regime, so-called Infrared Dark Clouds (IRDCs) have been studied
regularly since a few years (e.g.,
\citealt{egan1998,carey2000,bacmann2000}), in high-mass star formation
the focus has shifted to these objects only more recently (e.g.,
\citealt{sridharan2005,hill2005,klein2005,simon2006,beltran2006,rathborne2006,pillai2006}).

Massive IRDCs (in short IRDCs from now) are not a well defined class,
but they likely harbor various evolutionary stages, from genuine
High-Mass Starless Cores (HMSCs) via High-Mass Cores harboring
accreting Low/Intermediate-Mass Protostars to the youngest HMPOs
\citep{beuther2006b}. While the first stage provides good targets to
study the initial conditions of massive star formation prior to cloud
collapse, the other stages are important to understand the early
evolution of massive star-forming clumps. Here, we are targeting a
sample of massive IRDCs first identified as candidate HMSCs by
\citet{sridharan2005} to differentiate between the different
evolutionary stages existing within this sample. To facilitate this,
we observed the target sample with the IRAM 30\,m telescope in
different spectral lines sensitive to various physical processes like
molecular outflows (e.g., SiO, CO), dense gas (e.g., H$^{13}$CO$^+$,
CH$_3$CN, CH$_3$OH) or hot core conditions (CH$_3$CN, CH$_3$OH). Table
\ref{observing} lists the observed spectral lines, the upper level
excitation energies $E_{\rm{upper}}$ and the critical densities
$n_{\rm{crit}}$.

\section{Observations and Data Reduction}
\label{obs}

We observed a sample of 43 IRDCs selected from \citet{sridharan2005}
in August 2005 at the IRAM 30\,m telescope on Pico Veleta with four
receivers simultaneously. The flexible multi-receiver setup of the
30\,m telescope allowed us to observe usually a few lines within the
same spectral setup of each receiver ((1) H$^{13}$CO$^+$(1--0) \&
SiO(2--1), (2) $^{13}$CO(1--0) and CH$_3$CN, (3) $^{12}$CO(2--1), (4)
CH$_3$OH). Table \ref{observing} lists the corresponding frequencies,
the spectral resolution and the system temperatures for each line. The
observations were conducted in a position-switching mode with OFF
positions carefully chosen from the Massachusetts-Stony Brook Galactic
plane CO survey \citep{sanders1986}, usually approximately of the
order $1000''$ offset from the source positions.

\begin{table*}[htb]
\caption{Observing parameters}
\begin{center}
\begin{tabular}{lrrrrrr}
\hline \hline
Line & Freq. & $\Delta v$   & $T_{\rm{sys}}^a$ & $1\sigma$\,rms$^a$ & $E_{\rm{upper}}$ & $n_{\rm{crit}}^b$ \\
     & GHz   & km\,s$^{-1}$ & K                & K   &    K             & cm$^{-3}$\\
\hline
H$^{13}$CO$^+$(1--0) & 86.754  & 1.1 & 110 & 0.02 & 4.2  & 1.7e5\\
SiO(2--1)            & 86.847  & 1.1 & 110 & 0.02 & 6.3  & 7.3e5\\
$^{13}$CO(1--0)      & 110.201 & 0.9 & 220 & 0.04 & 5.3  & 2.2e3\\
CH$_3$CN$(6_0-5_0)$  & 110.384 & 0.9 & 220 & 0.03 & 18.5 & 1.5e6$^c$\\
CH$_3$CN$(6_1-5_1)$  & 110.381 & 0.9 & 220 & 0.03 & 25.7 & 1.5e6$^c$\\
CH$_3$CN$(6_2-5_2)$  & 110.375 & 0.9 & 220 & 0.03 & 47.1 & 1.6e6$^c$\\
$^{12}$CO(2--1)      & 230.538 & 1.6 & 660 & 0.30 & 16.6 & 1.0e4\\
CH$_3$OH$(5_1-4_1)$E & 241.879 & 1.2 & 760 & 0.13 & 55.9 & 1.1e6\\
CH$_3$OH$(5_0-4_0)$A+& 241.791 & 1.2 & 760 & 0.13 & 34.8 & $^d$\\
CH$_3$OH$(5_{-1}-4_{-1})$E&241.767&1.2&760 & 0.13 & 40.4 & 9.5e5\\
CH$_3$OH$(5_0-4_0)$E & 241.700 & 1.2 & 760 & 0.13 & 47.9 & 1.0e6\\
\hline \hline
\end{tabular}
\end{center}
$^a$ Average system temperature and $1\sigma$\,rms values.\\
$^b$ Calculated from $A=0.3\lambda_{100}^{-3}\mu ^2$ ($\lambda_{100}$ in units of 
100\,$\mu$m and $\mu=3.9$\,debye) and $\gamma(\rm{CH_3CN}(6-5))_{20K}$ from \citet{pei1995a,pei1995b}.\\
$^c$ $n_{\rm{crit}} = A/\gamma$ with the Einstein coefficient $A$ and 
the collisional rate $\gamma$. Except for CH$_3$CN, $A$ and $\gamma$ 
were taken from LAMBDA \citep{schoeier2005} at $T=20$\,K.\\
$^d$ No data for this transition in LAMBDA.
\label{observing}
\end{table*}

\section{Results and discussion}

\subsection{$^{12}$CO(2--1), $^{13}$CO(1--0) and H$^{13}$CO$^+$(1--0)}

\begin{figure*}[h]
\includegraphics[angle=-90,width=5.4cm]{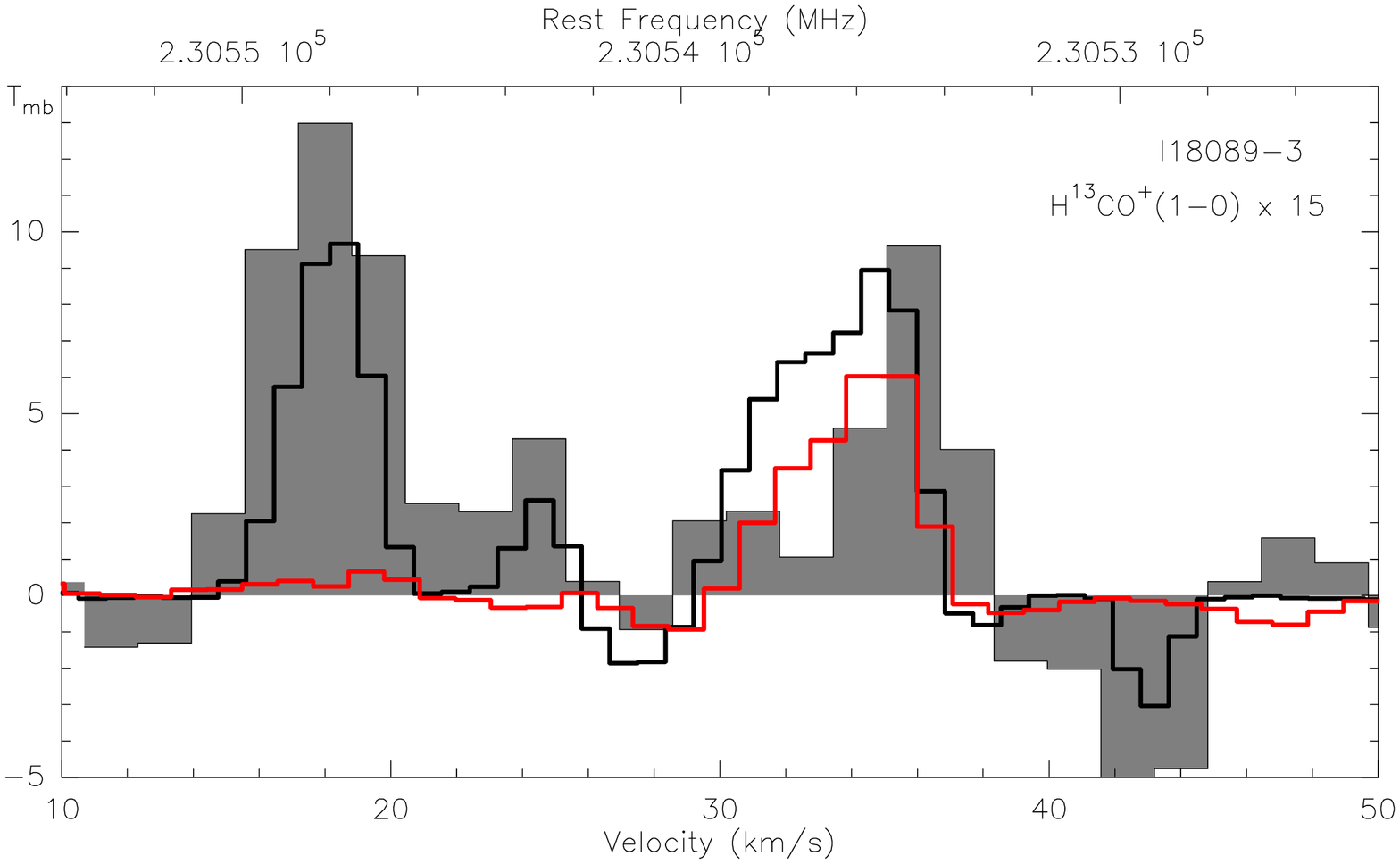}
\includegraphics[angle=-90,width=5.4cm]{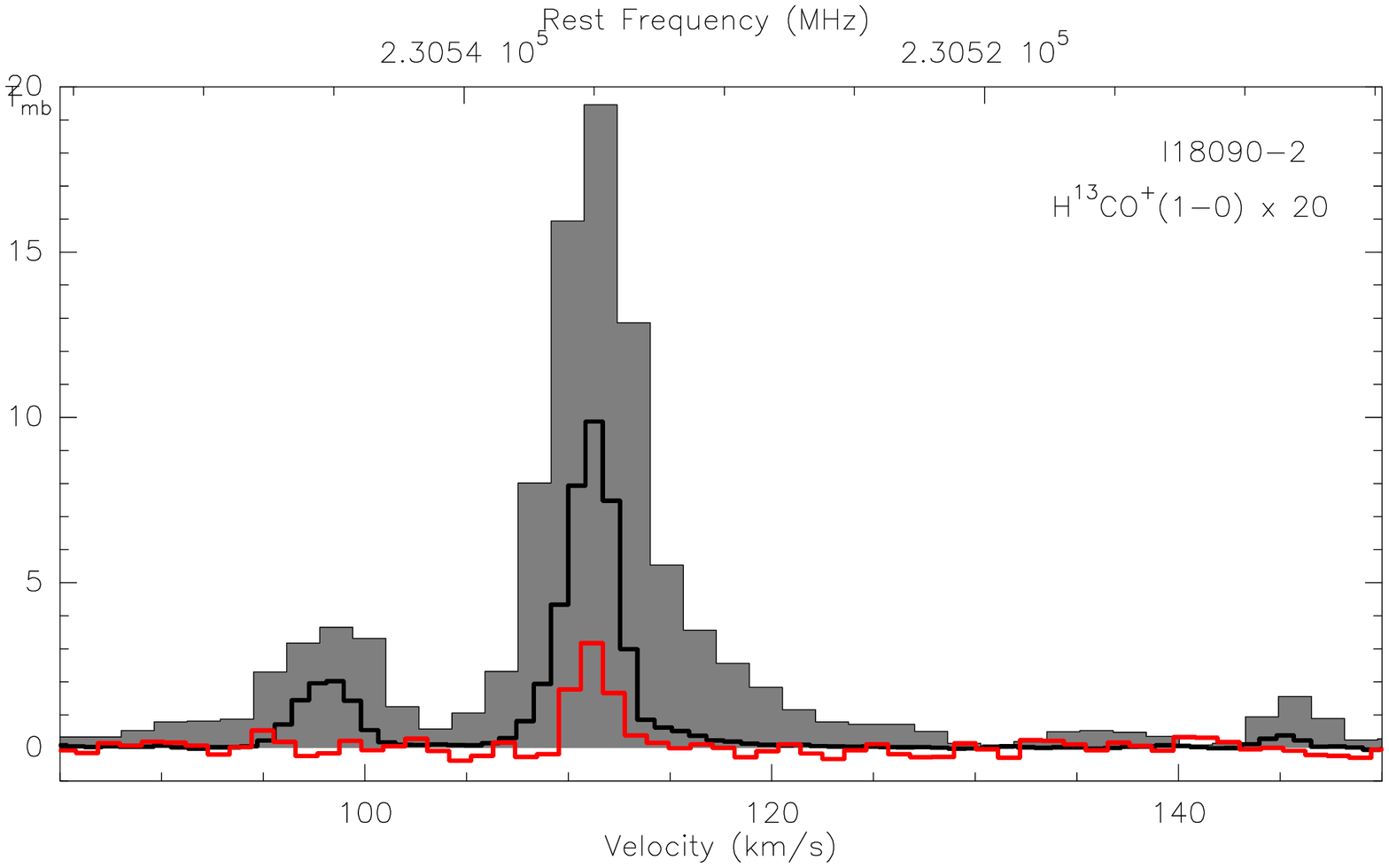}
\includegraphics[angle=-90,width=5.4cm]{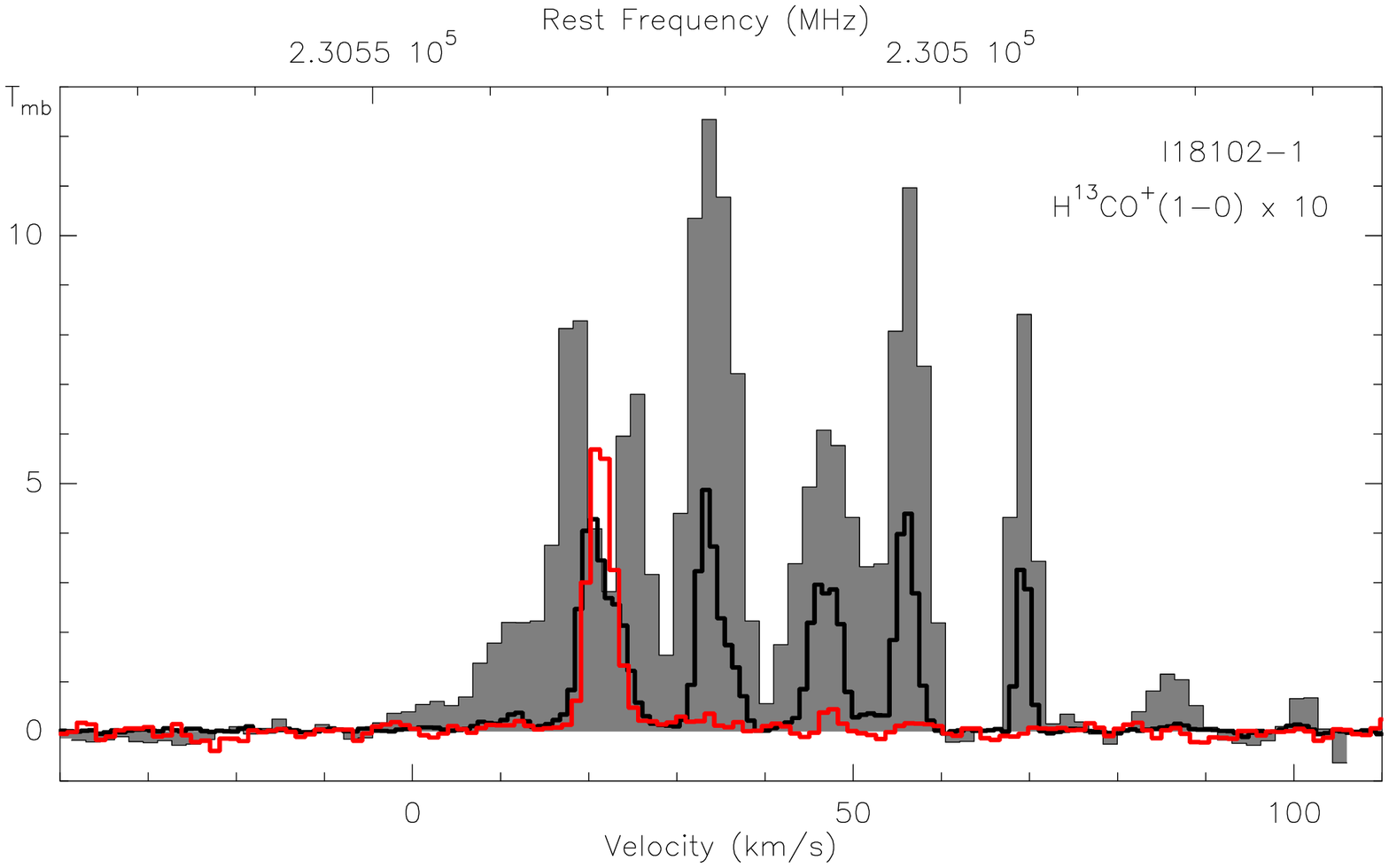}\\
\includegraphics[angle=-90,width=5.4cm]{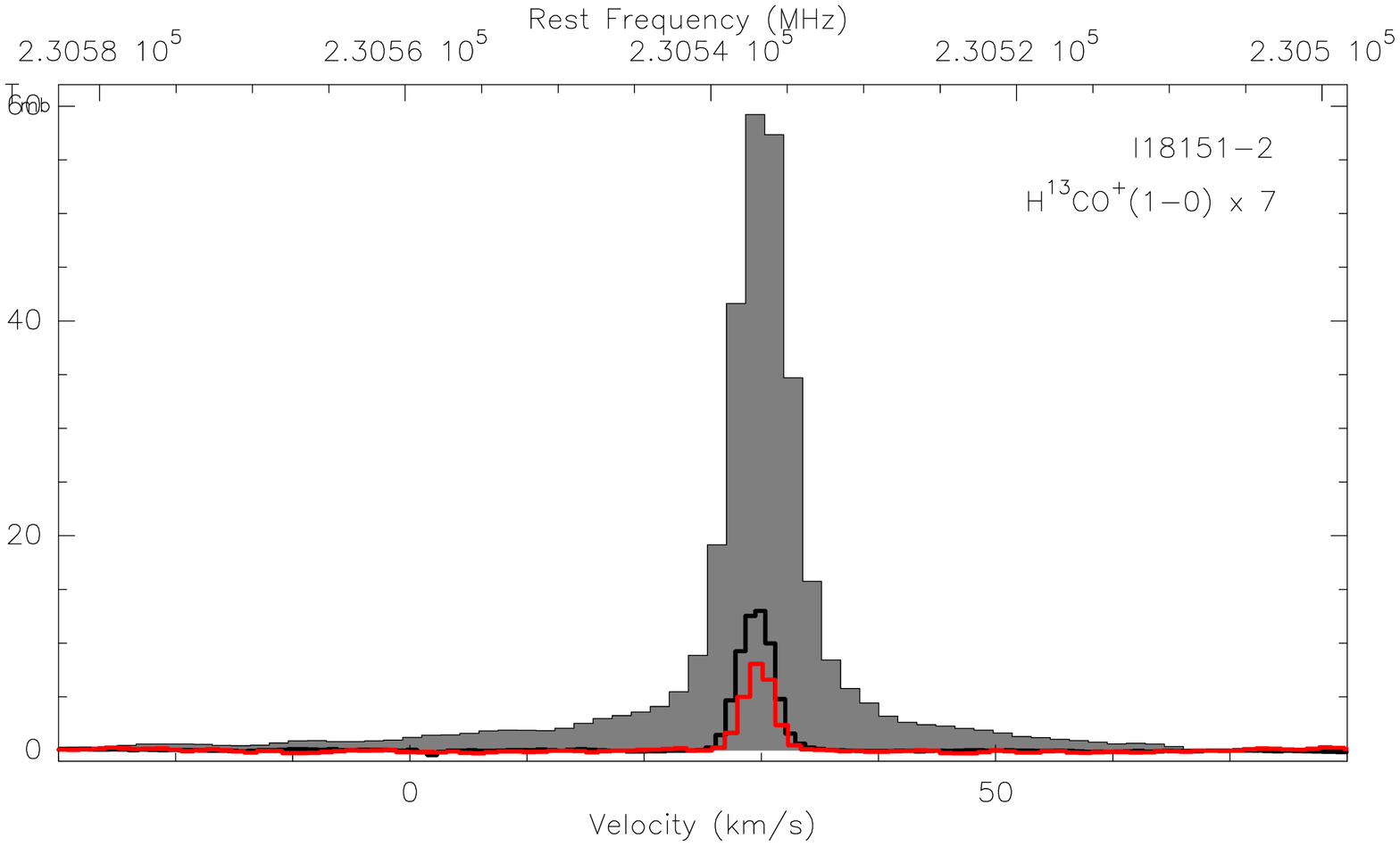}
\includegraphics[angle=-90,width=5.4cm]{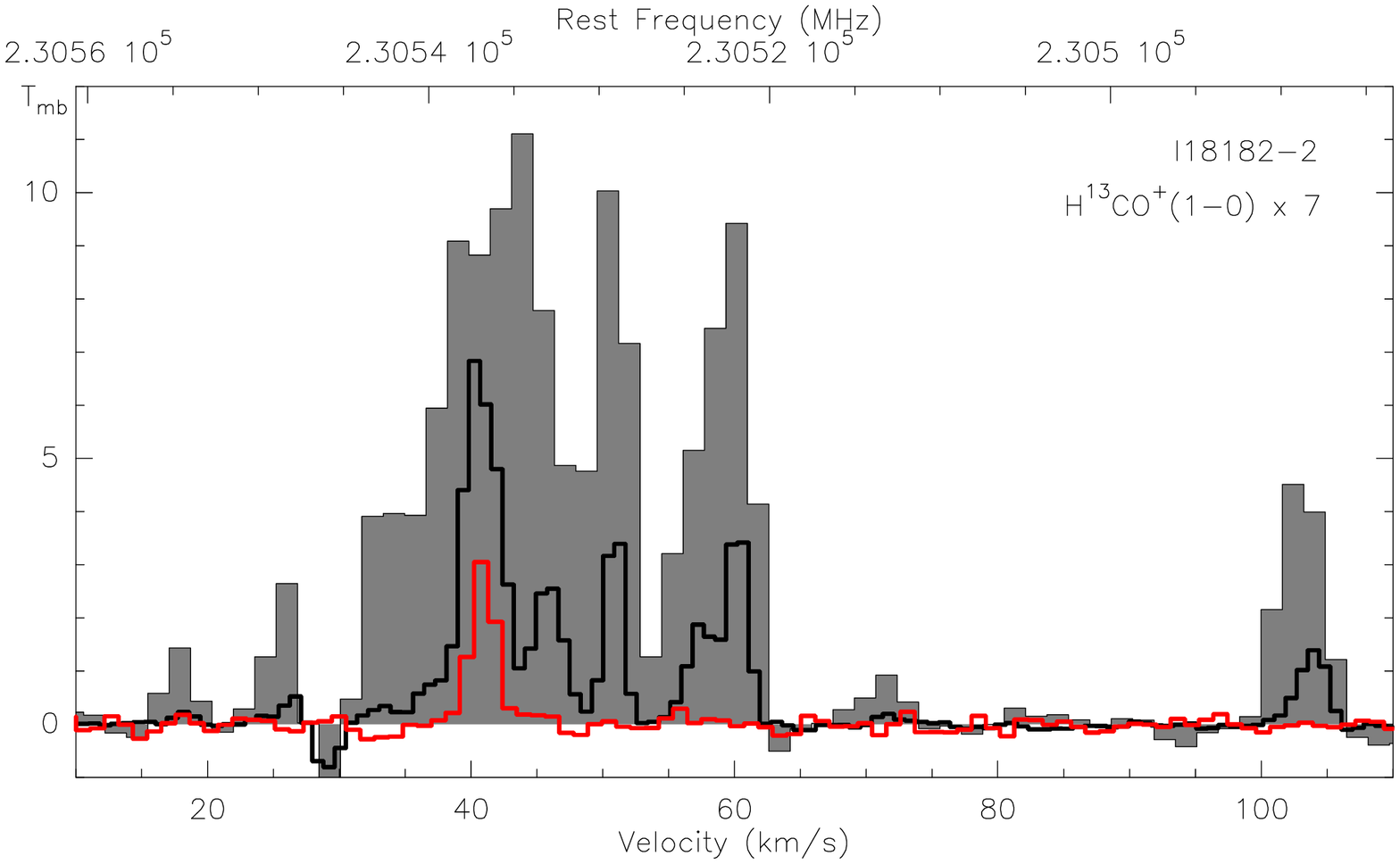}
\includegraphics[angle=-90,width=5.4cm]{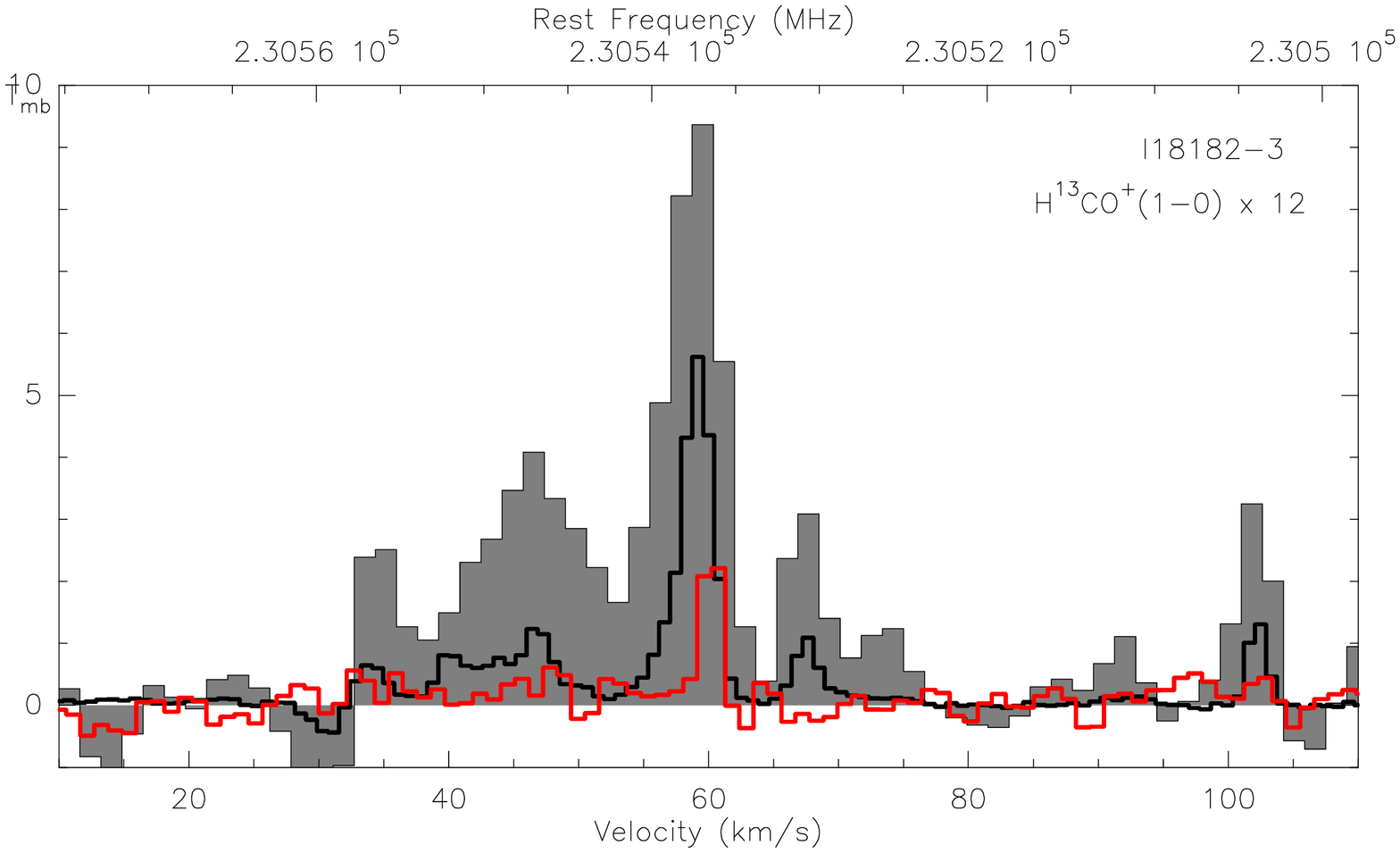}\\
\includegraphics[angle=-90,width=5.4cm]{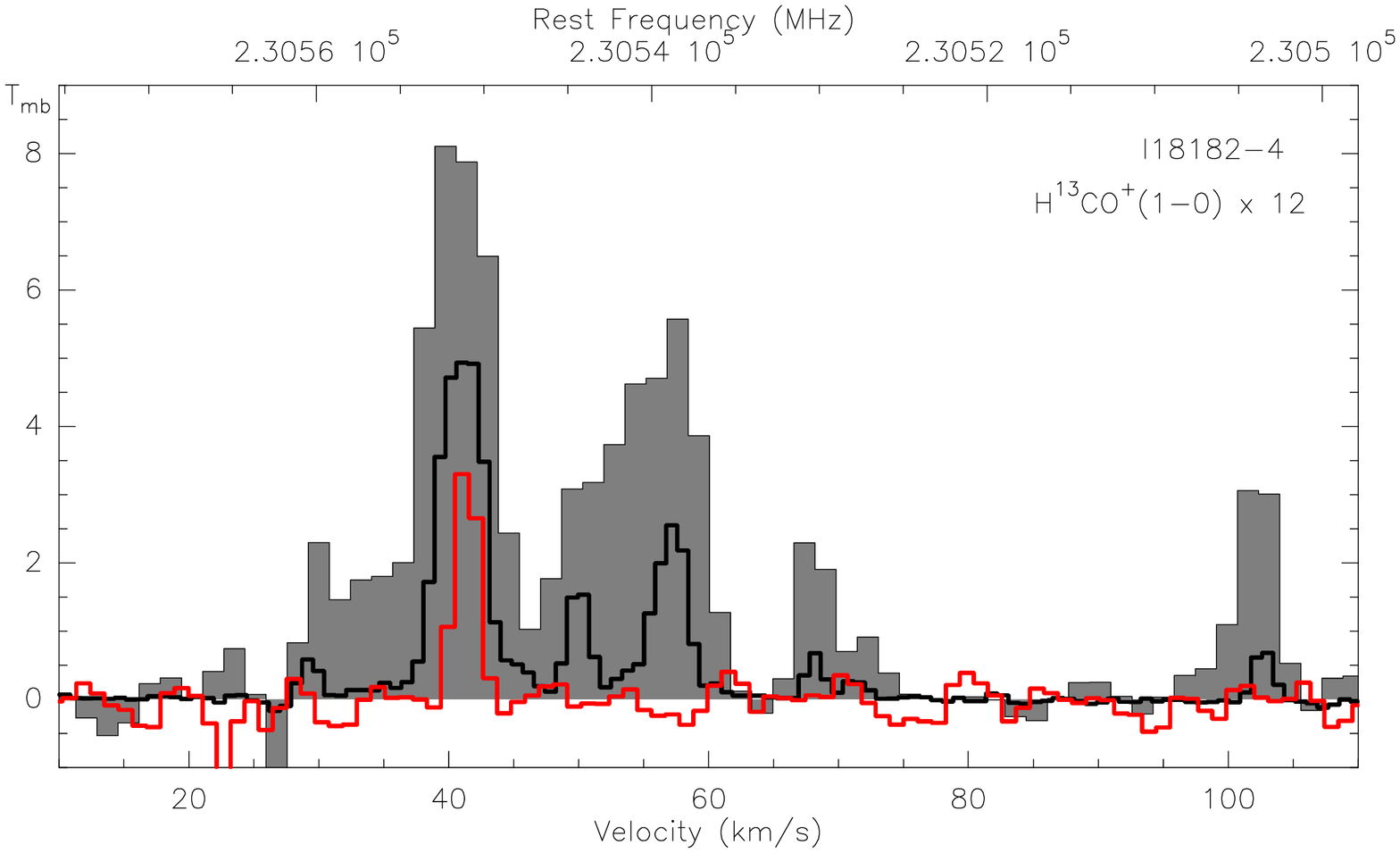}
\includegraphics[angle=-90,width=5.4cm]{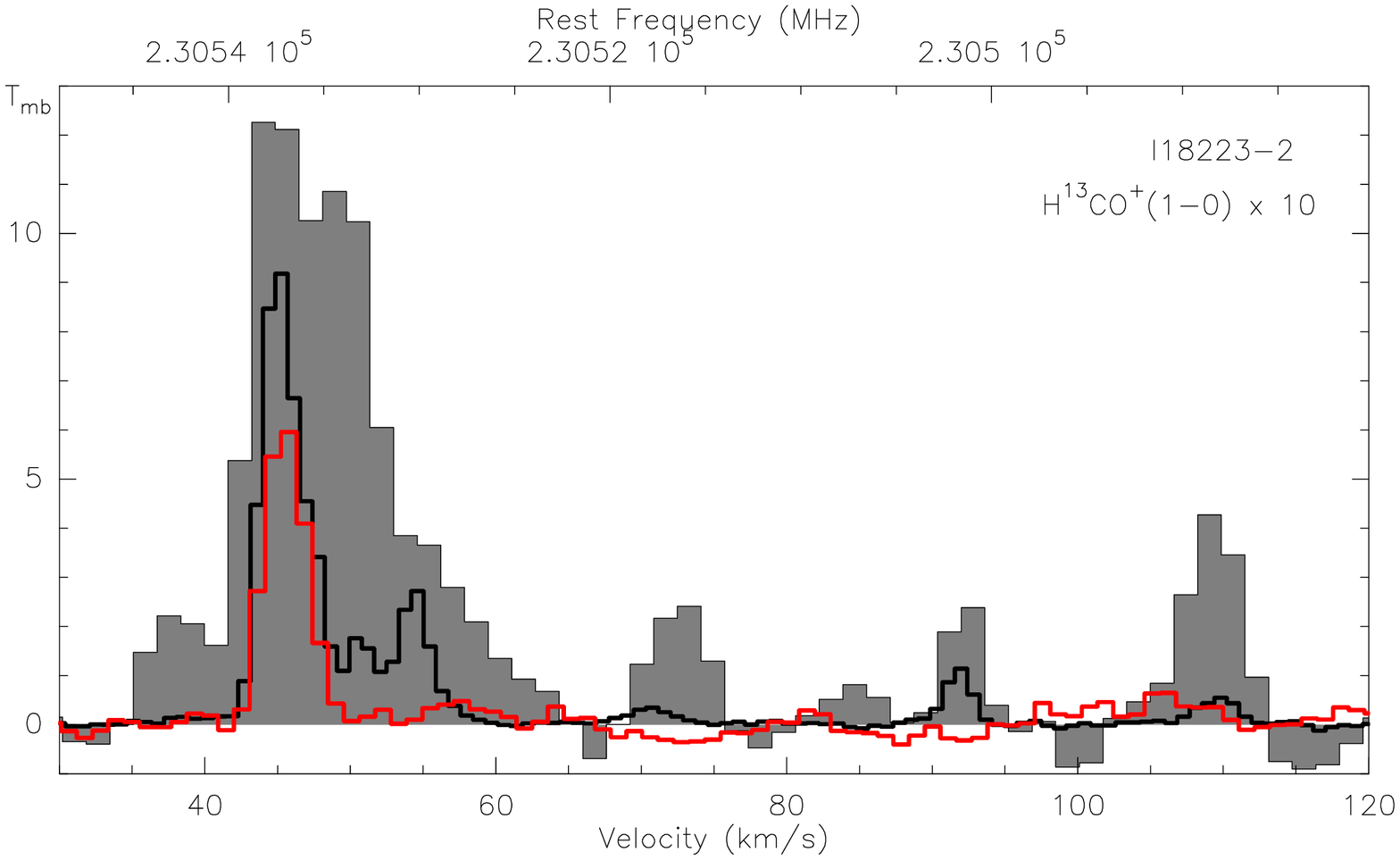}
\includegraphics[angle=-90,width=5.4cm]{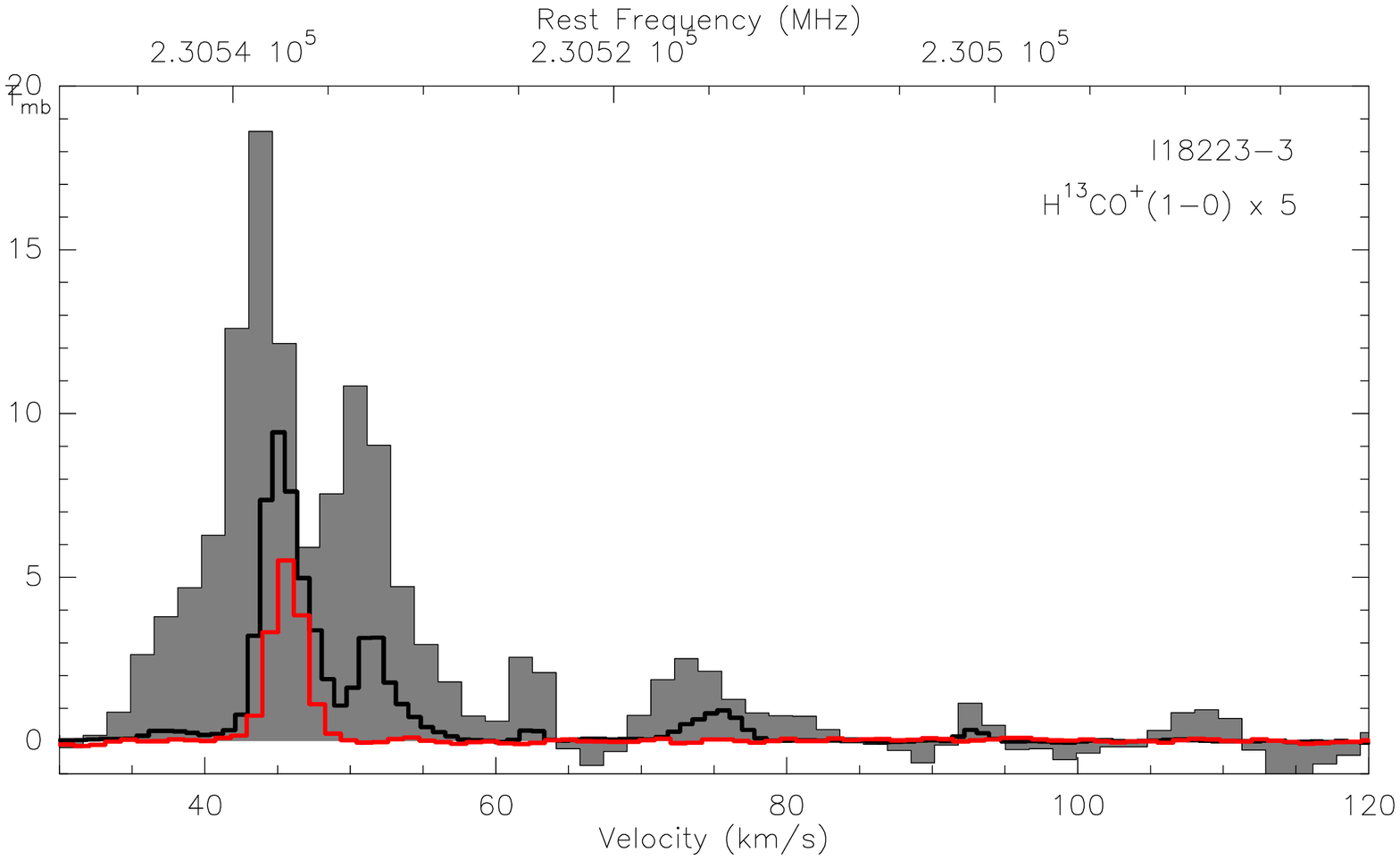}\\
\includegraphics[angle=-90,width=5.4cm]{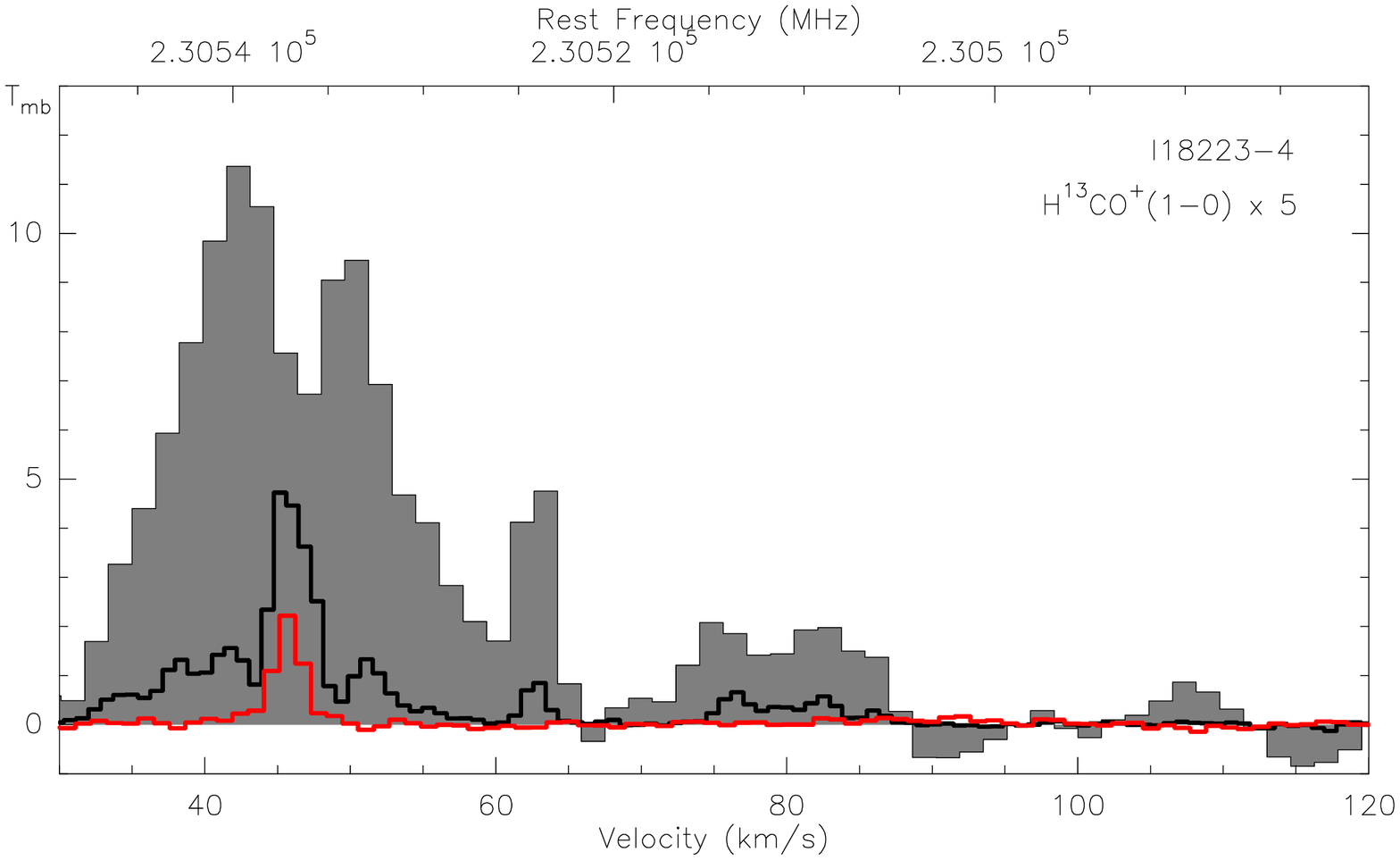}
\includegraphics[angle=-90,width=5.4cm]{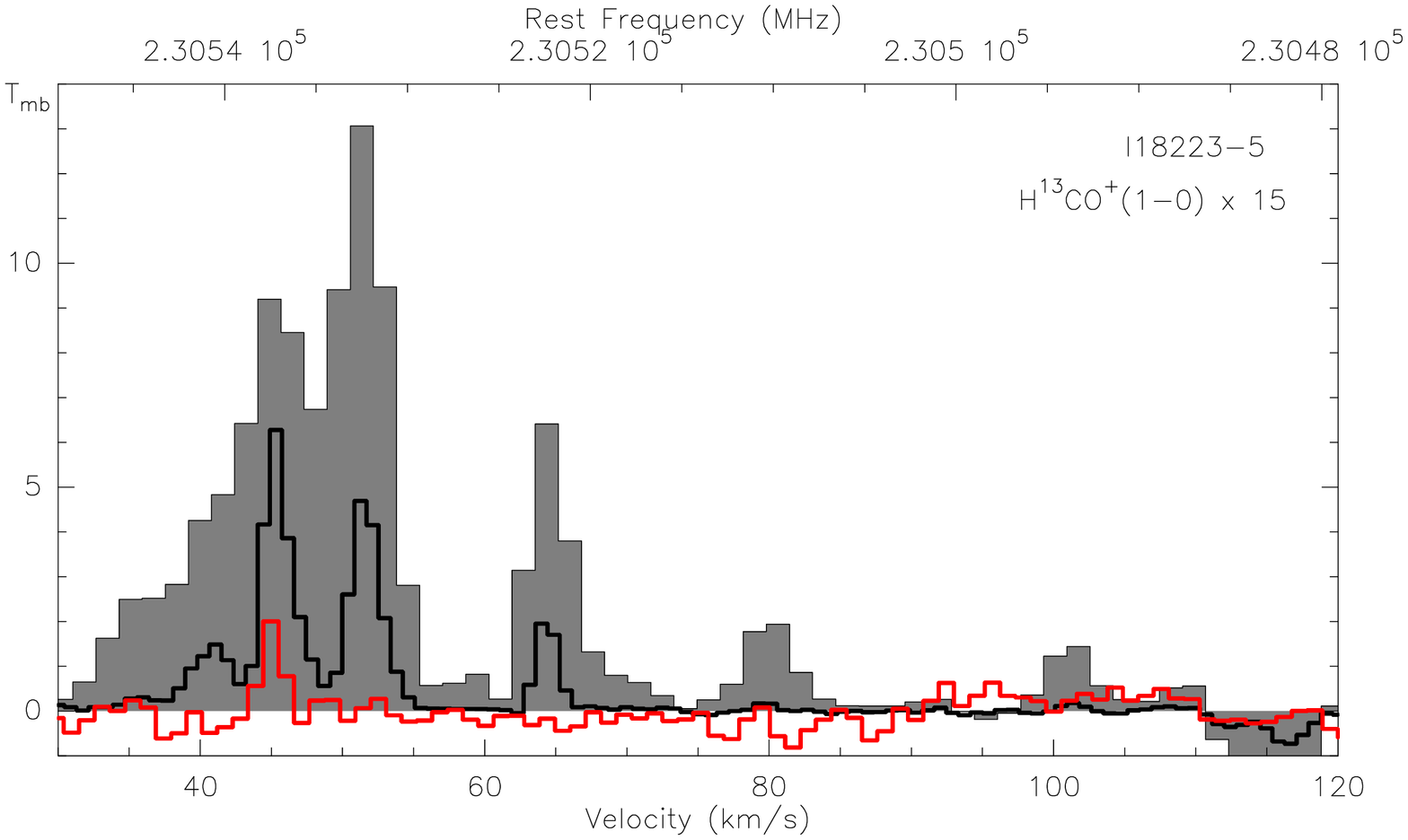}
\includegraphics[angle=-90,width=5.4cm]{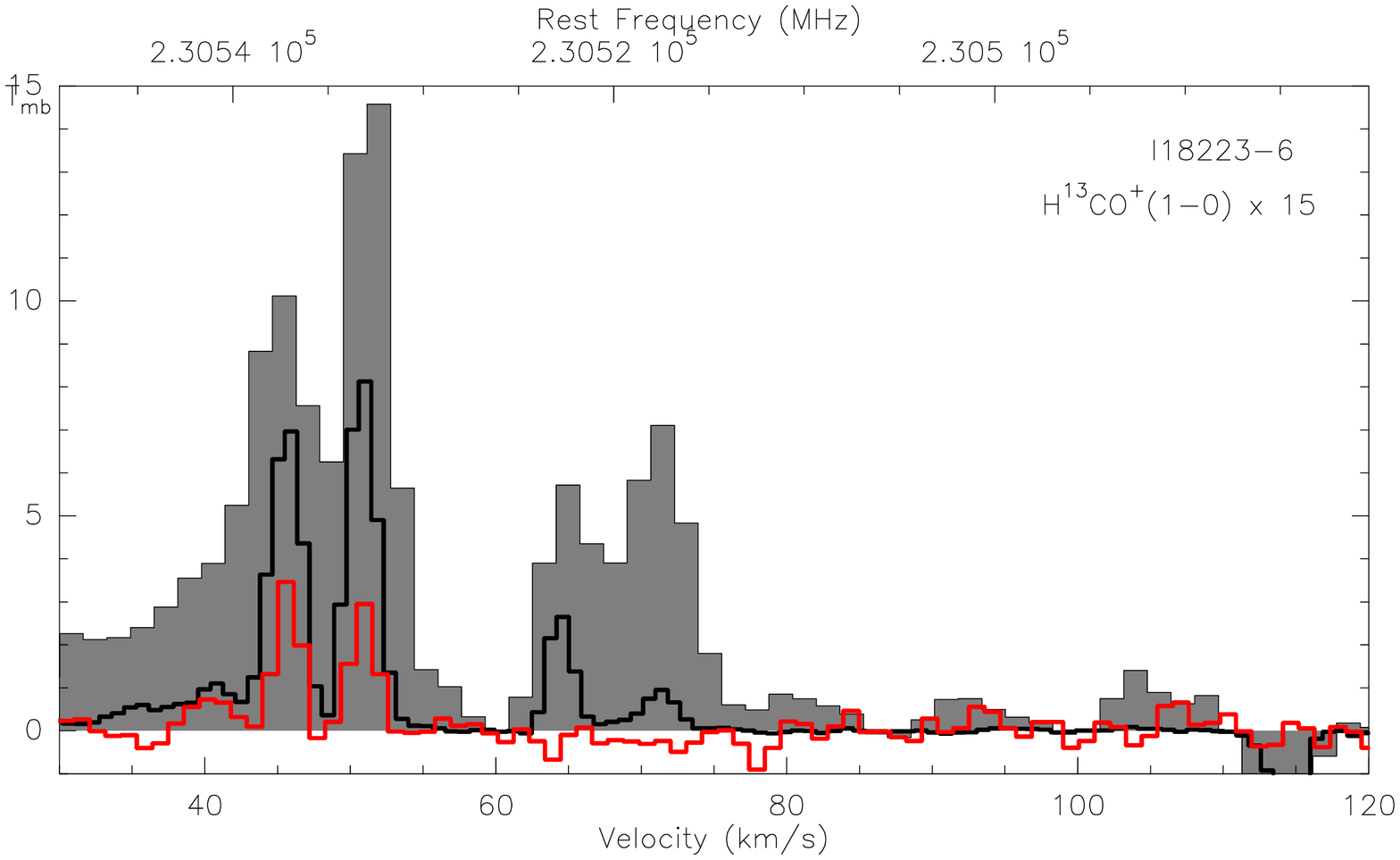}\\
\includegraphics[angle=-90,width=5.4cm]{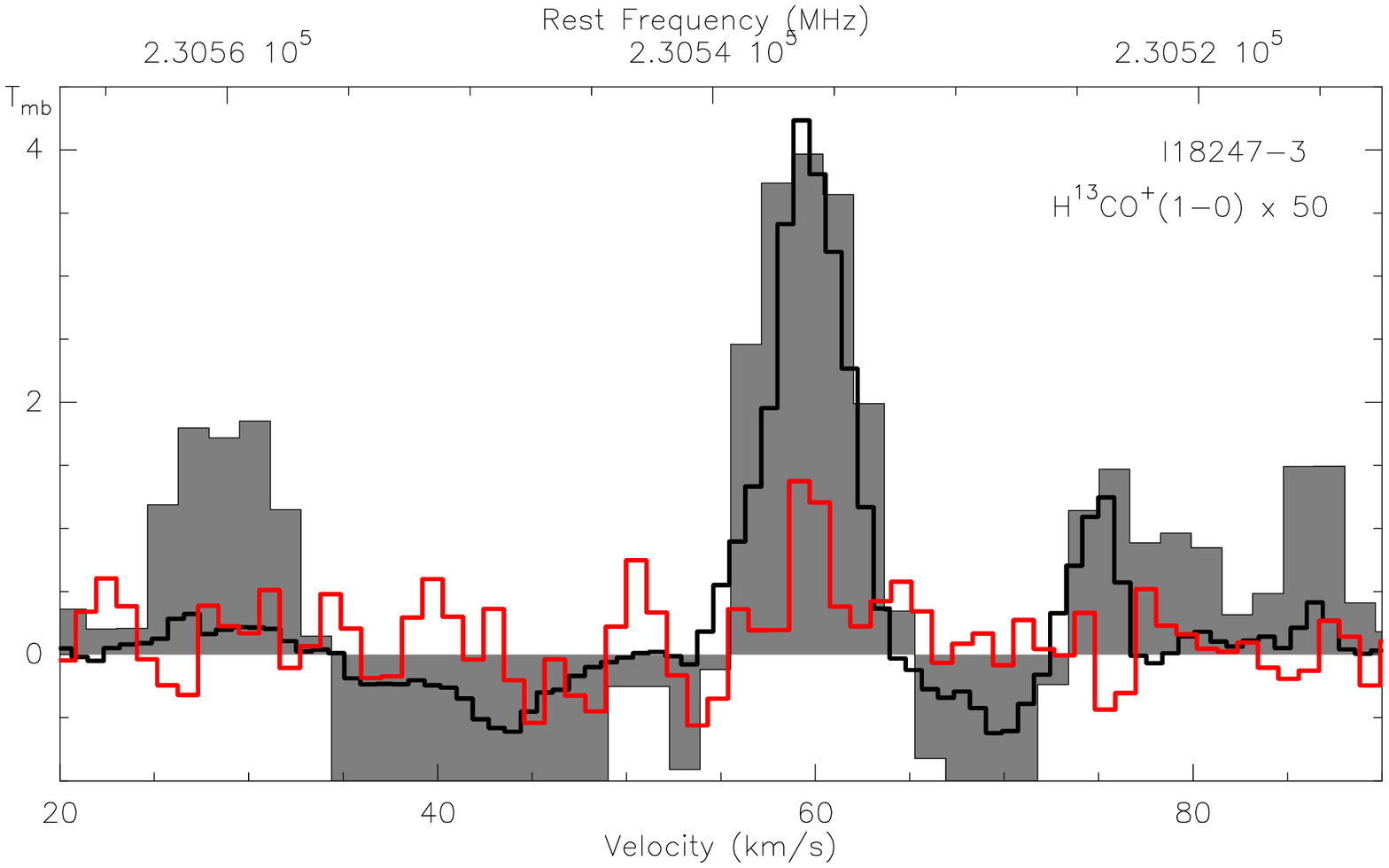}
\includegraphics[angle=-90,width=5.4cm]{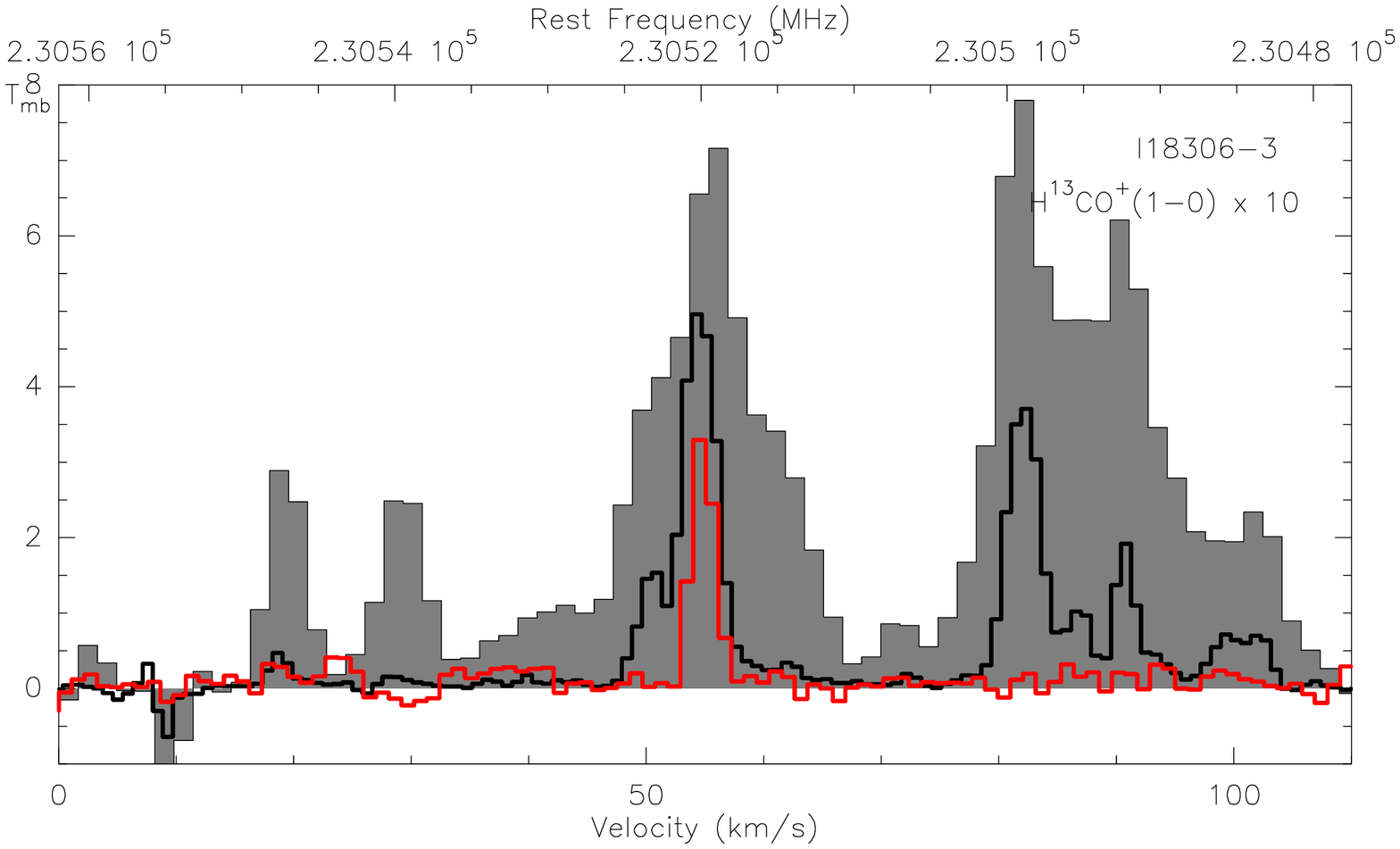}
\includegraphics[angle=-90,width=5.4cm]{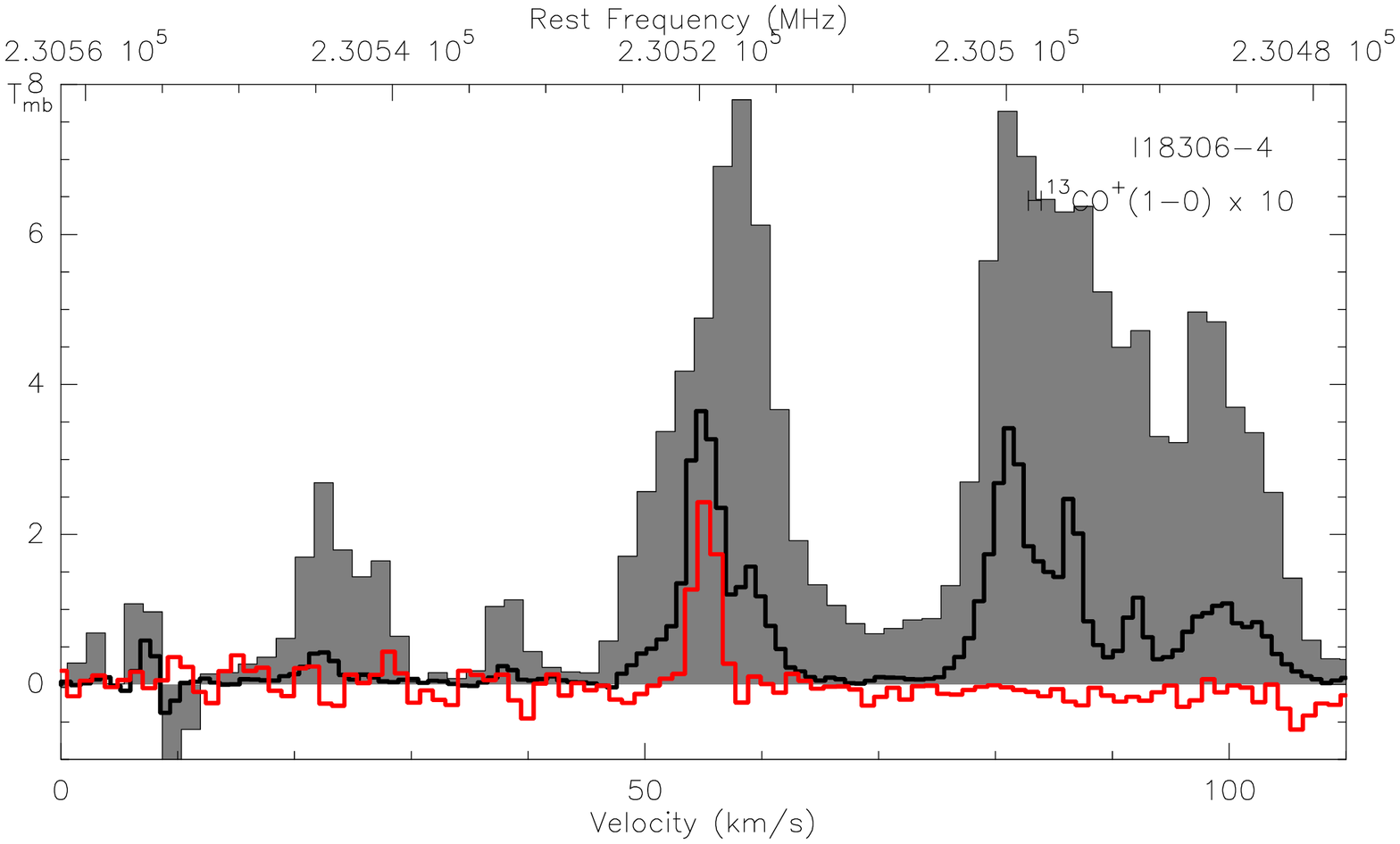}\\
\includegraphics[angle=-90,width=5.4cm]{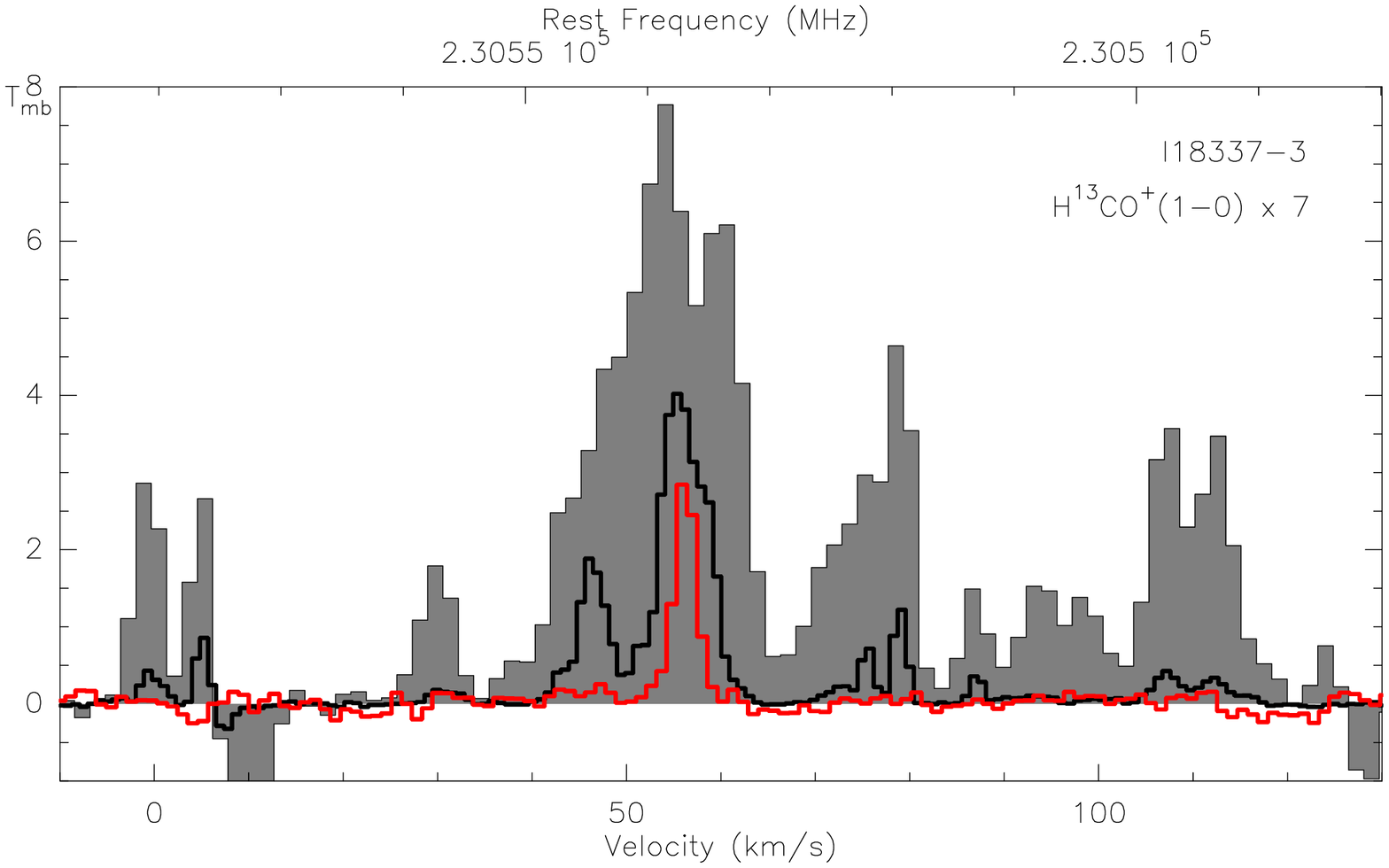}
\includegraphics[angle=-90,width=5.4cm]{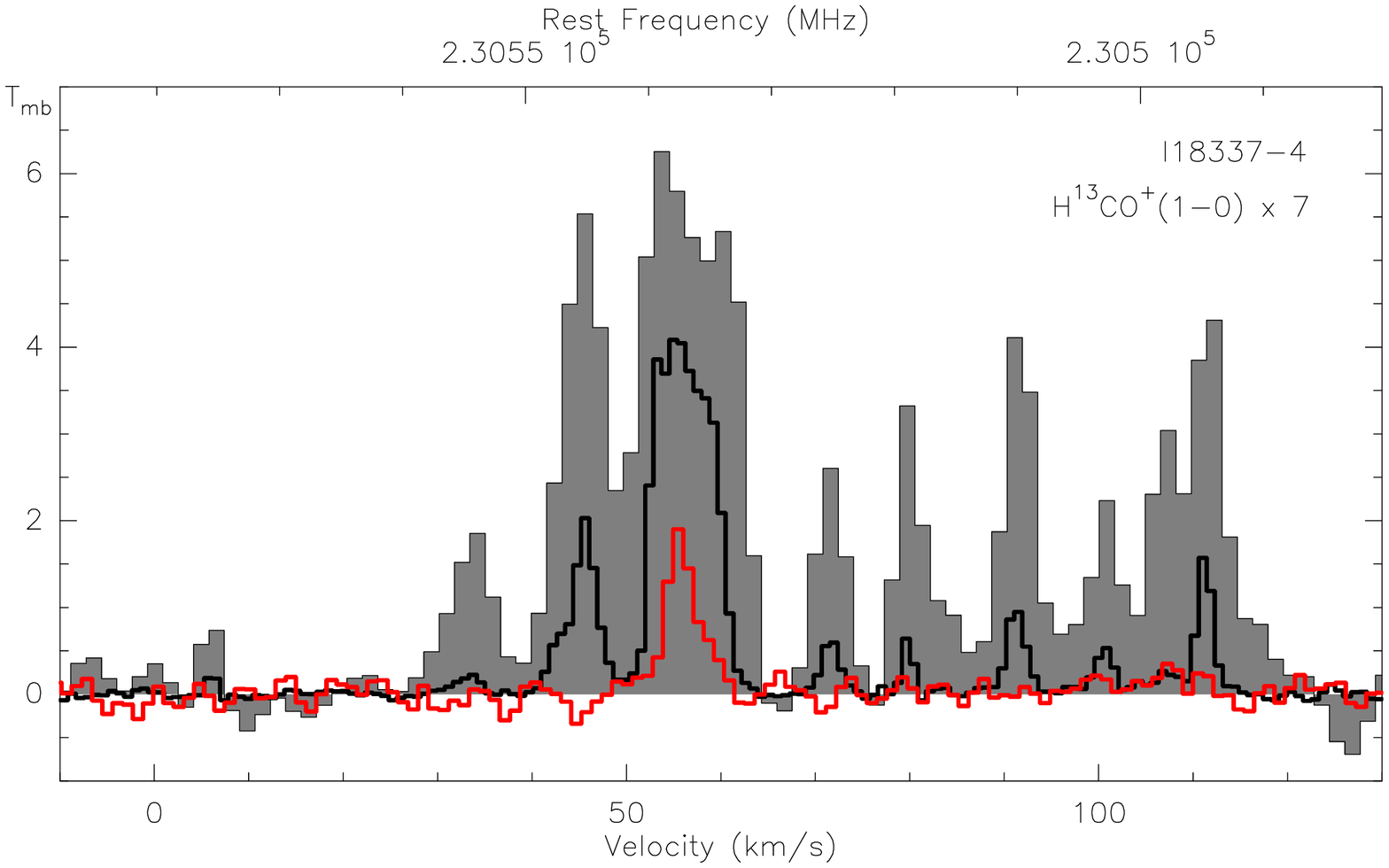}
\includegraphics[angle=-90,width=5.4cm]{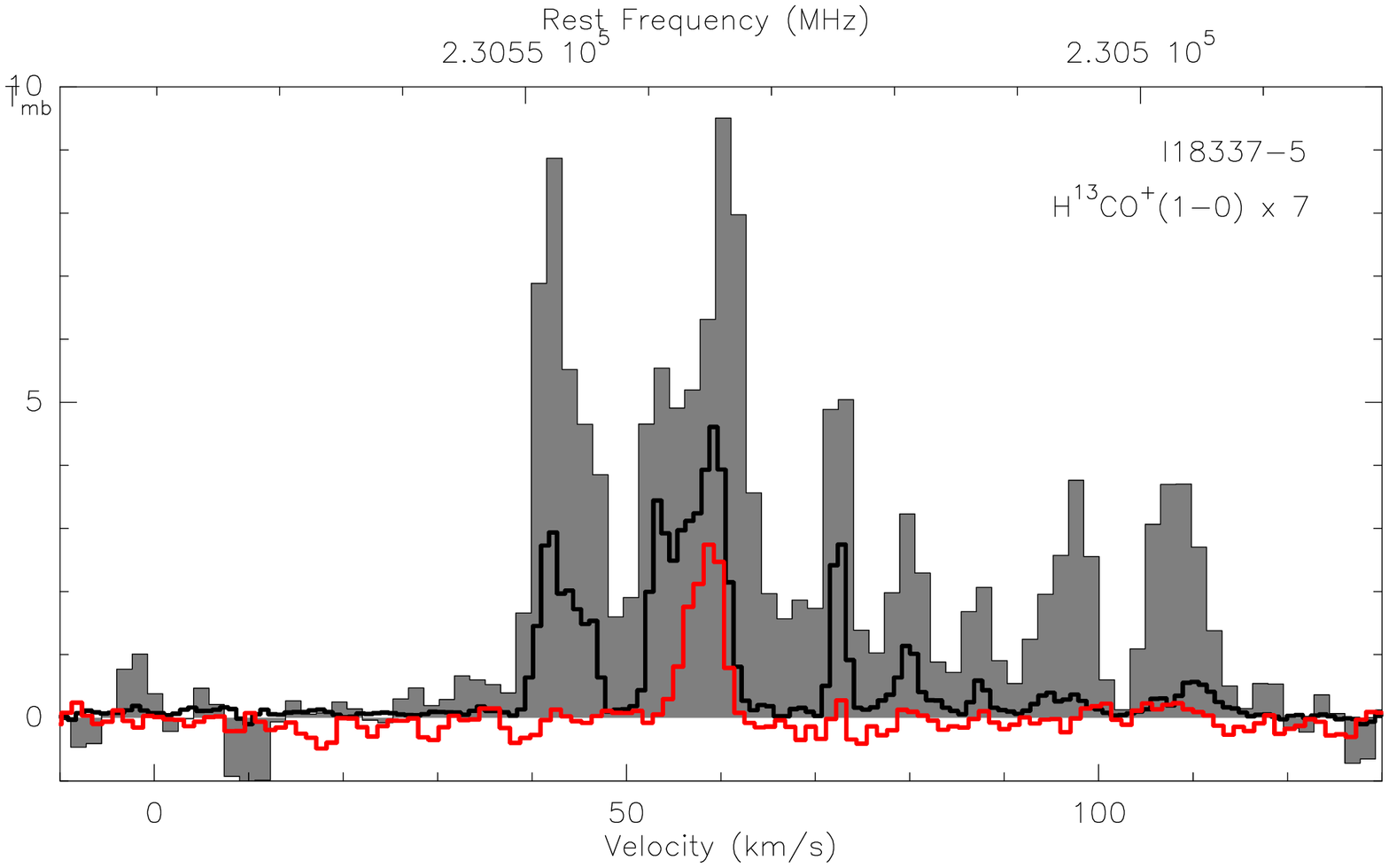}
\caption{$^{12}$CO(2--1) (grey), $^{13}$CO(1--0) (black line) and
  H$^{13}$CO$^+$(1--0) (red line) spectra toward the whole sample of
  IRDCs. The H$^{13}$CO$^+$(1--0) are scaled up by the factor labeled
  in each panel for presentation purposes.}
\label{sample1}
\end{figure*}

\begin{figure*}[h]
\includegraphics[angle=-90,width=5.4cm]{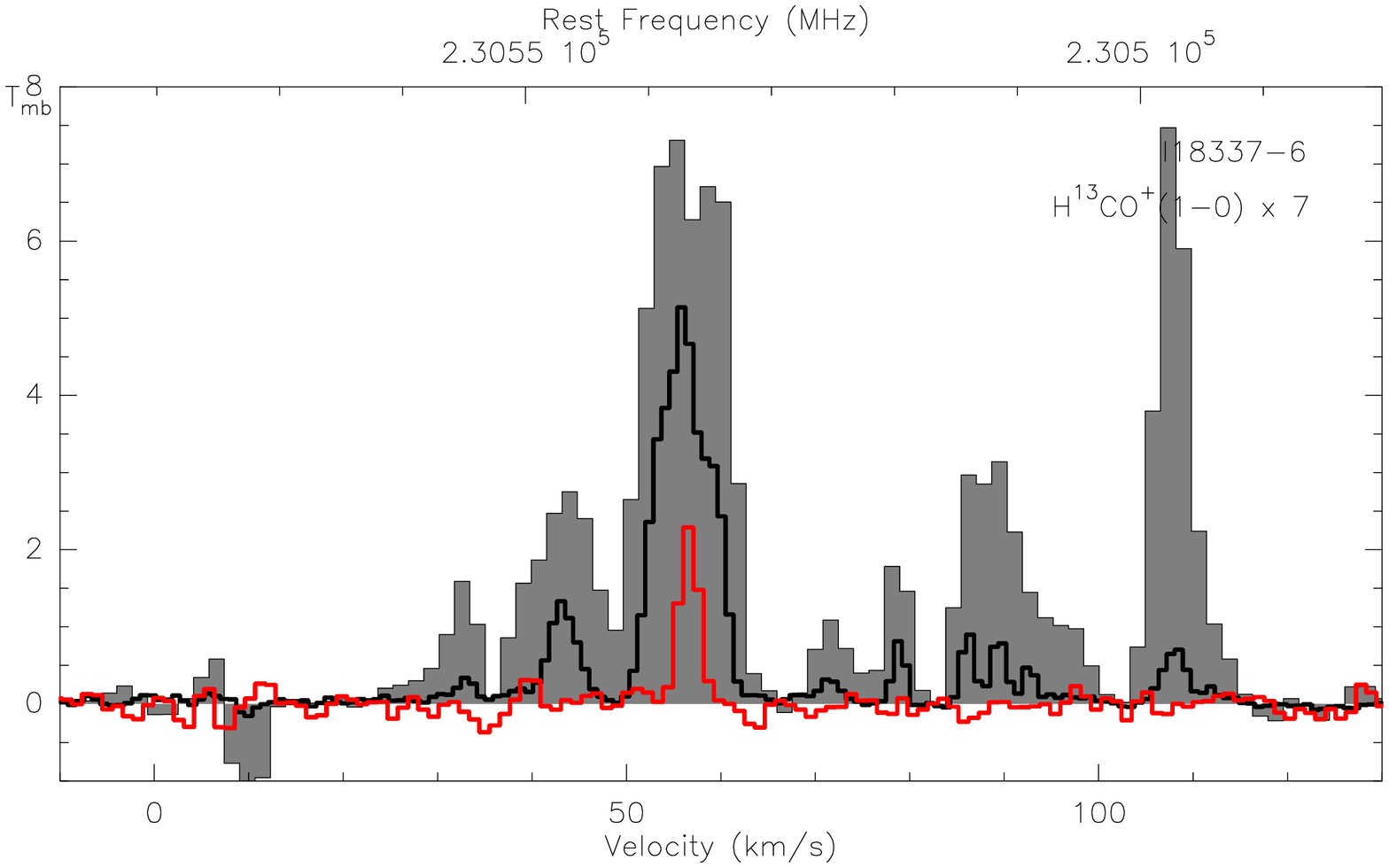}
\includegraphics[angle=-90,width=5.4cm]{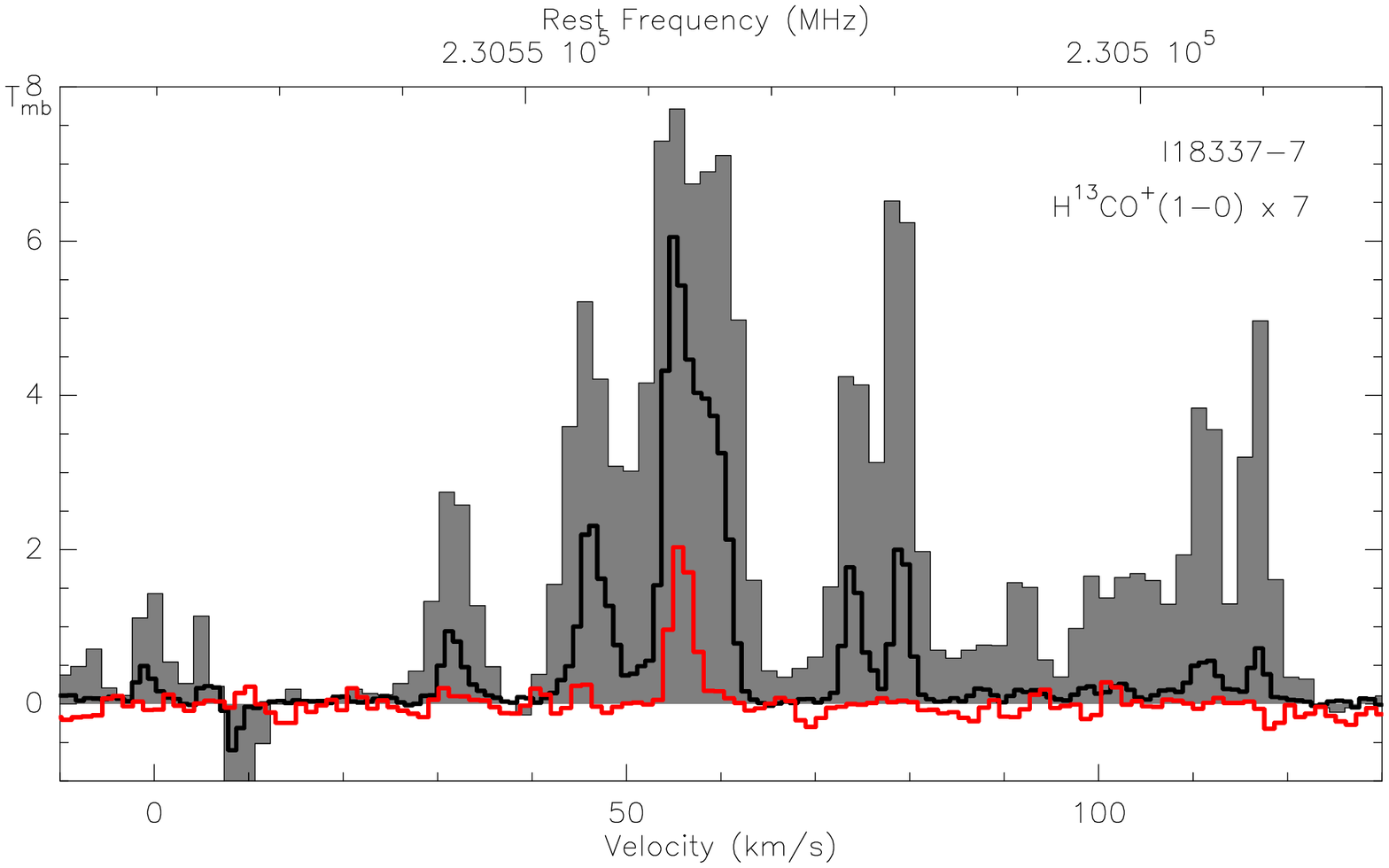}
\includegraphics[angle=-90,width=5.4cm]{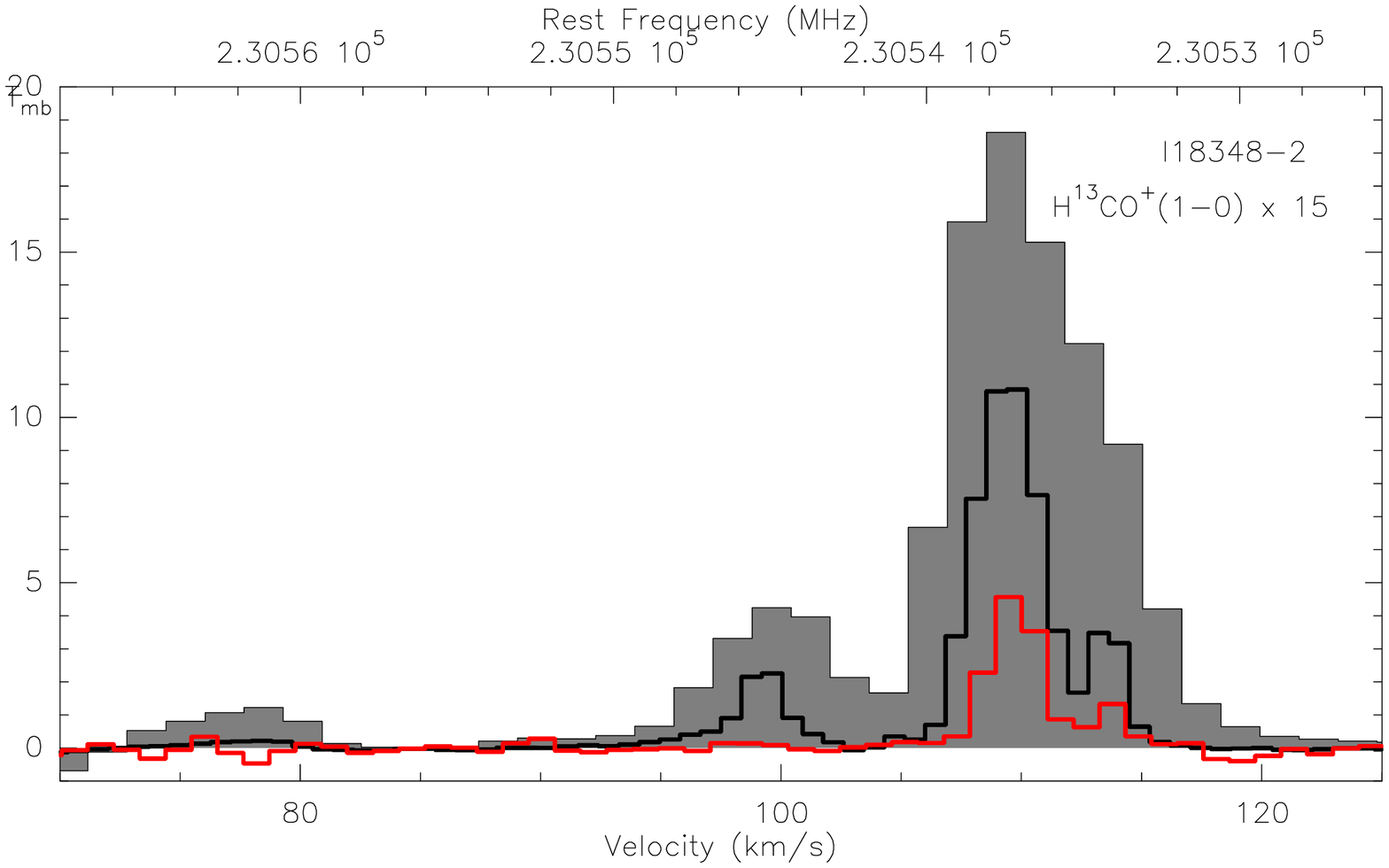}\\
\includegraphics[angle=-90,width=5.4cm]{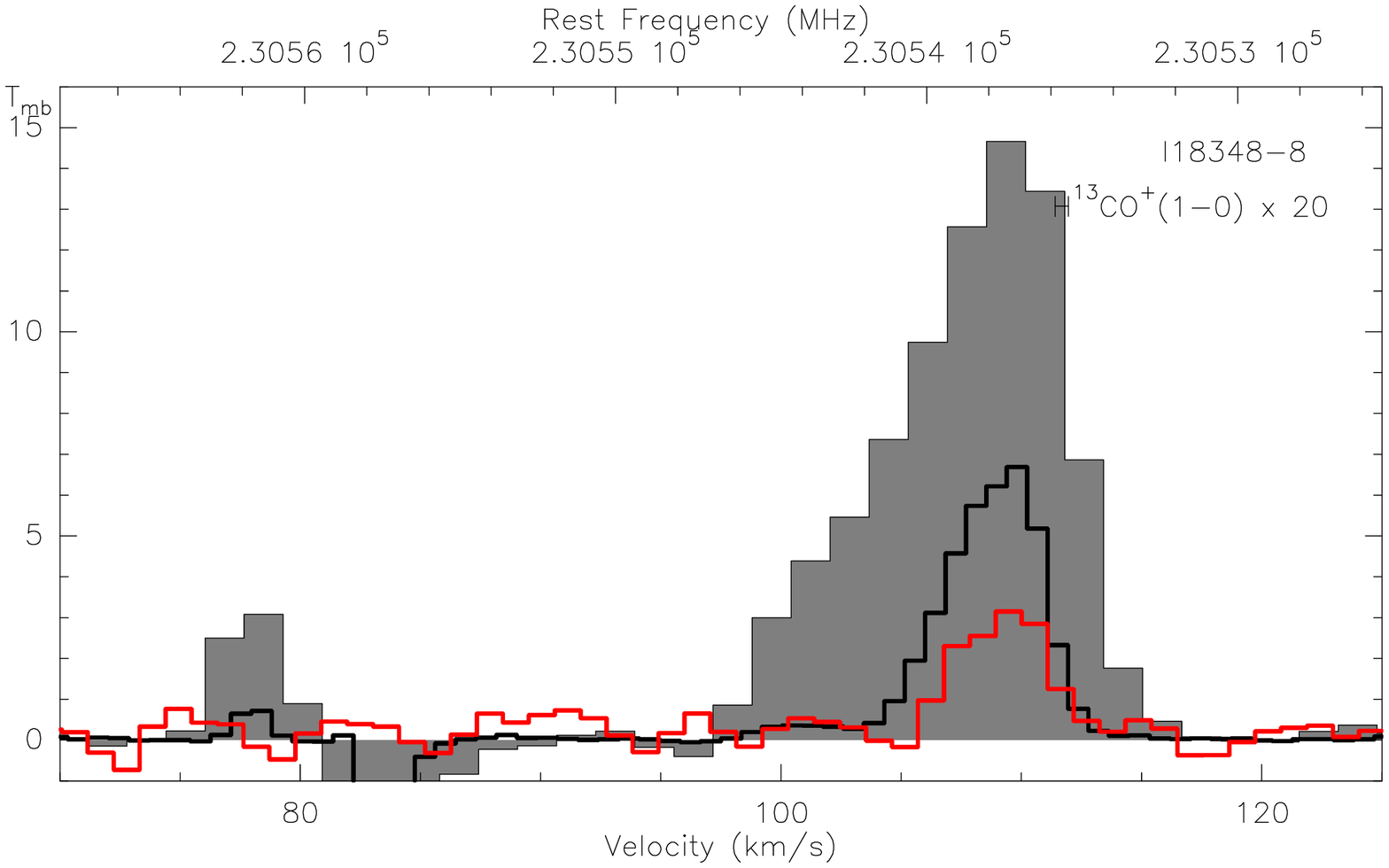}
\includegraphics[angle=-90,width=5.4cm]{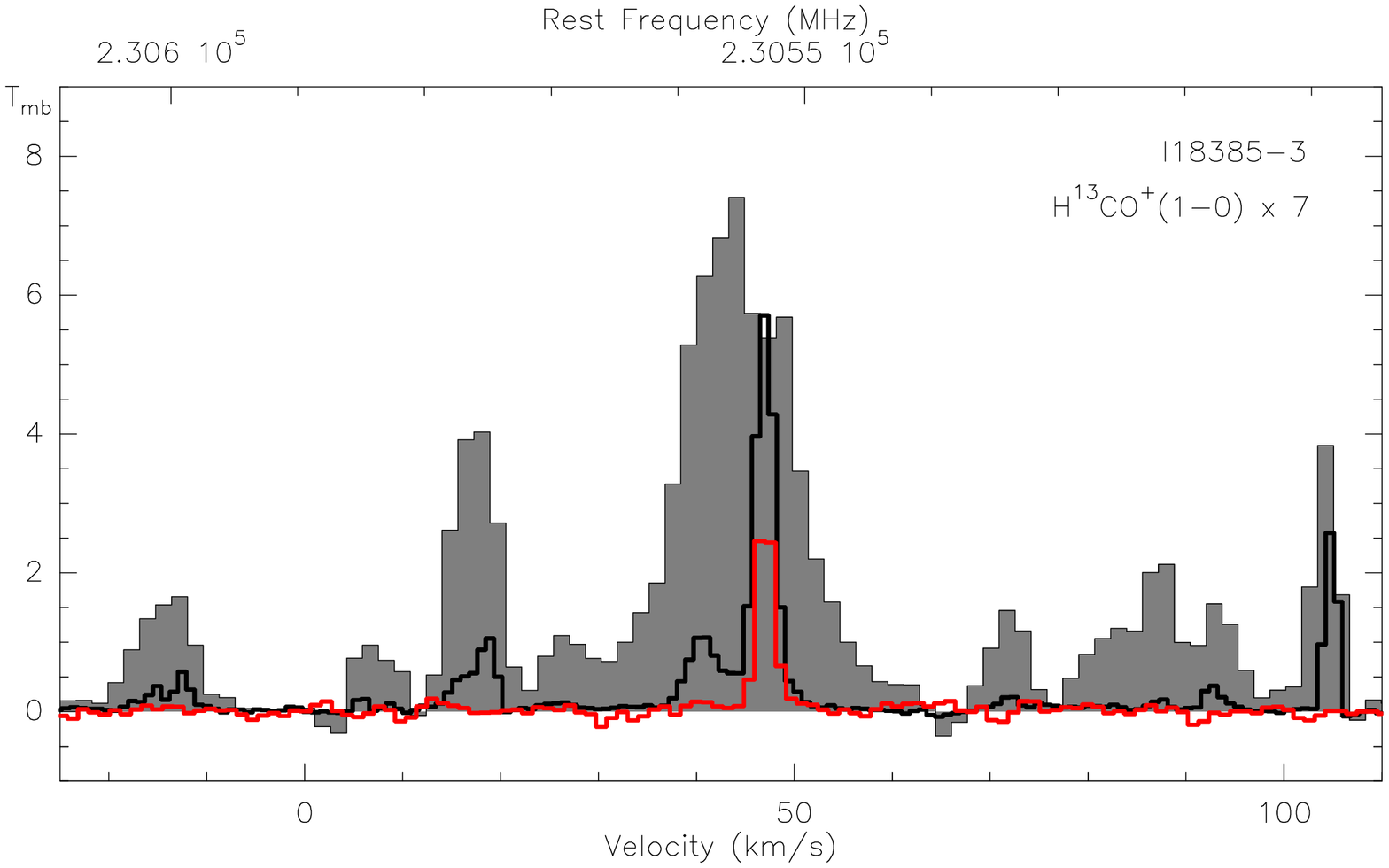}
\includegraphics[angle=-90,width=5.4cm]{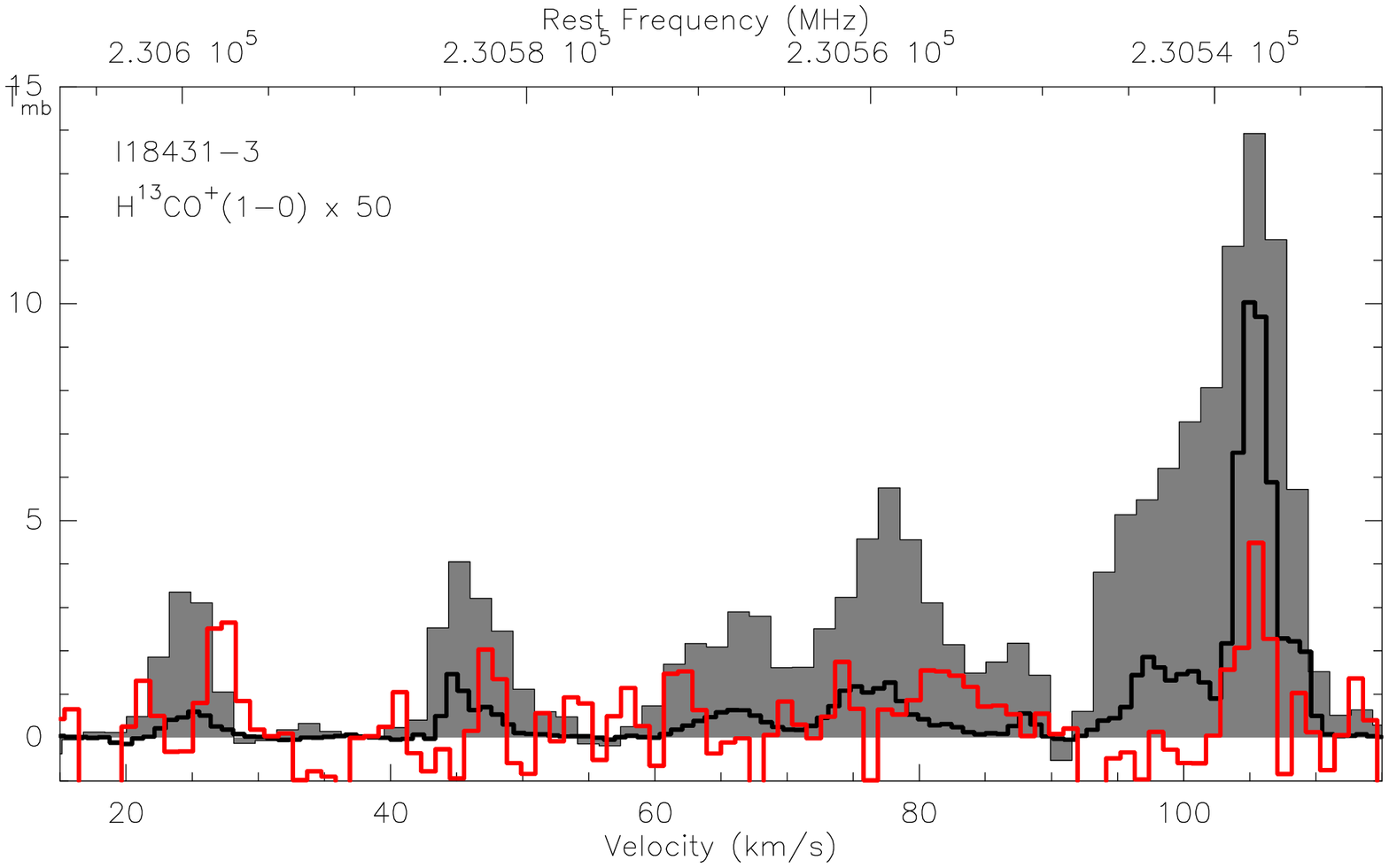}\\
\includegraphics[angle=-90,width=5.4cm]{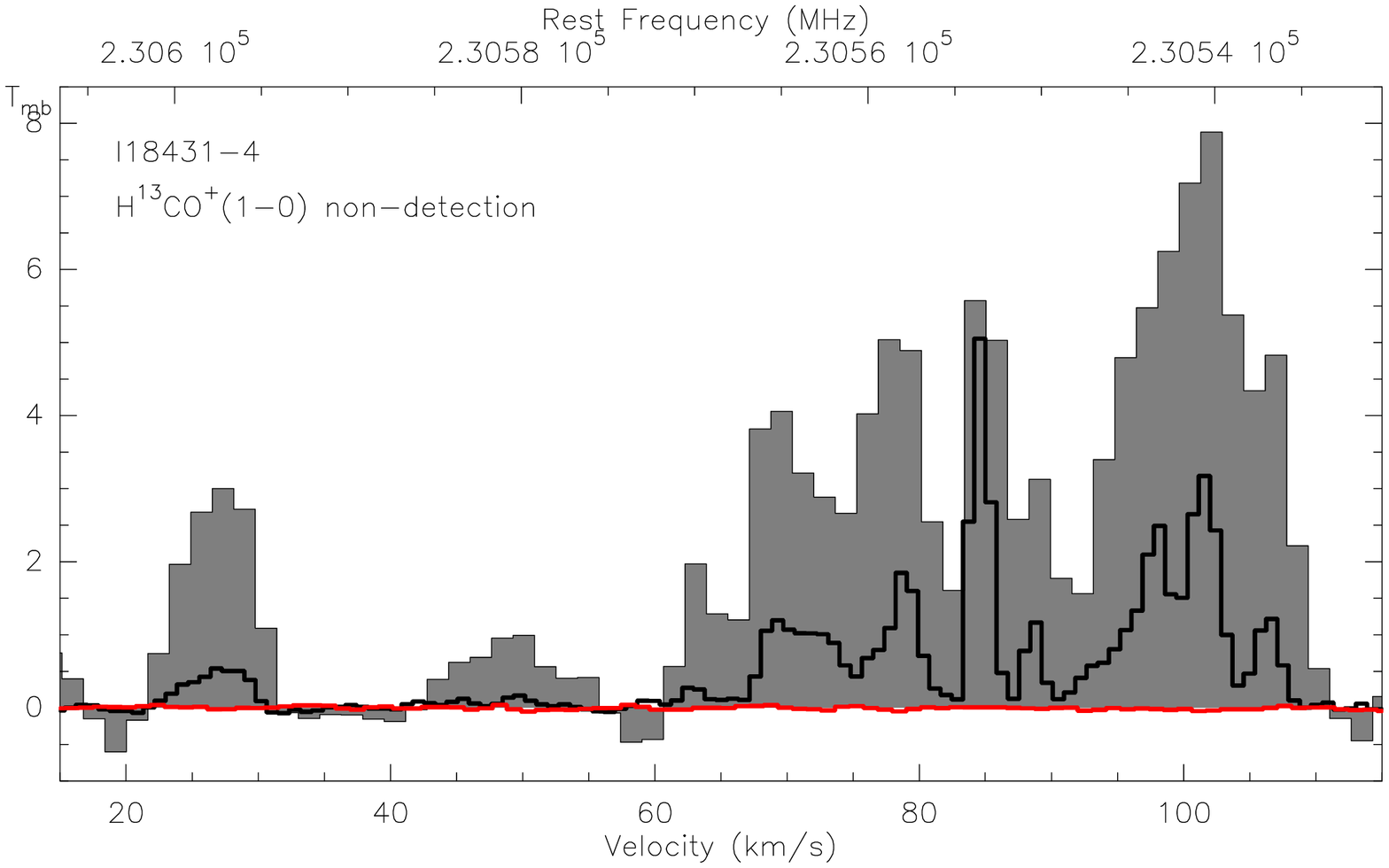}
\includegraphics[angle=-90,width=5.4cm]{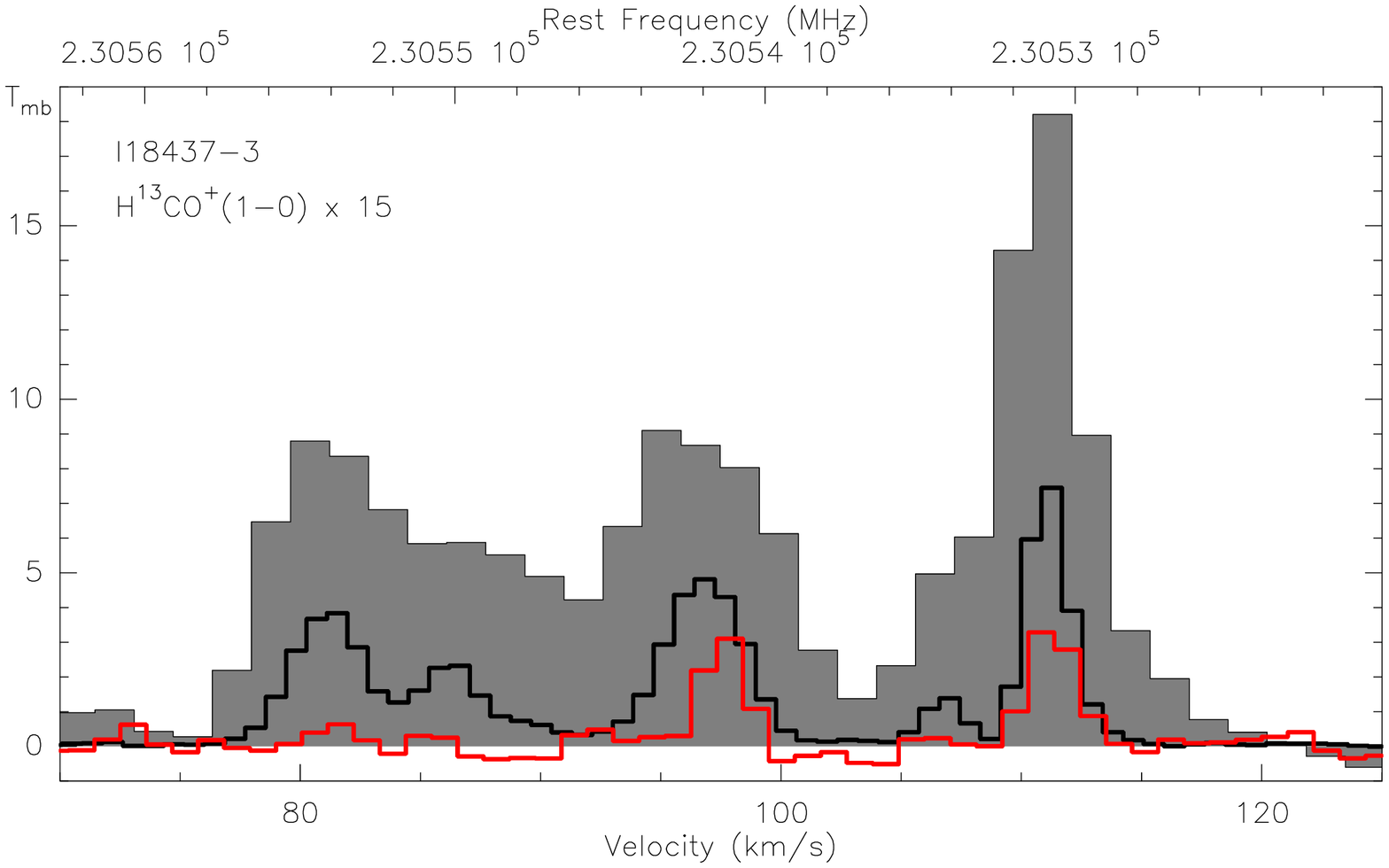}
\includegraphics[angle=-90,width=5.4cm]{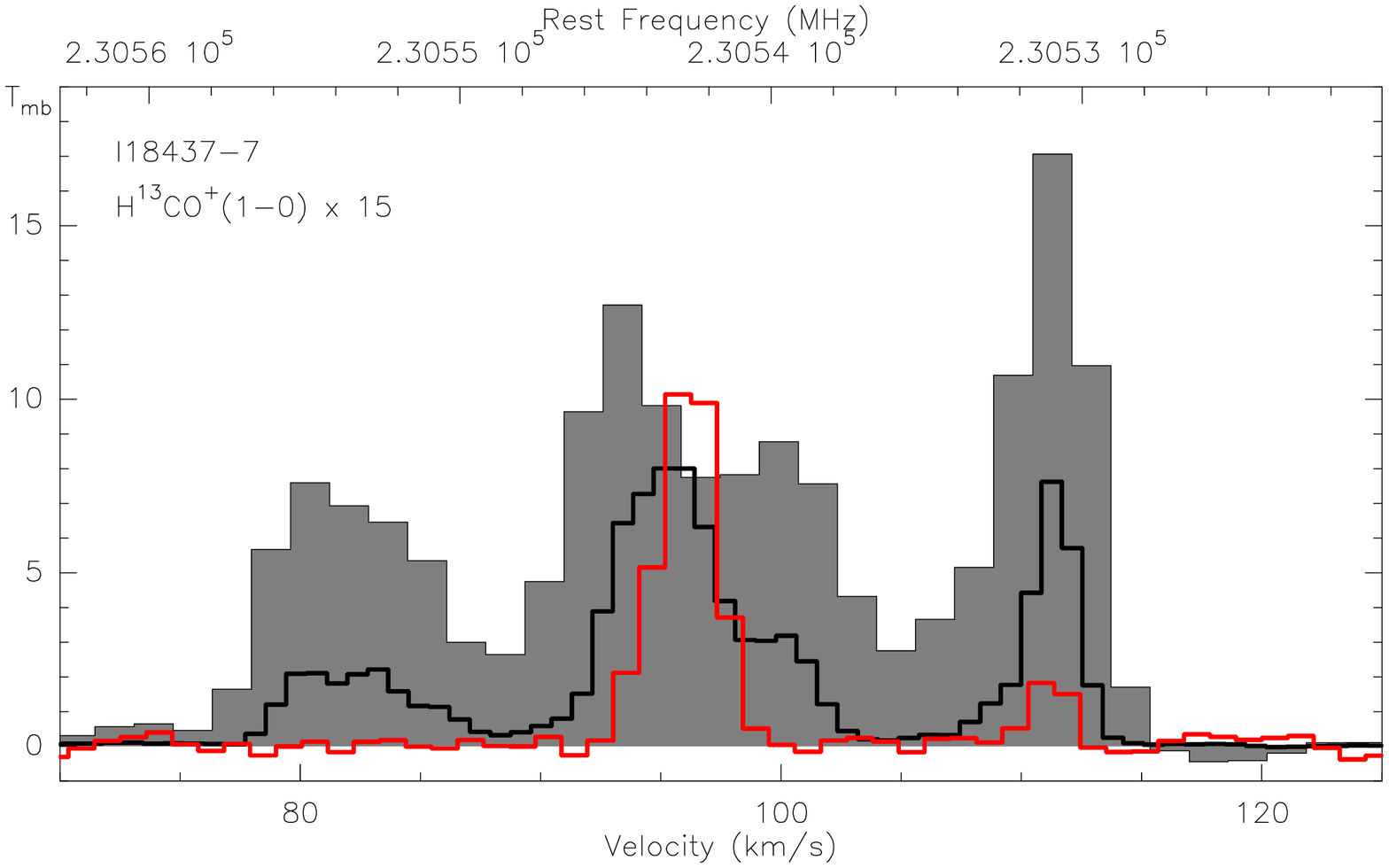}\\
\includegraphics[angle=-90,width=5.4cm]{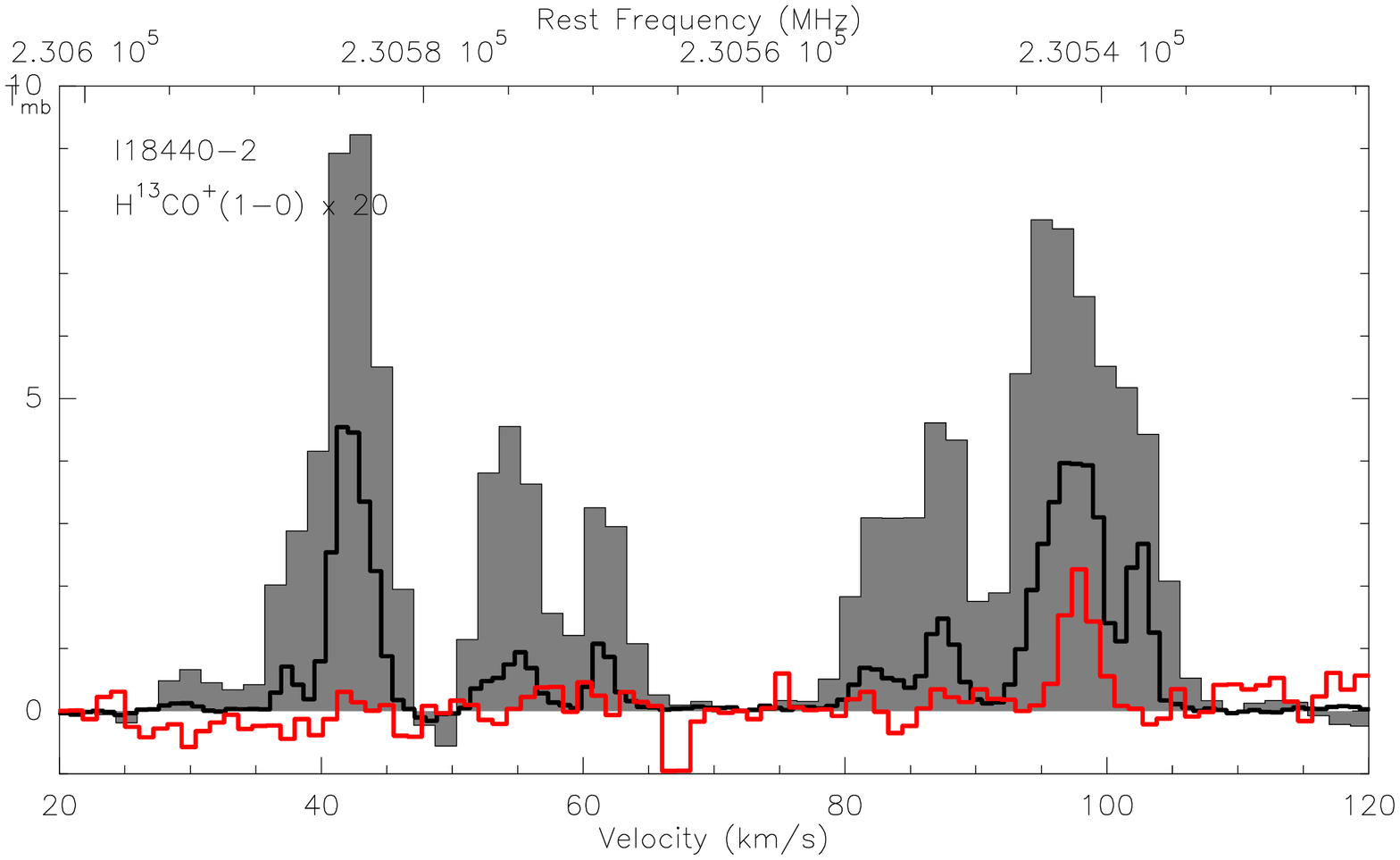}
\includegraphics[angle=-90,width=5.4cm]{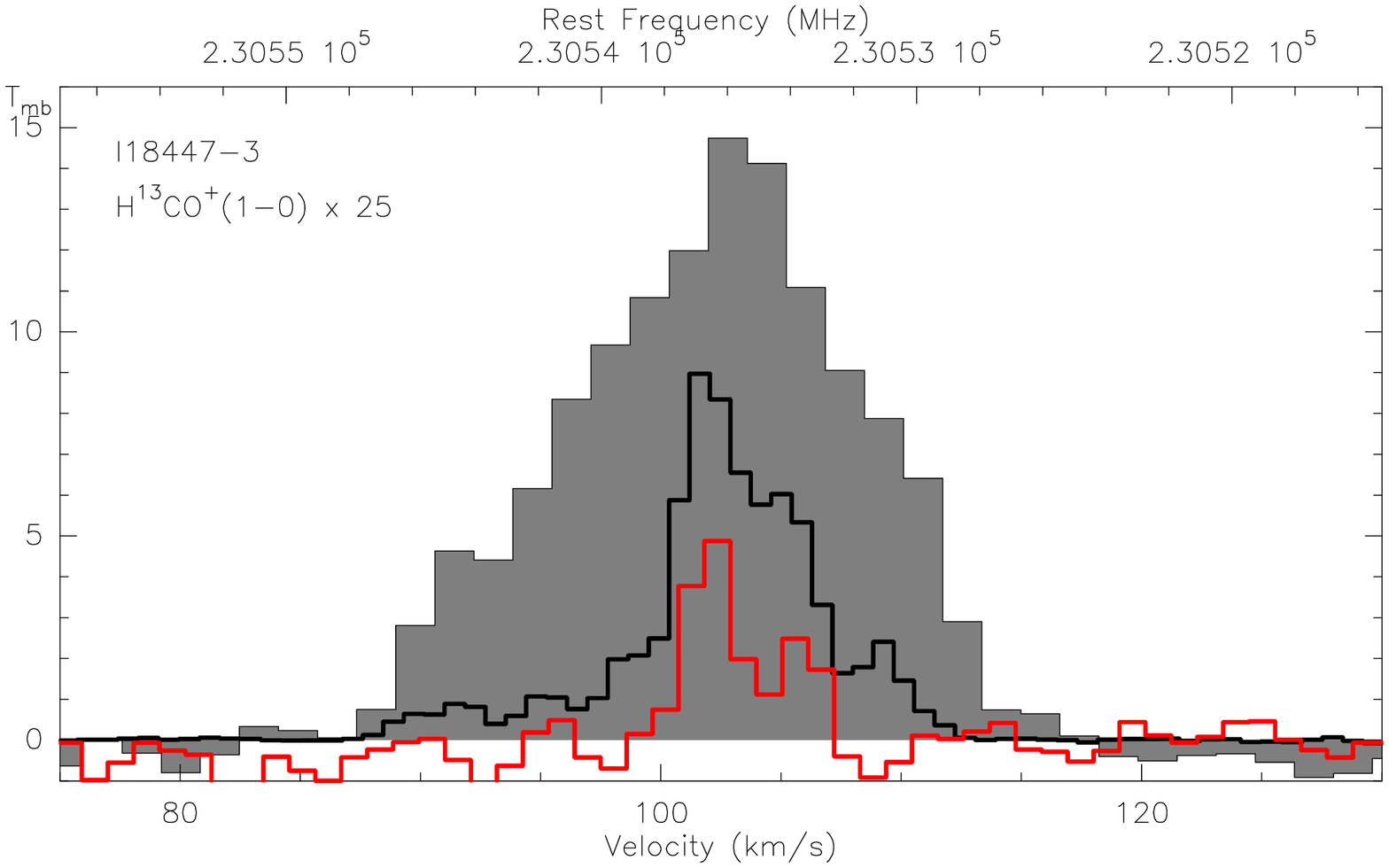}
\includegraphics[angle=-90,width=5.4cm]{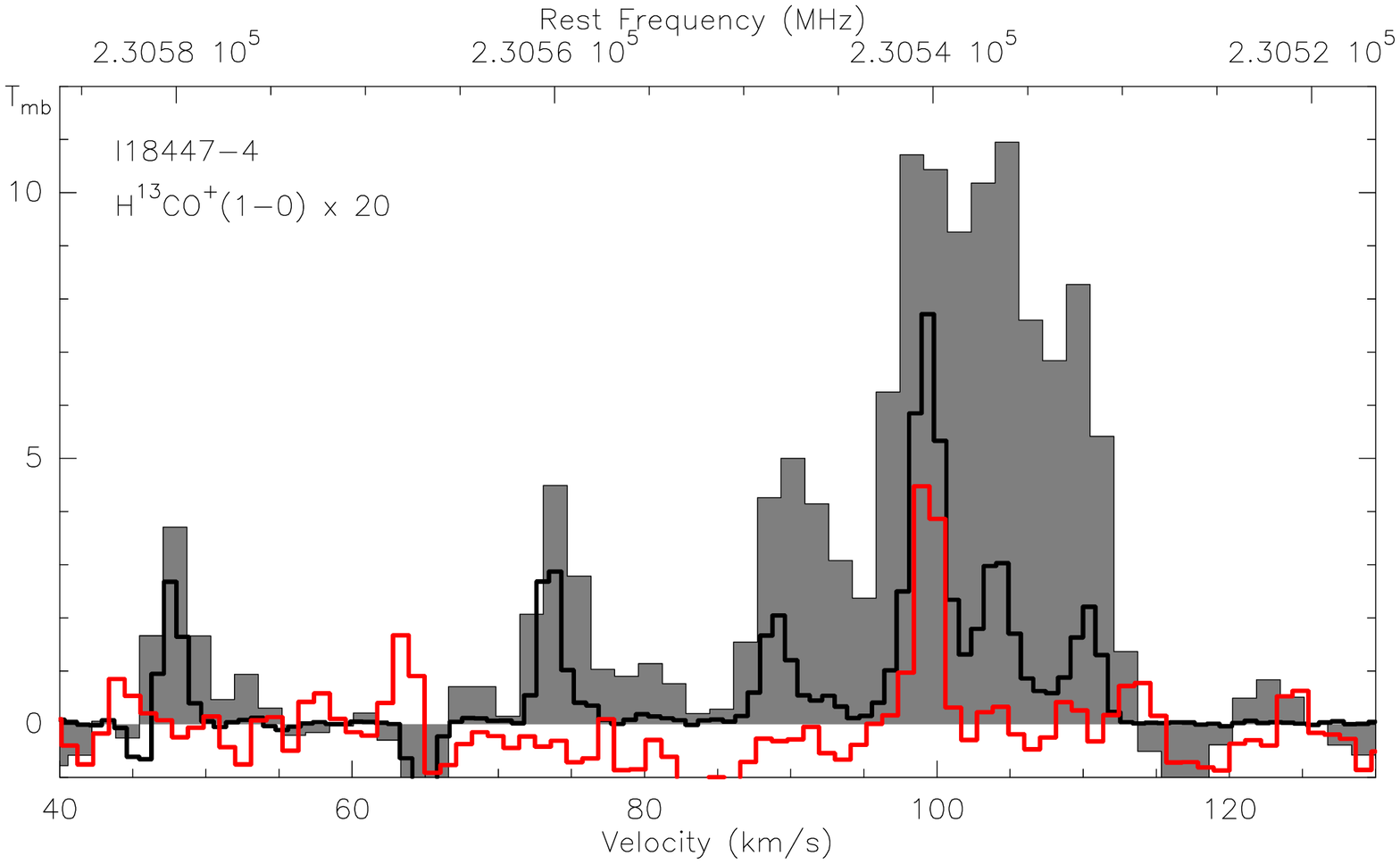}\\
\includegraphics[angle=-90,width=5.4cm]{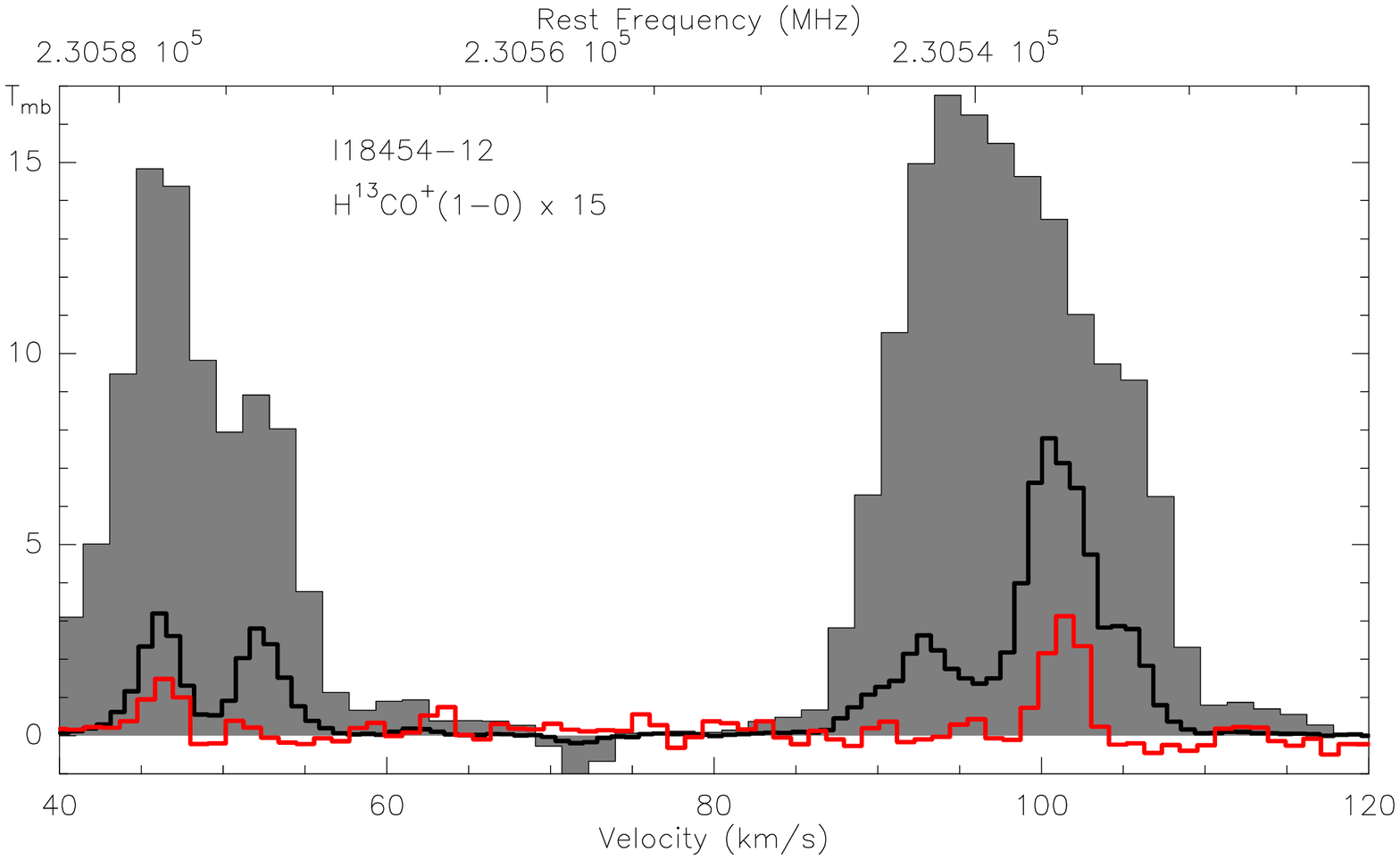}
\includegraphics[angle=-90,width=5.4cm]{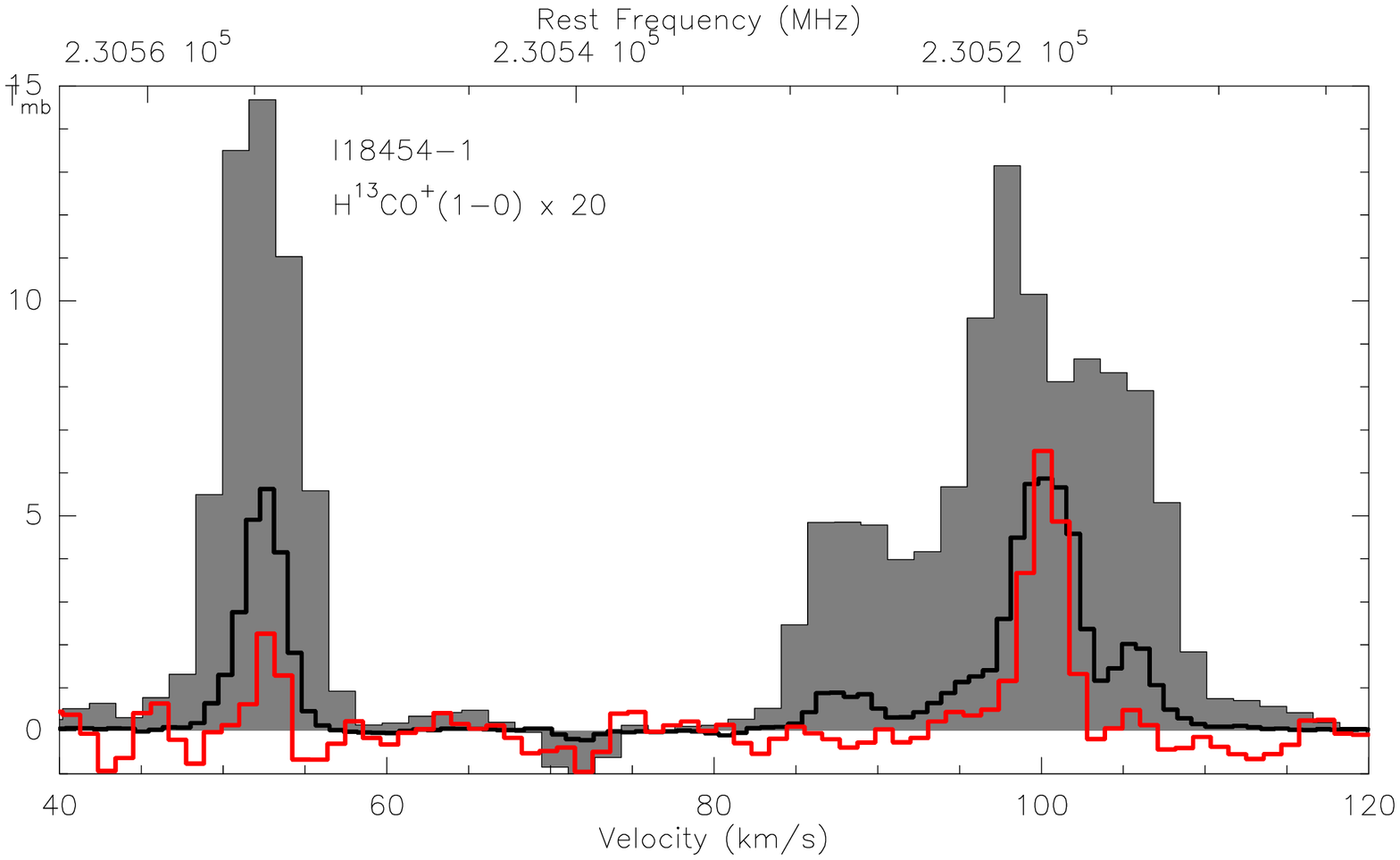}
\includegraphics[angle=-90,width=5.4cm]{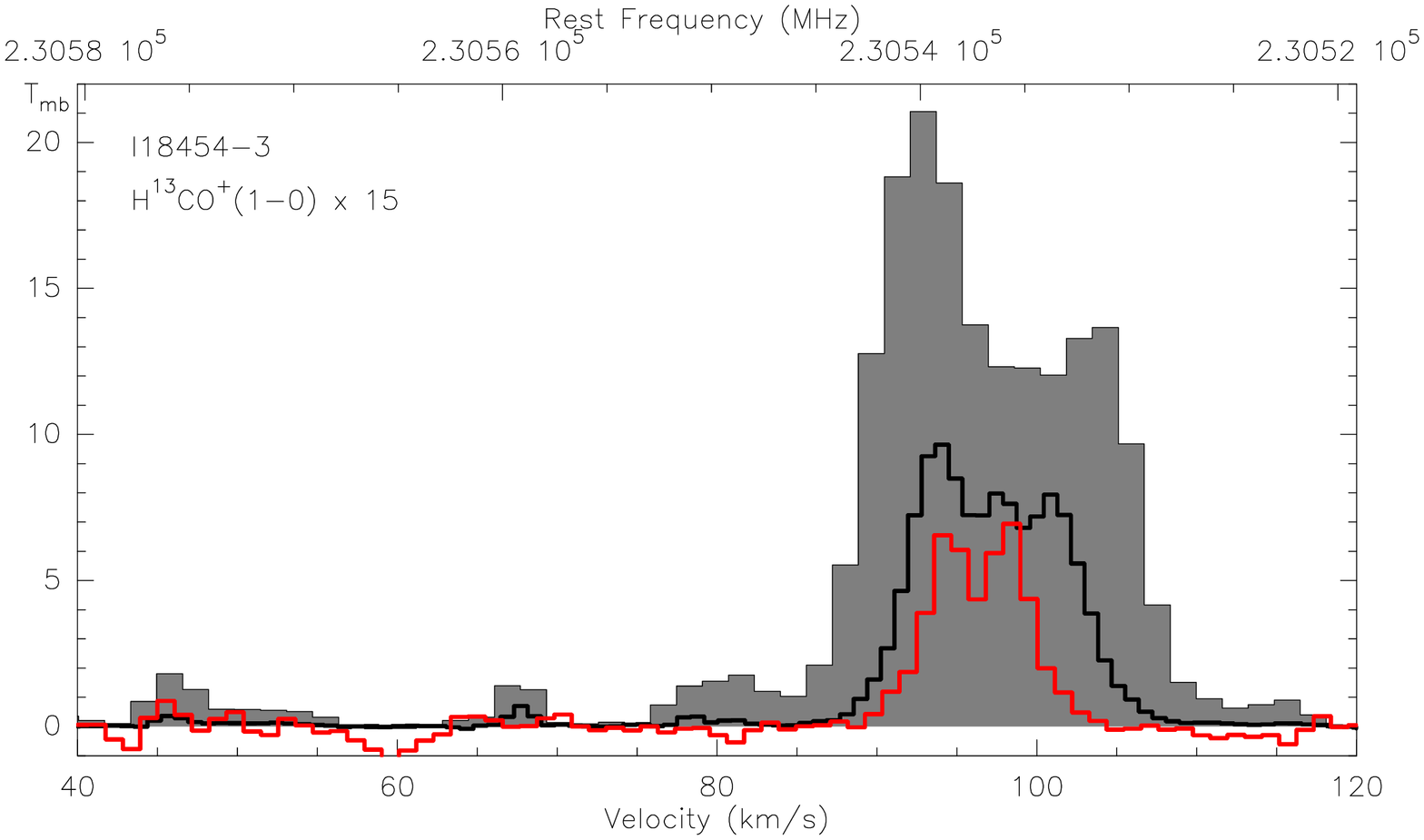}\\
\includegraphics[angle=-90,width=5.4cm]{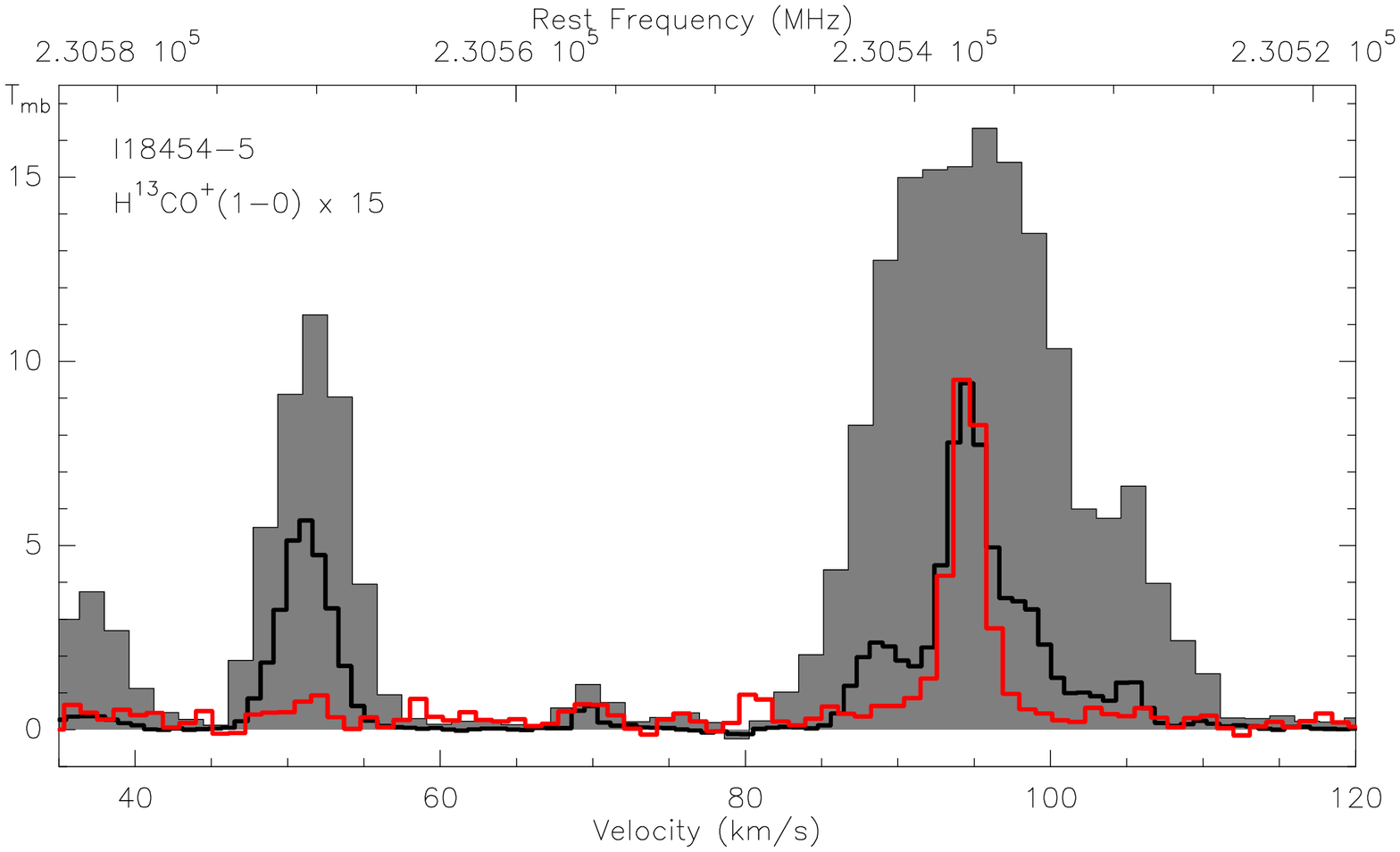}
\includegraphics[angle=-90,width=5.4cm]{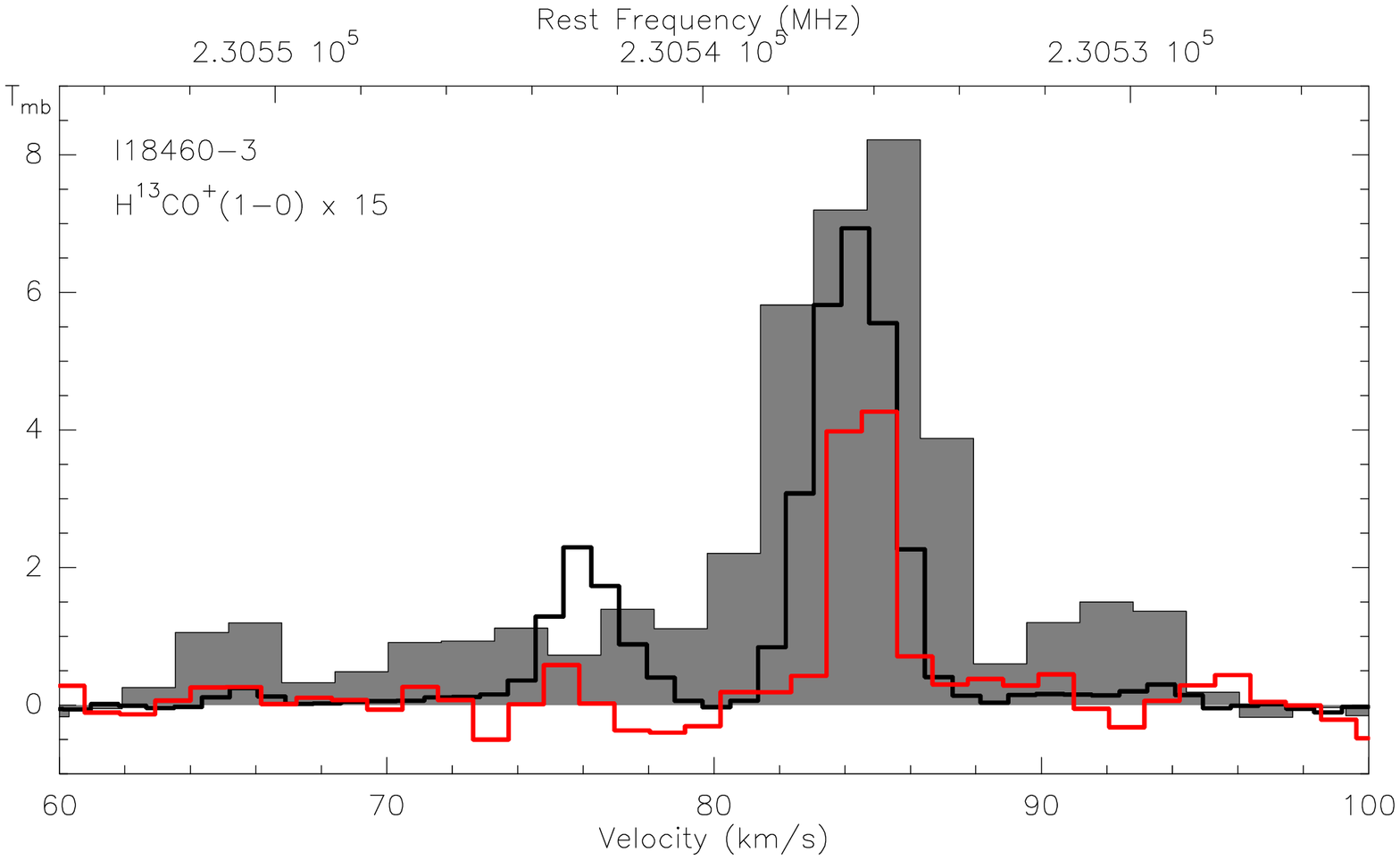}
\includegraphics[angle=-90,width=5.4cm]{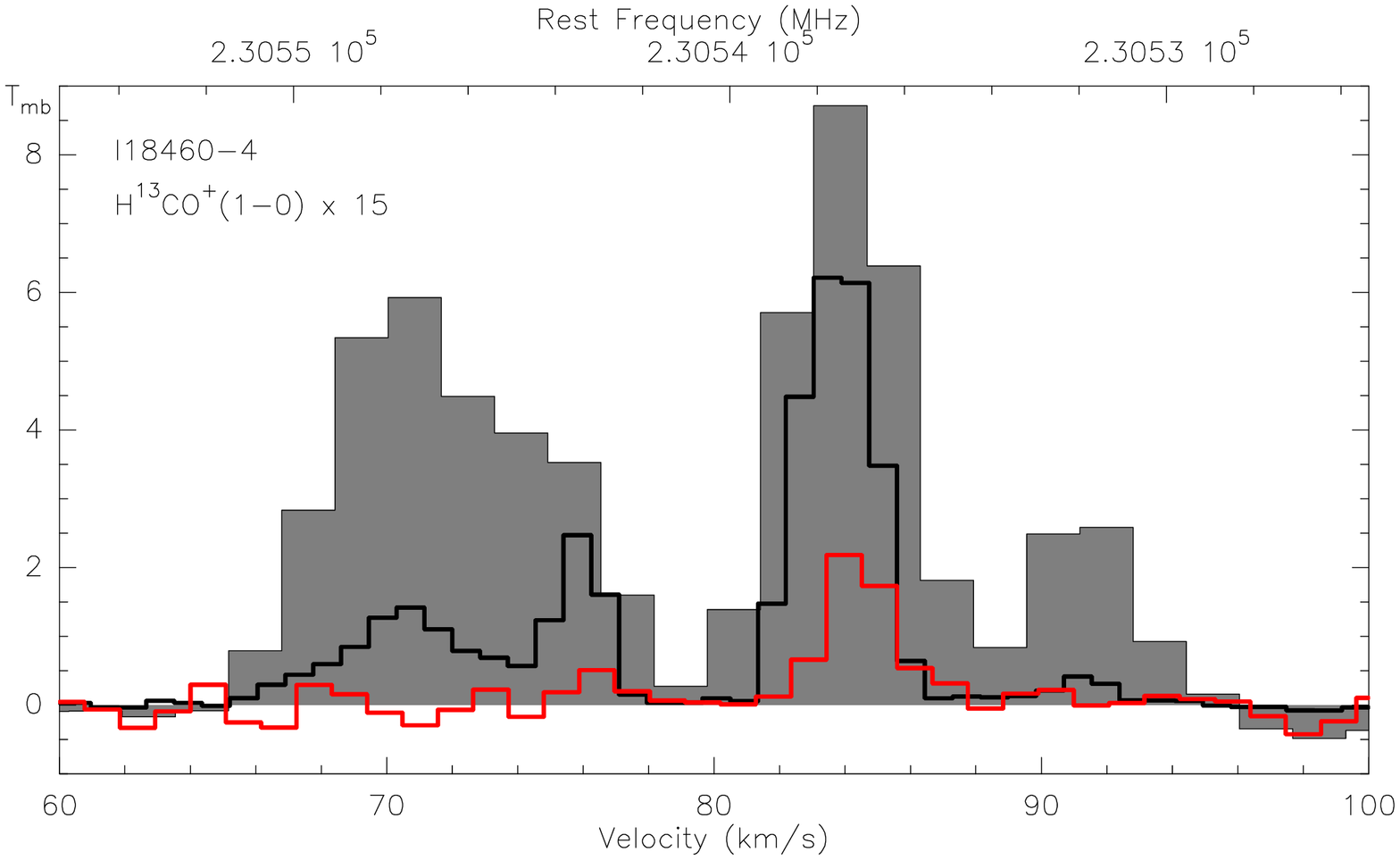}
\caption{Figure \ref{sample1} continued.}
\label{sample2}
\end{figure*}

\begin{figure*}[h]
\includegraphics[angle=-90,width=5.4cm]{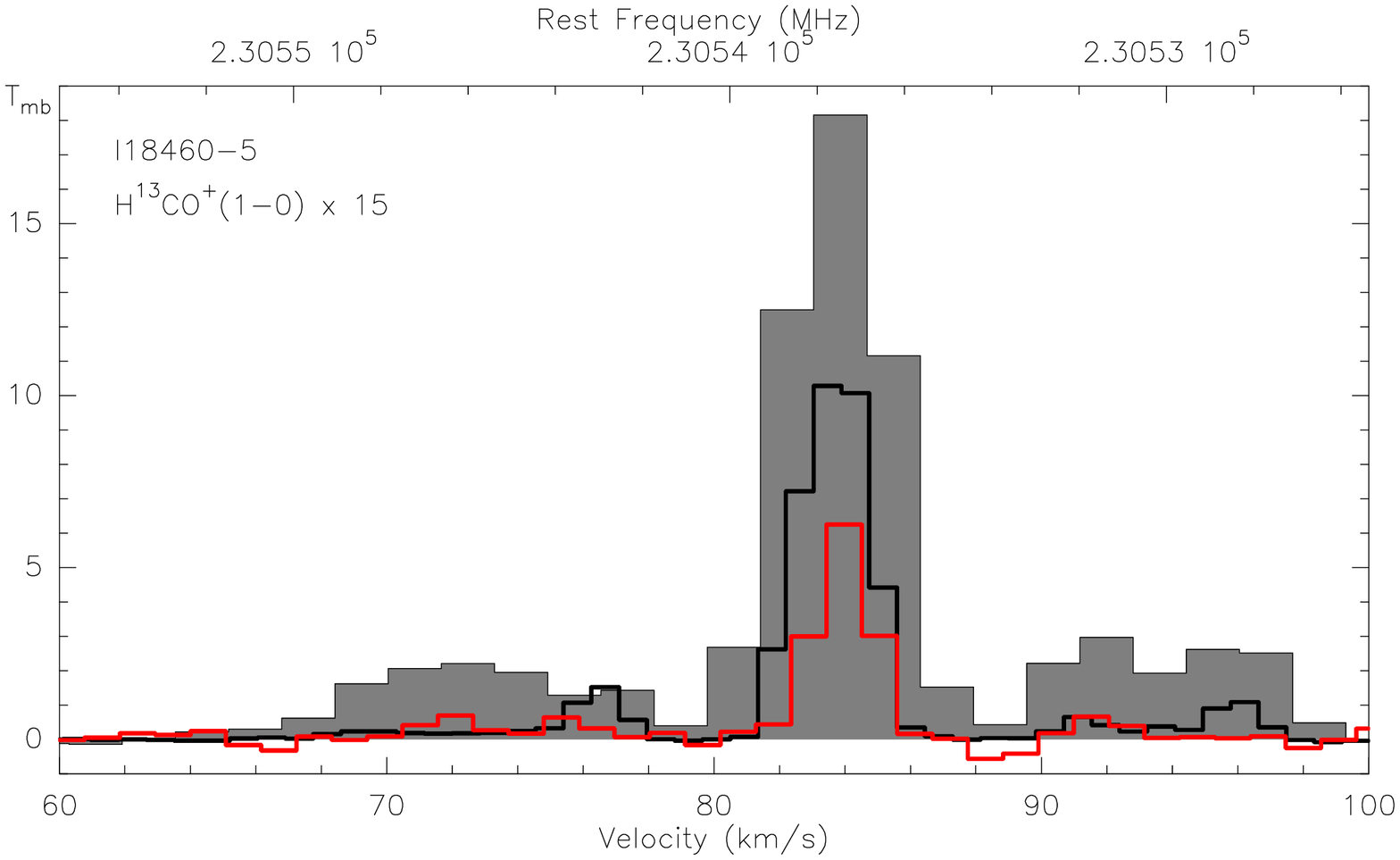}
\includegraphics[angle=-90,width=5.4cm]{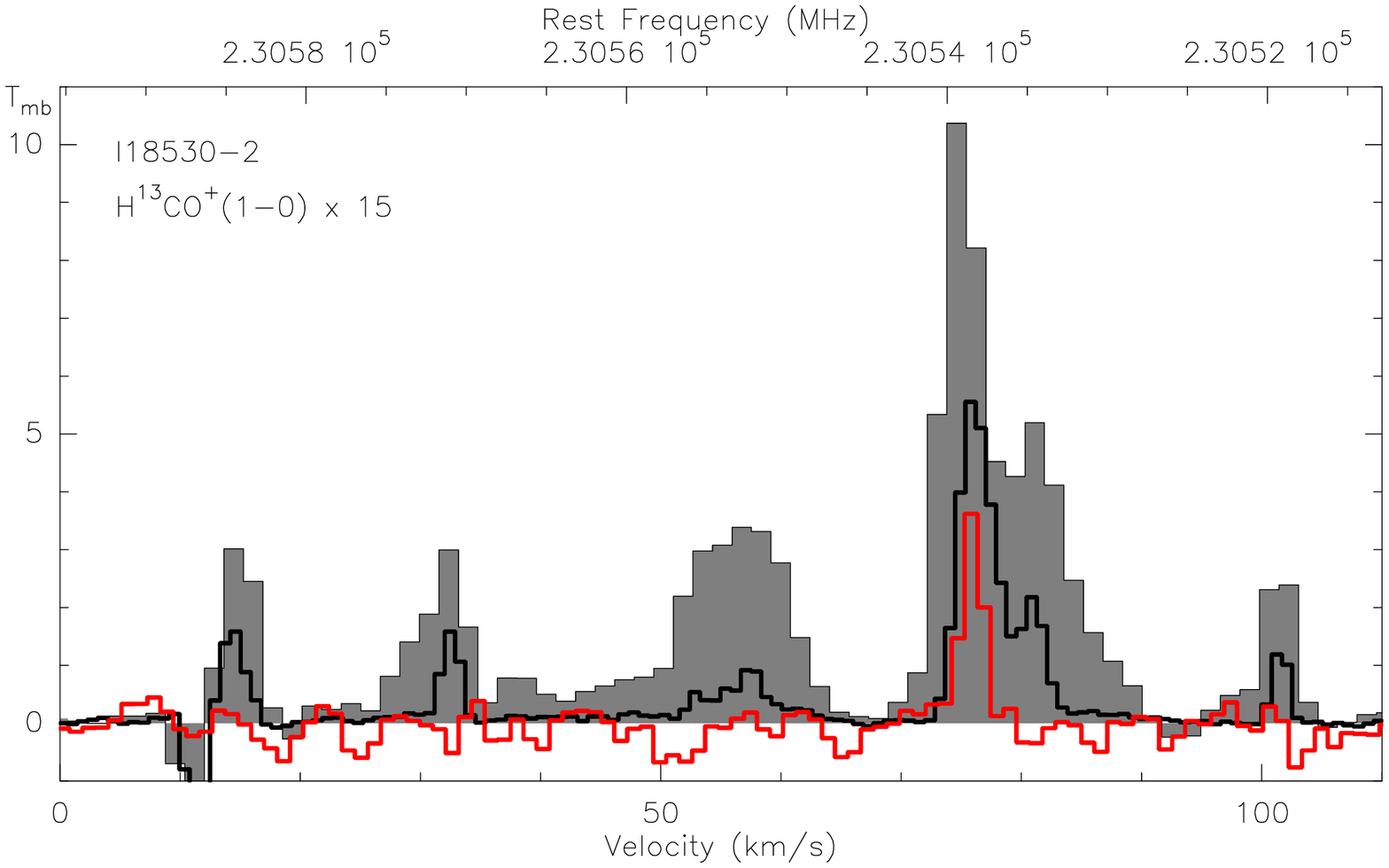}
\includegraphics[angle=-90,width=5.4cm]{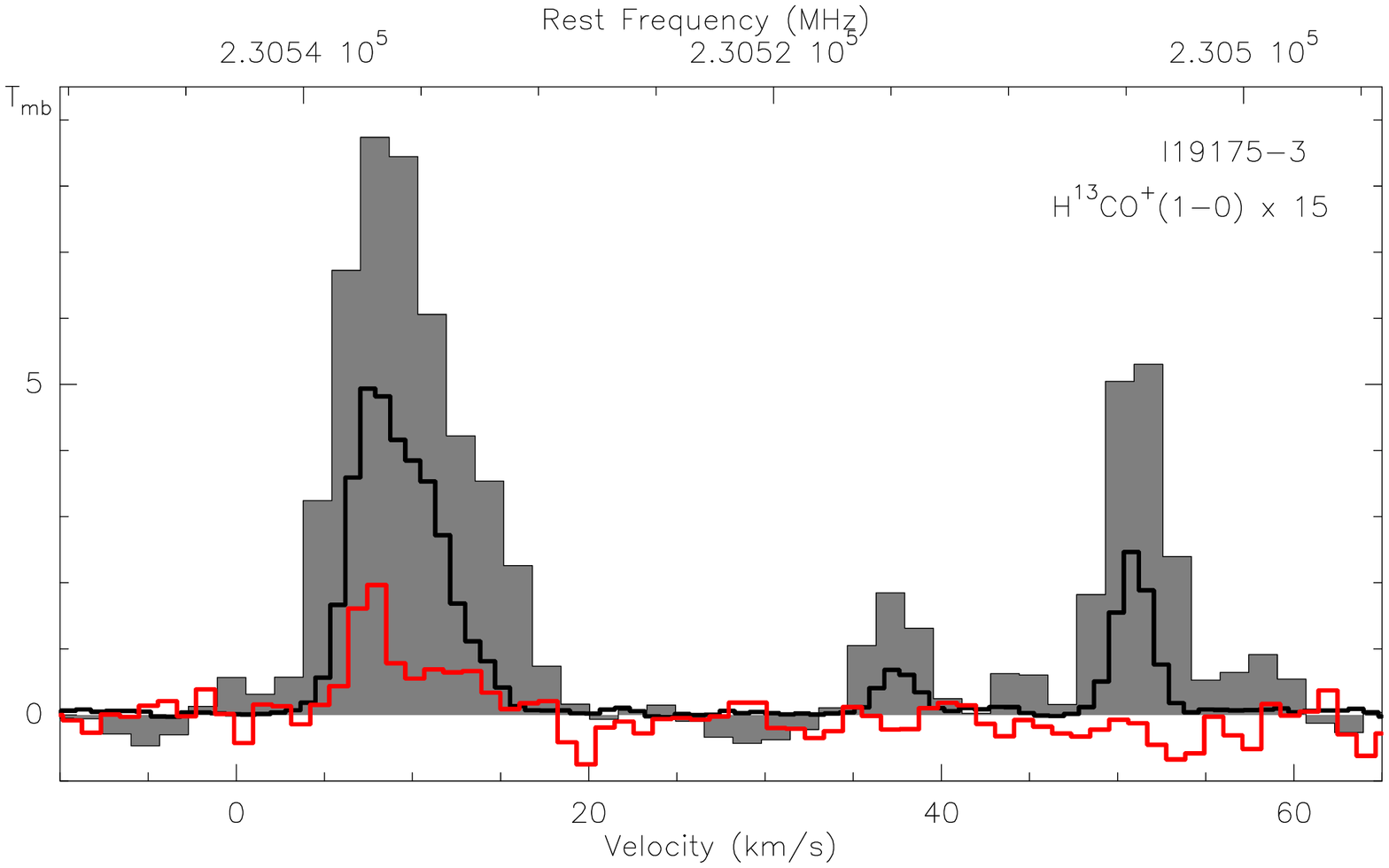}\\
\includegraphics[angle=-90,width=5.4cm]{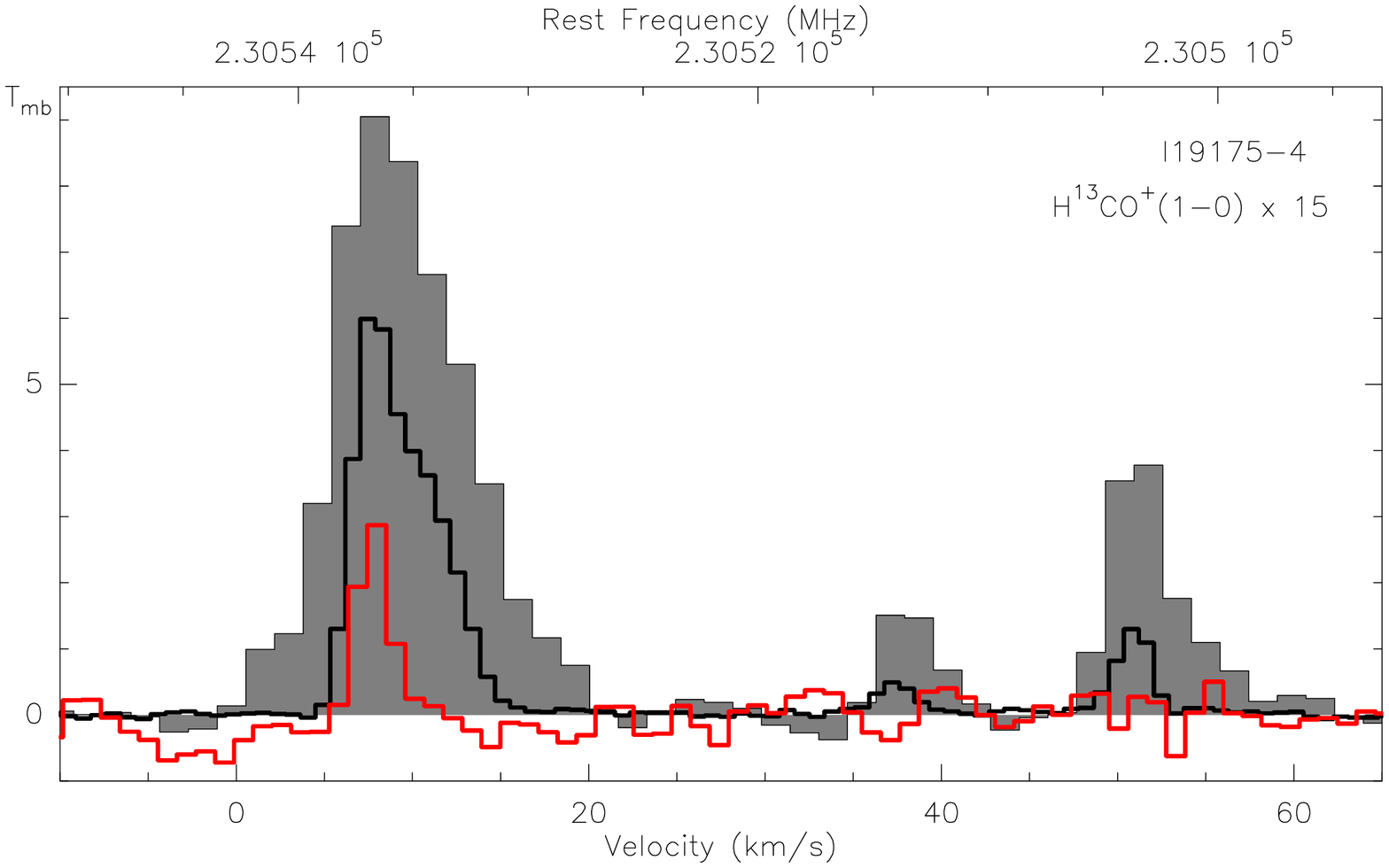}
\includegraphics[angle=-90,width=5.4cm]{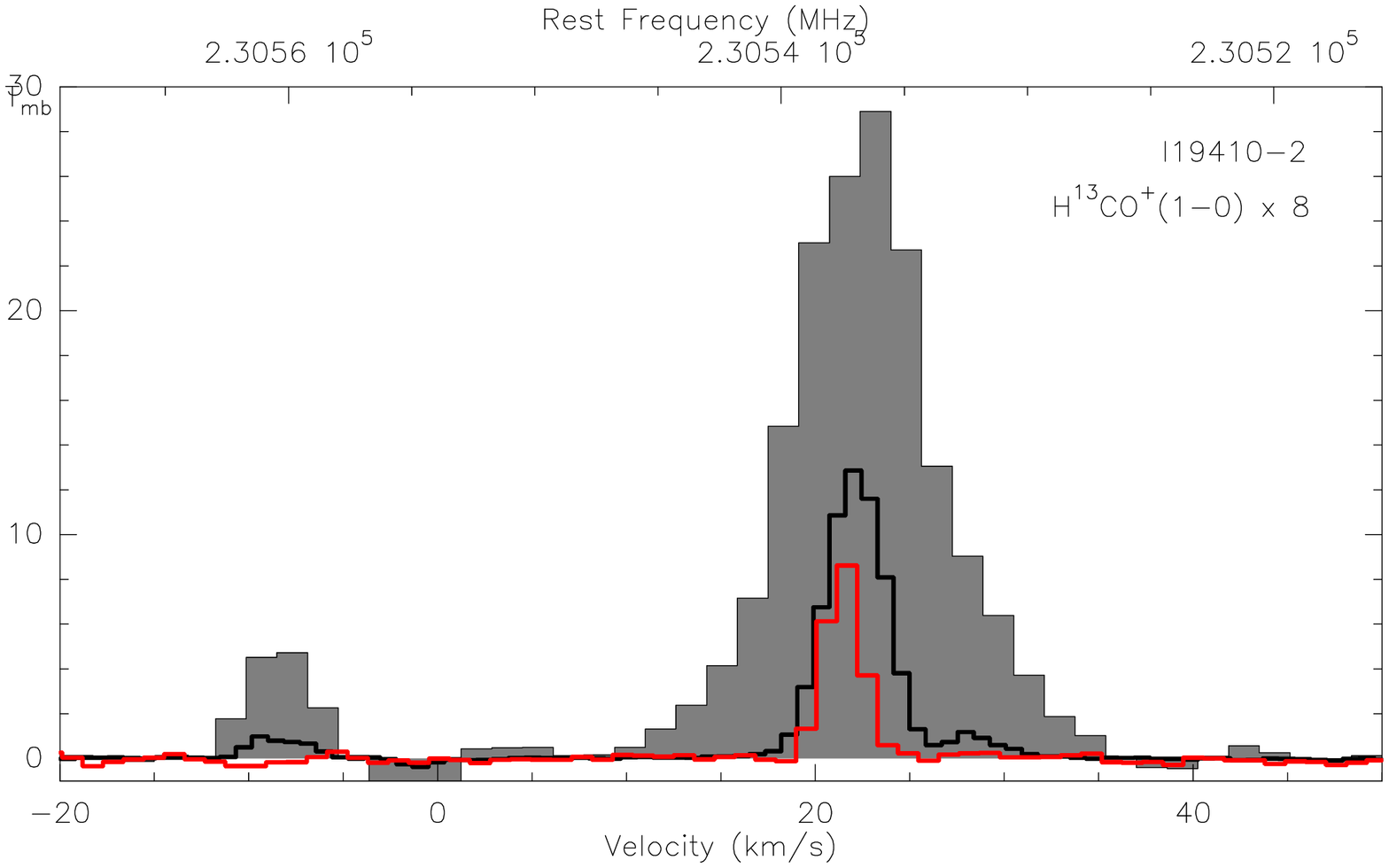}
\includegraphics[angle=-90,width=5.4cm]{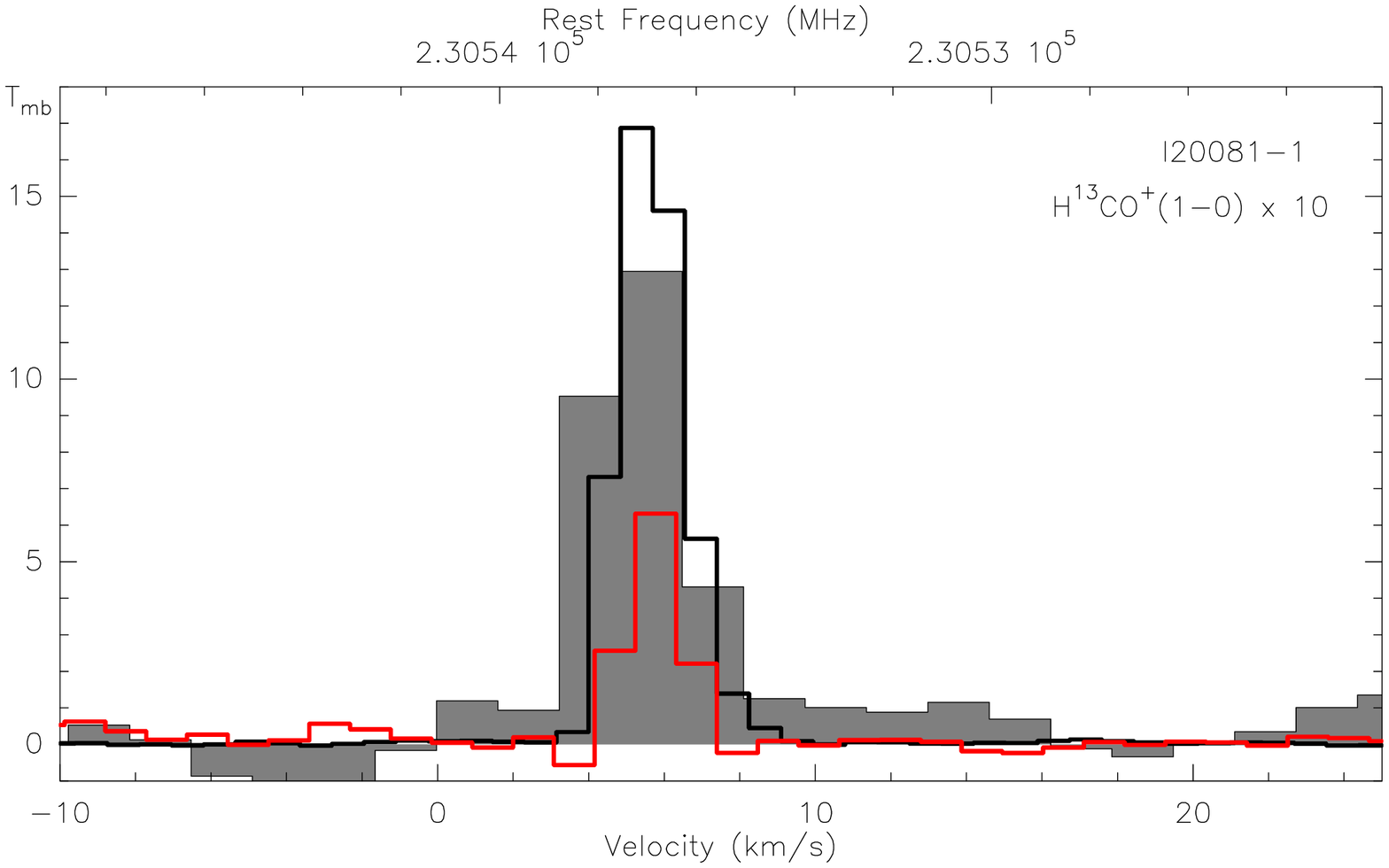}\\
\includegraphics[angle=-90,width=5.4cm]{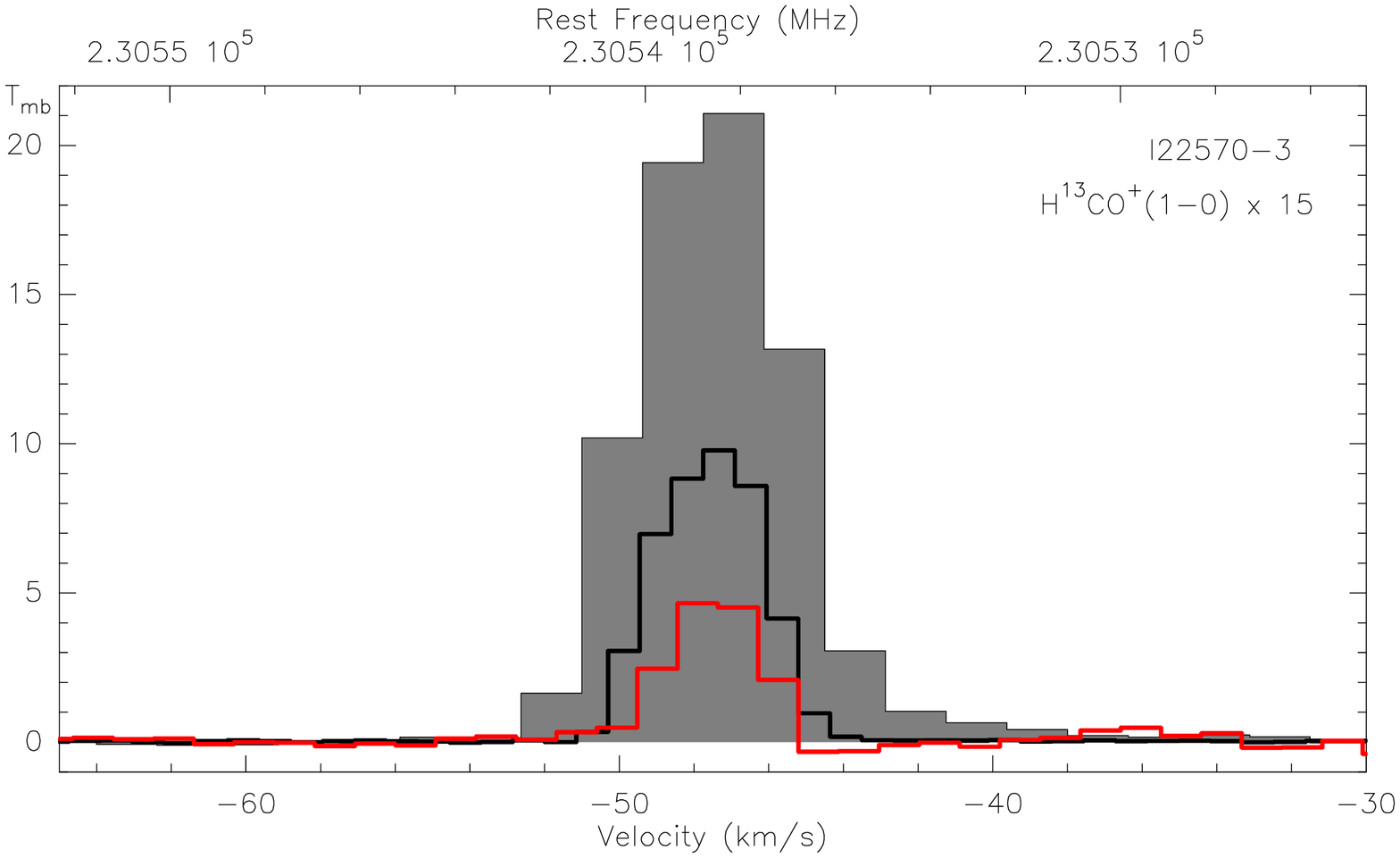}
\caption{Figures \ref{sample1} \& \ref{sample2} continued.}
\label{sample3}
\end{figure*}

Figures \ref{sample1}, \ref{sample2} and \ref{sample3} show the
$^{12}$CO(2--1), $^{13}$CO(1--0) and H$^{13}$CO$^+$(1--0) spectra for
the whole source sample. Except of IRDC\,18431-0312-4, where the
H$^{13}$CO$^+$(1--0) line remained undetected, toward all other
sources the three lines were detected. The most obvious difference
between $^{12}$CO/$^{13}$CO on the one side and H$^{13}$CO$^+$ on the
other side is that the CO isotopologues are detected at many different
velocities whereas H$^{13}$CO$^+$ usually only shows a single spectral
peak (except of a few double-peaked structures listed in Table
\ref{sourceparameters}). This difference can be explained
straightforwardly by the location of the sources within the Galactic
plane because CO traces more or less all molecular clouds along the
line of sight, hence contributions from various Galactic spiral arms,
whereas H$^{13}$CO$^+$ has significantly higher critical densities
(Table \ref{observing}) and thus traces only the dense cores we are
interested in. Therefore, if one wants to know the velocity of a
specific region CO observations are mostly not useful and one has to
refer to high density tracers. The peak velocities $v$ of
H$^{13}$CO$^+$ (Table \ref{sourceparameters}) correspond well to the
previously derived velocities from NH$_3$ observations of the same
sample \citep{sridharan2005}. Identifying the correct associated
velocity component to derive accurate velocities is an important step
to get distance estimates from the Galactic rotation curve as reported
in \citet{brand1993} and \citet{sridharan2005}.

\begin{figure}[h]
\includegraphics[angle=-90,width=8.6cm]{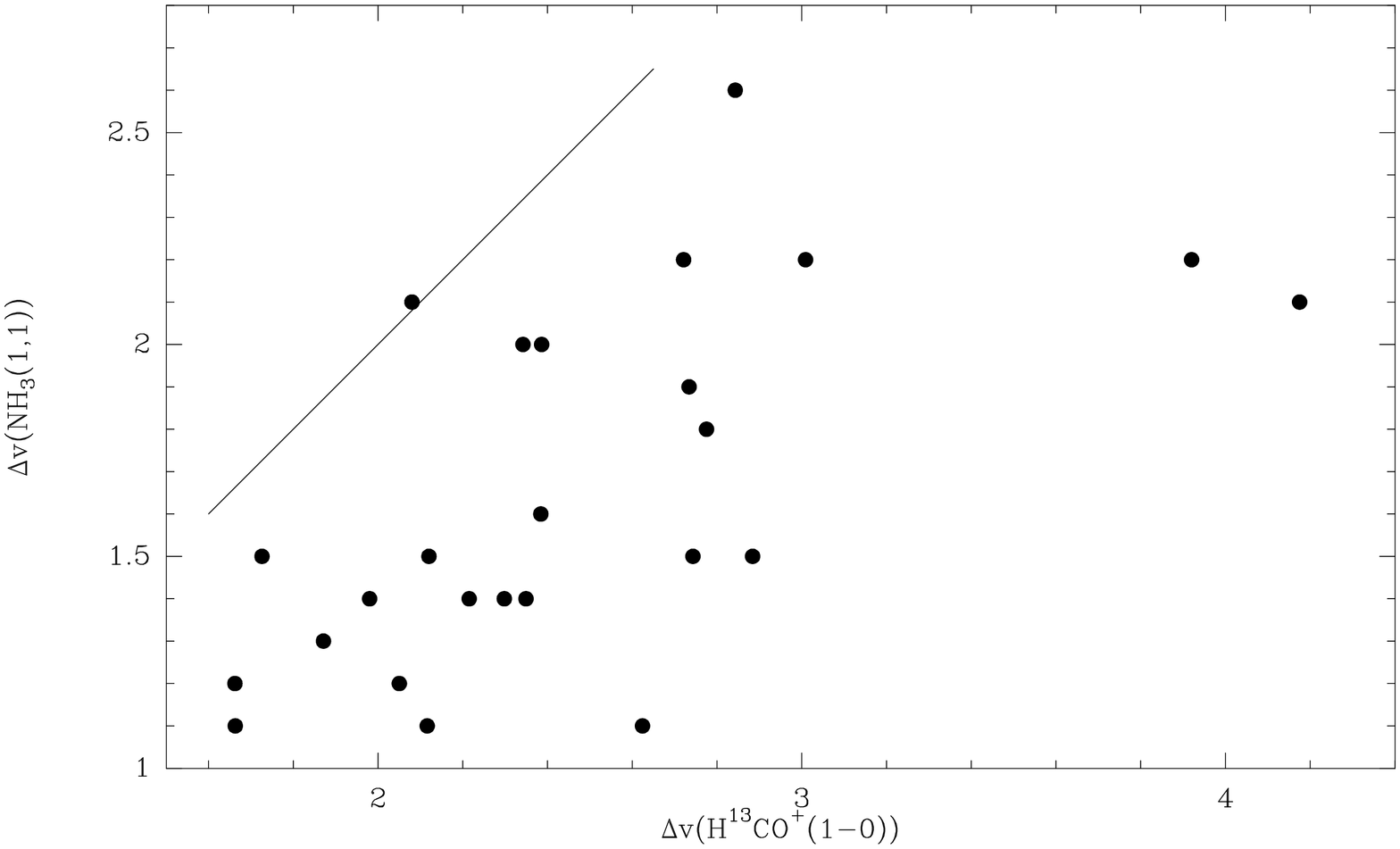}
\caption{Comparison of line-widths between H$^{13}$CO$^+$(1--0) (this
  work) and NH$_3$(1,1) \citep{sridharan2005}.}
\label{dv}
\end{figure}

Table \ref{sourceparameters} also lists the Full Width Half Maximum
(FWHM) line-widths $\Delta v(\rm{H^{13}CO^+})$ of the observed
H$^{13}$CO$^+$ lines from Gaussian fits, the average H$^{13}$CO$^+$
line-width is 2.6\,km\,s$^{-1}$.  Figure \ref{dv} shows a comparison
of the H$^{13}$CO$^+$ line-width with the previously derived
line-width from the NH$_3$(1,1) observations from
\citep{sridharan2005}. We find that the H$^{13}$CO$^+$ line-width
nearly always exceeds the line-width from the NH$_3$(1,1) with an
average line-width ratio $\frac{\Delta v(\rm{H^{13}CO^+(1-0)})}{\Delta
  v(\rm{NH_3(1,1)})}=1.5$. This is interesting since the critical
densities for both molecules differ by about 2 orders of magnitude
($n_{\rm{crit}}(\rm{NH_3}(1,1))\sim 2\times 10^3$\,cm$^{-3}$, e.g.
\citealt{zhou1989}, $n_{\rm{crit}}(\rm{H^{13}CO^+(1-0)})\sim 1.7\times
10^5$\,cm$^{-3}$, Table \ref{observing}), whereas the thermal
line-widths ($\Delta
v_{\rm{th}}=\sqrt{8{\rm{ln}}2\frac{kT}{m_{\rm{mol}}}}$ with the
molecular mass $m_{\rm{mol}}$) are similar ($\Delta
v_{\rm{th}}(\rm{NH_3}(1,1))_{15\rm{K}}\sim 0.2$\,km\,s$^{-1}$ and
$\Delta v_{\rm{th}}(\rm{H^{13}CO^+(1-0)})_{15\rm{K}}\sim
0.15$\,km\,s$^{-1}$), $\Delta v_{\rm{th}}(\rm{H^{13}CO^+(1-0)})$ being
even smaller than $\Delta v_{\rm{th}}(\rm{NH_3}(1,1))$ because of its
larger molecular weight. Therefore, differences in the non-thermal
motions have to be responsible for the line-width variations
between both species.  Similar results were obtained for low-mass
cores in a comparison of CS with NH$_3$ line-widths \citep{zhou1989}.

There exist several potential ways to explain such a line-width
difference. For example, there is the well-known line-width--size
relation $\Delta v \propto R^{0.5}$ \citep{larson1981}. This relation
would imply that our target regions would be on average 2.3 times
larger in H$^{13}$CO$^+$ than they are in NH$_3$. Although we have no
maps of the regions, this appears very unlikely. A different
possibility was discussed by \citet{zhou1989} and \citet{myers1983}:
If the stability of a core is determined by the entire
line-broadening, for a gravitationally stable core the critical
density should relate to the line-width and size of a molecular core
as
$$n_{\rm{crit}}\propto (\Delta v/R)^2 \Longleftrightarrow \Delta v \propto R\sqrt{n_{\rm{crit}}}.$$
To fulfill this criterion the H$^{13}$CO$^+$ source size would need
to be $\sim$7 times smaller than that of NH$_3$ which
again appears to be very unlikely. In this scenario, one could then
conclude that molecular clumps are not gravitationally stable but
maybe collapsing already. However, overall collapse is not sure at all
from such an analysis, and the clumps could also be fragmenting or
only partially collapsing (see discussion in \citealt{zhou1989}).

As discussed below in \S\ref{sio}, a significant fraction of sources
exhibits broad line wings in the jet/outflow tracer SiO, which is
indicative of early outflow and hence early star formation activity.
Since H$^{13}$CO$^+$ has a larger critical density it traces denser
regions of the gas clumps. With ongoing star formation activity, one
expects additional line broadening effects from various physical
processes like outflows, infall and potential rotation of embedded
accretion disks, all spatially unresolved by the current observations.
Most of these effects are likely contributing more strongly in the
dense gas components and hence could explain the excess in
H$^{13}$CO$^+$ line-width.

\subsection{SiO(2--1)}
\label{sio}

\begin{figure*}[h]
\includegraphics[angle=-90,width=5.4cm]{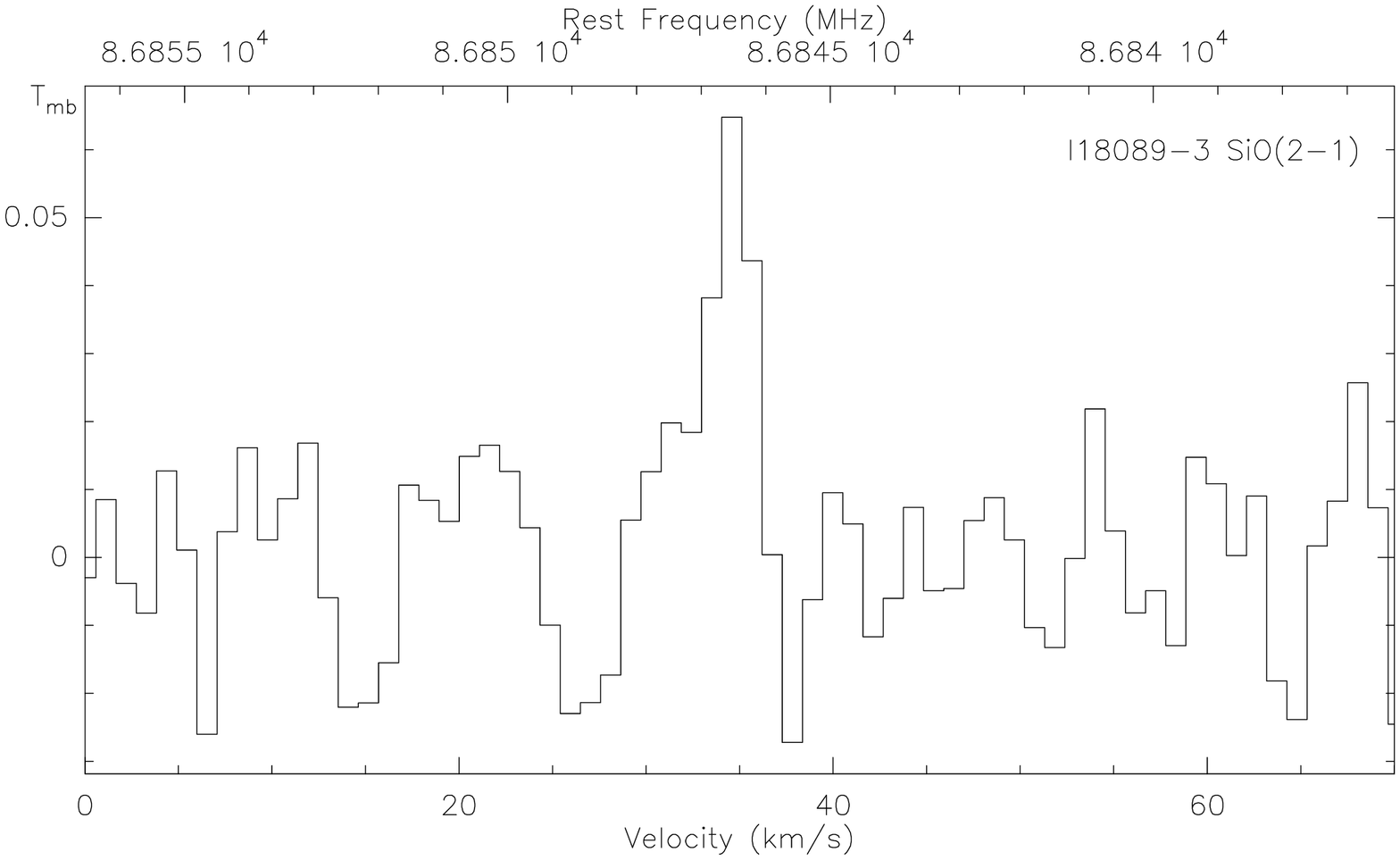}
\includegraphics[angle=-90,width=5.4cm]{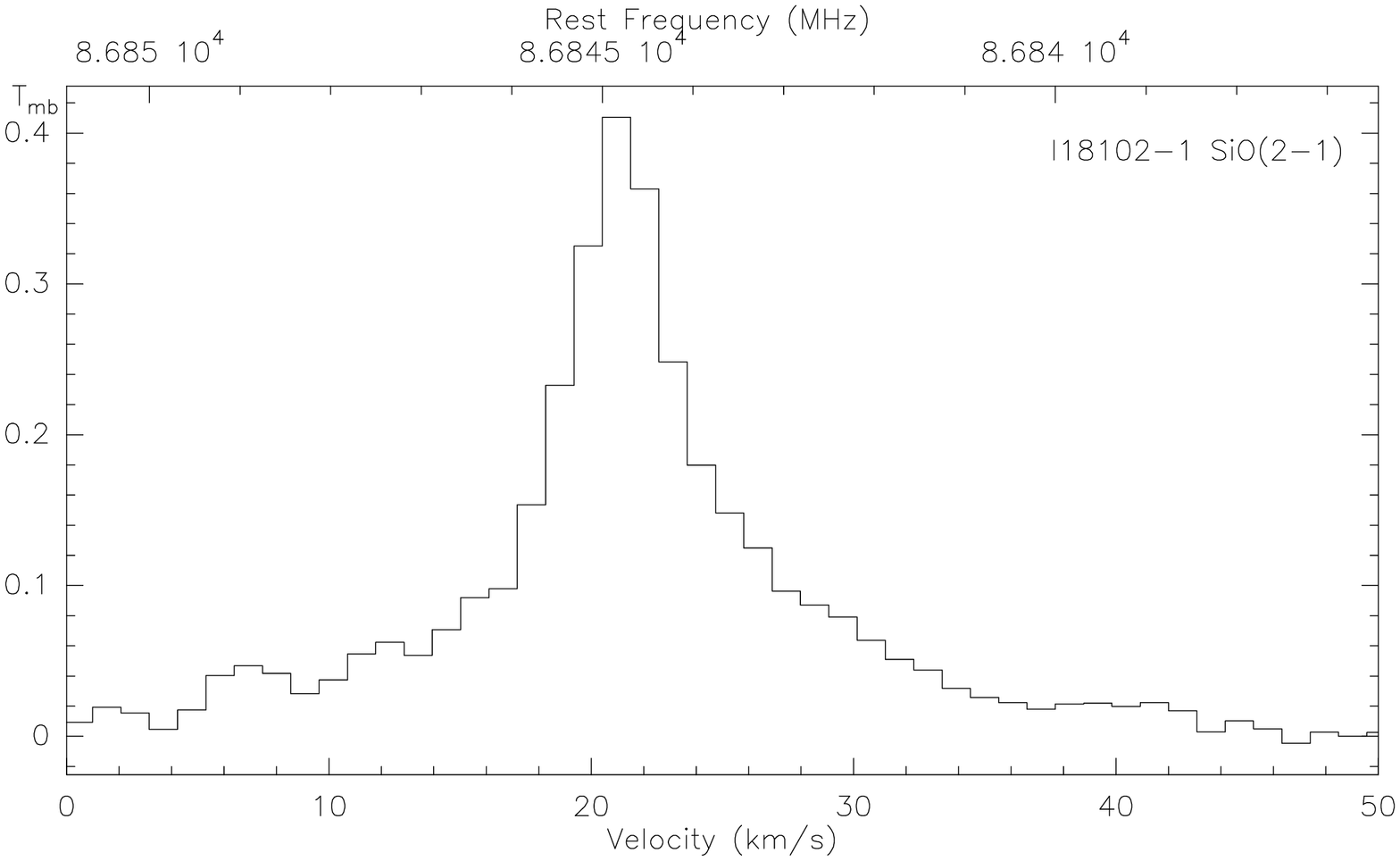}
\includegraphics[angle=-90,width=5.4cm]{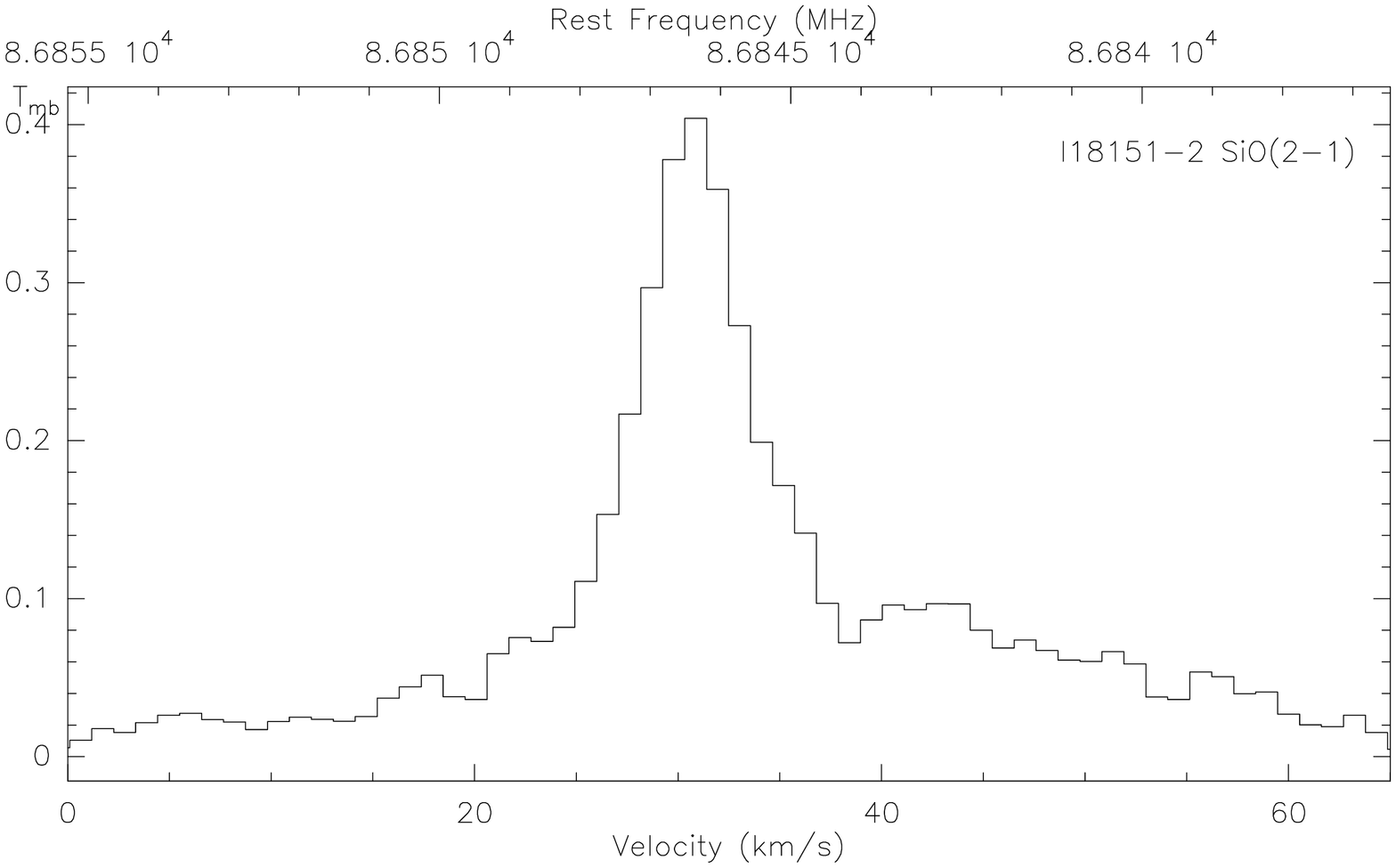}\\
\includegraphics[angle=-90,width=5.4cm]{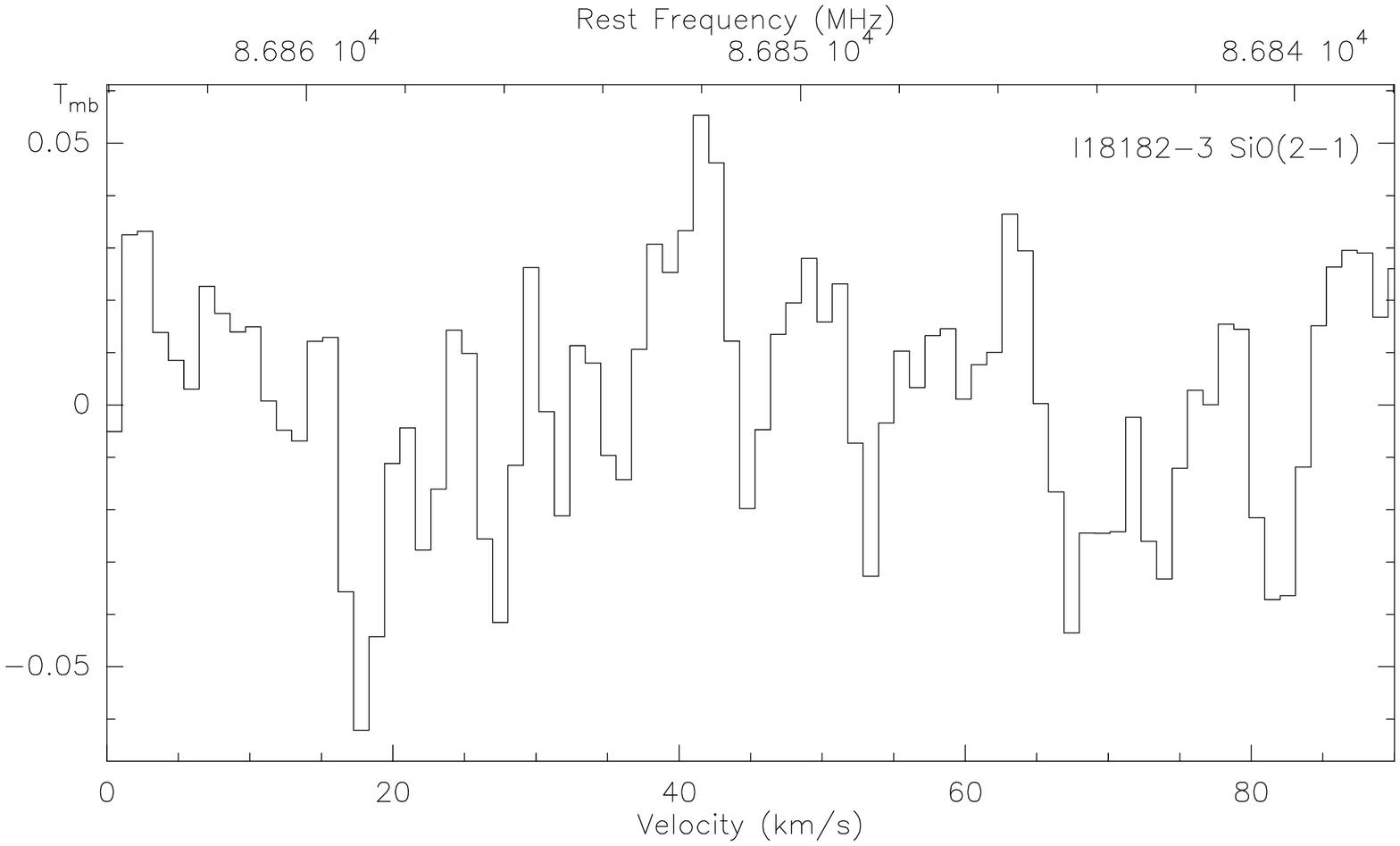}
\includegraphics[angle=-90,width=5.4cm]{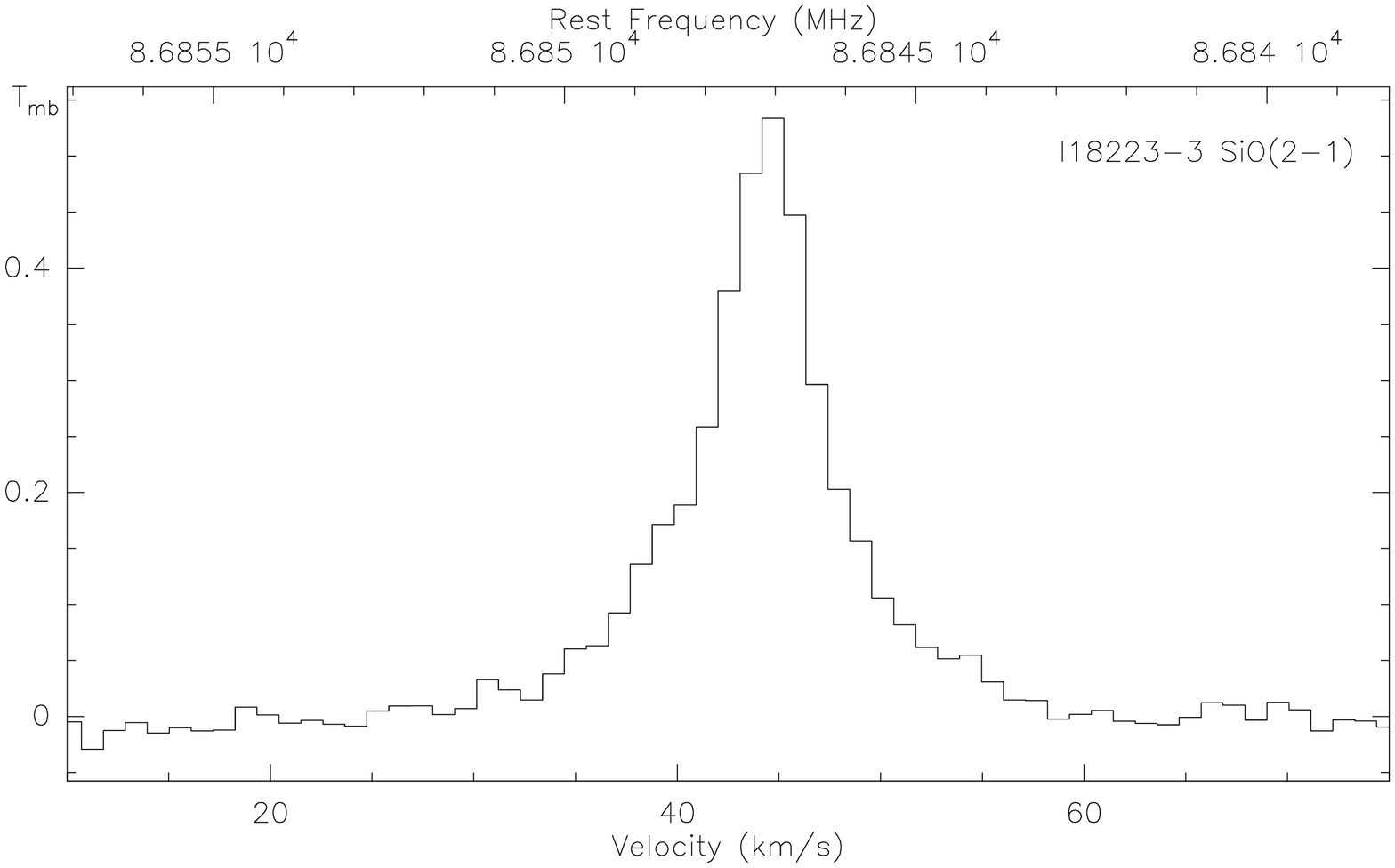}
\includegraphics[angle=-90,width=5.4cm]{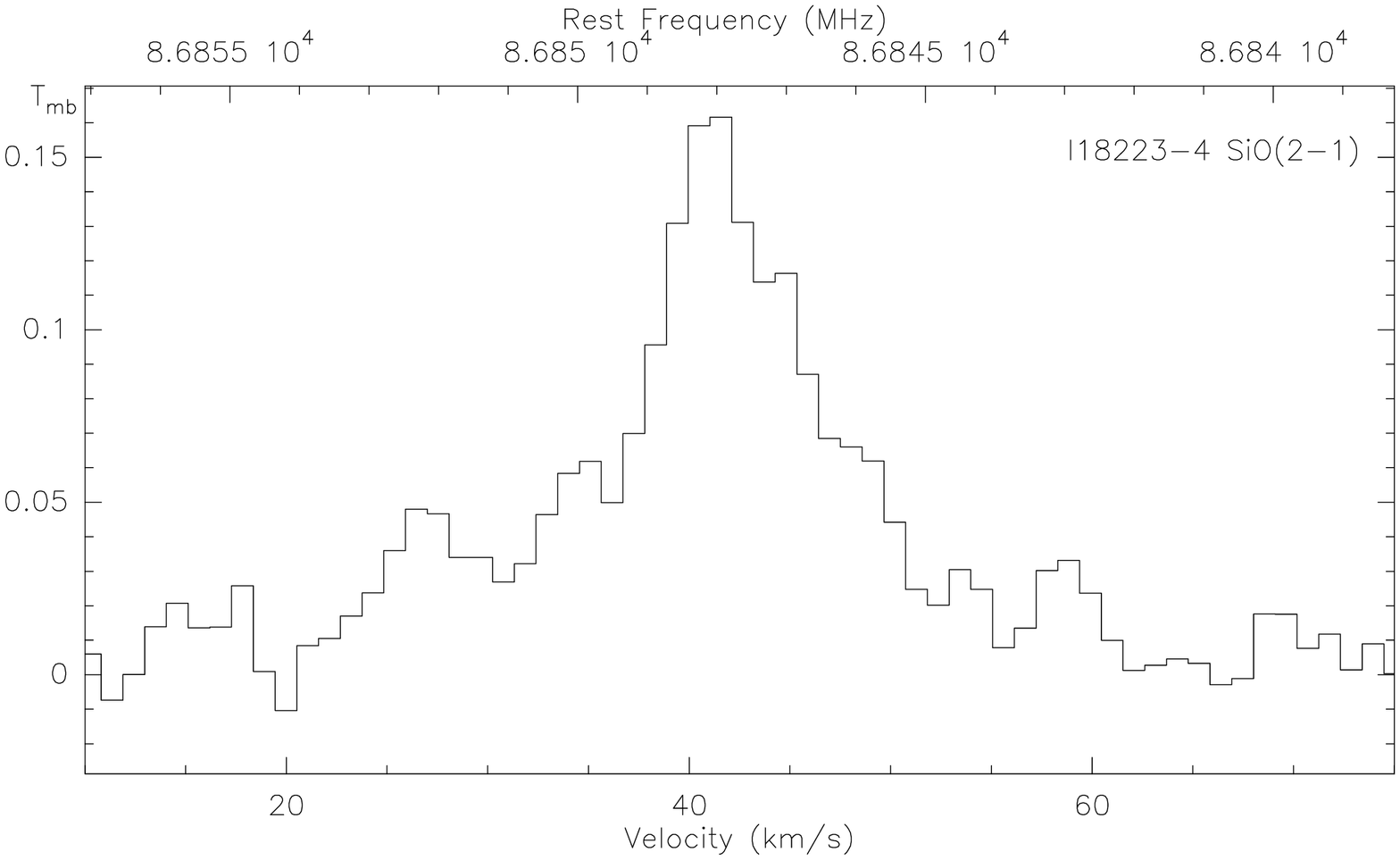}\\
\includegraphics[angle=-90,width=5.4cm]{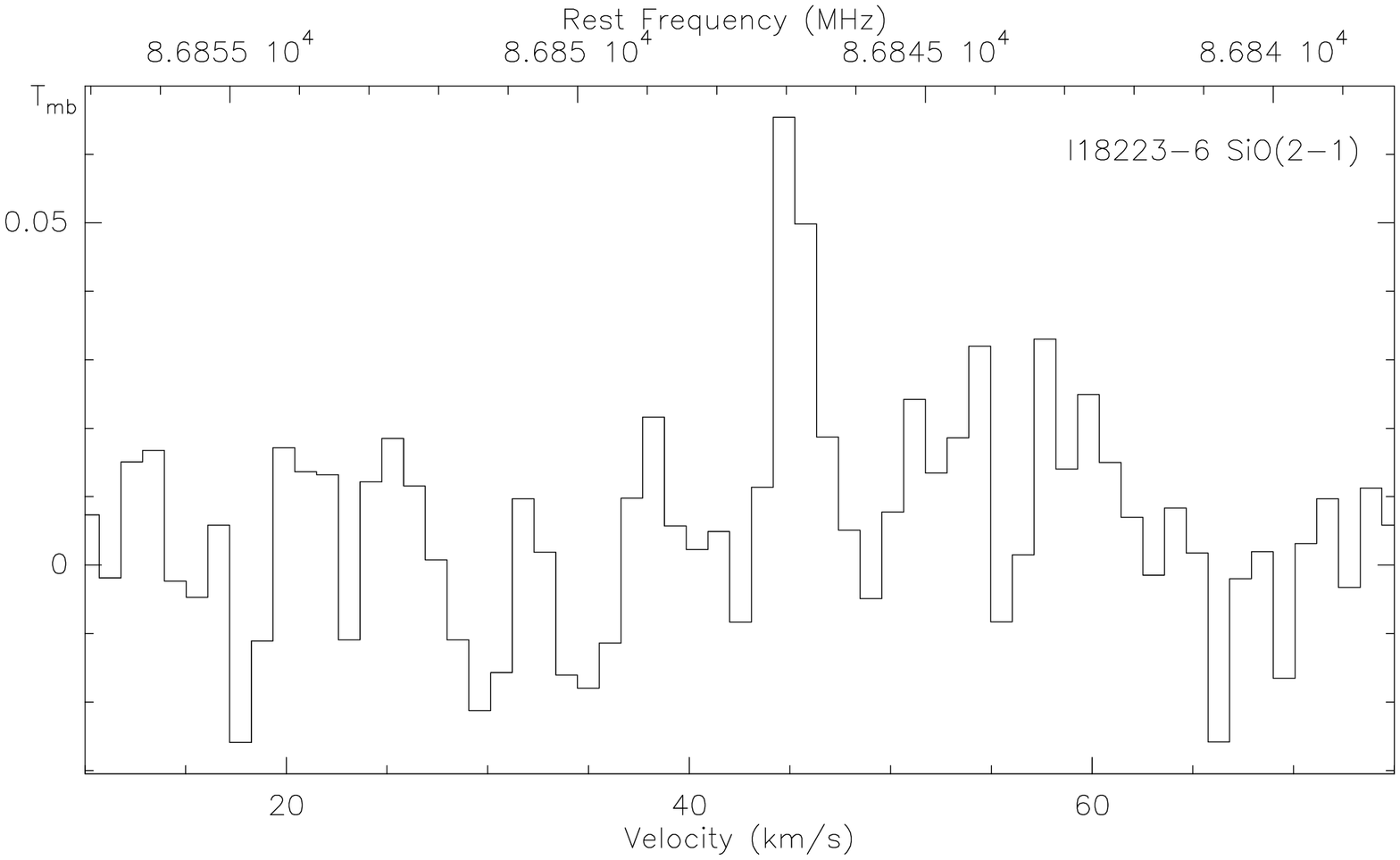}
\includegraphics[angle=-90,width=5.4cm]{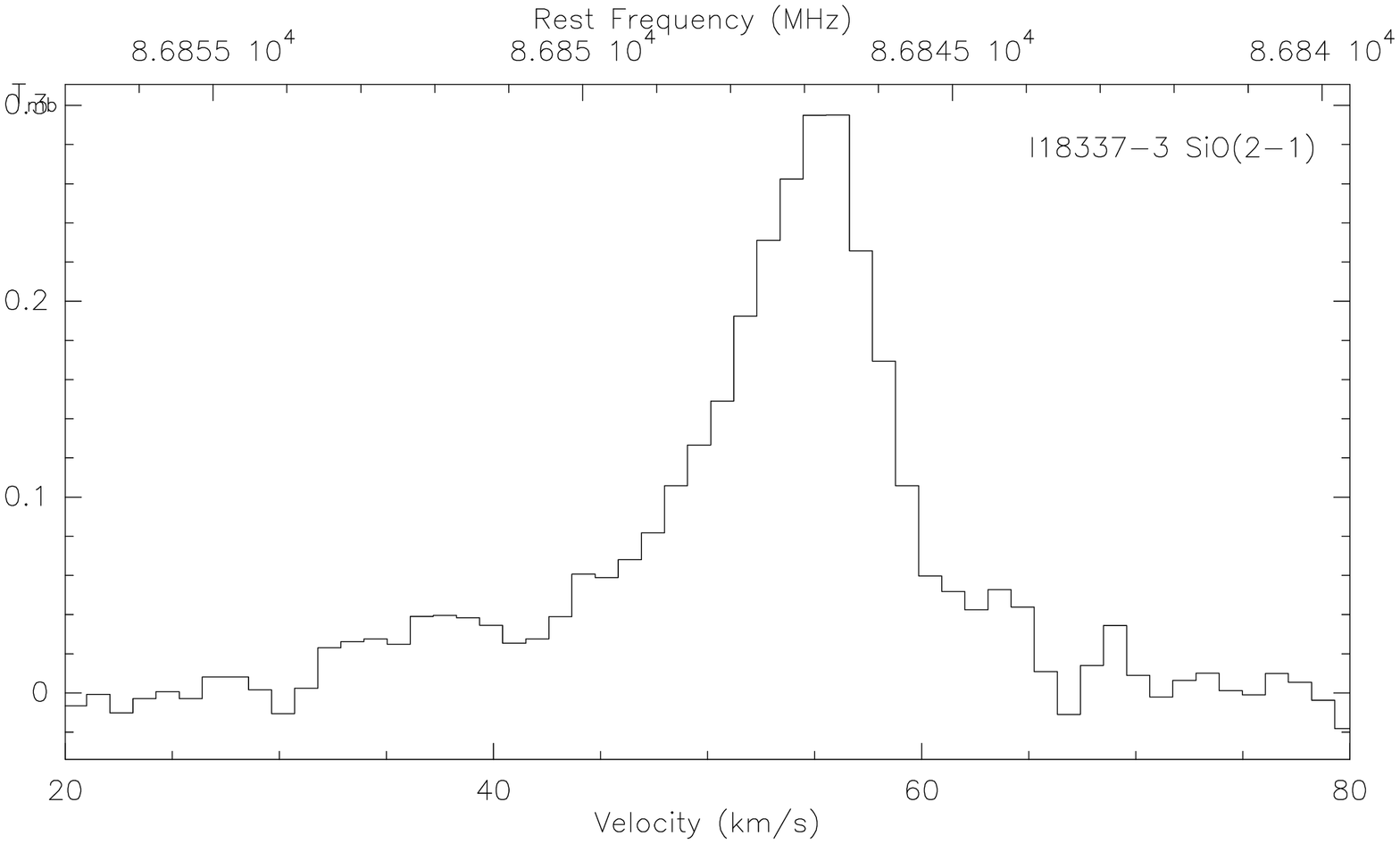}
\includegraphics[angle=-90,width=5.4cm]{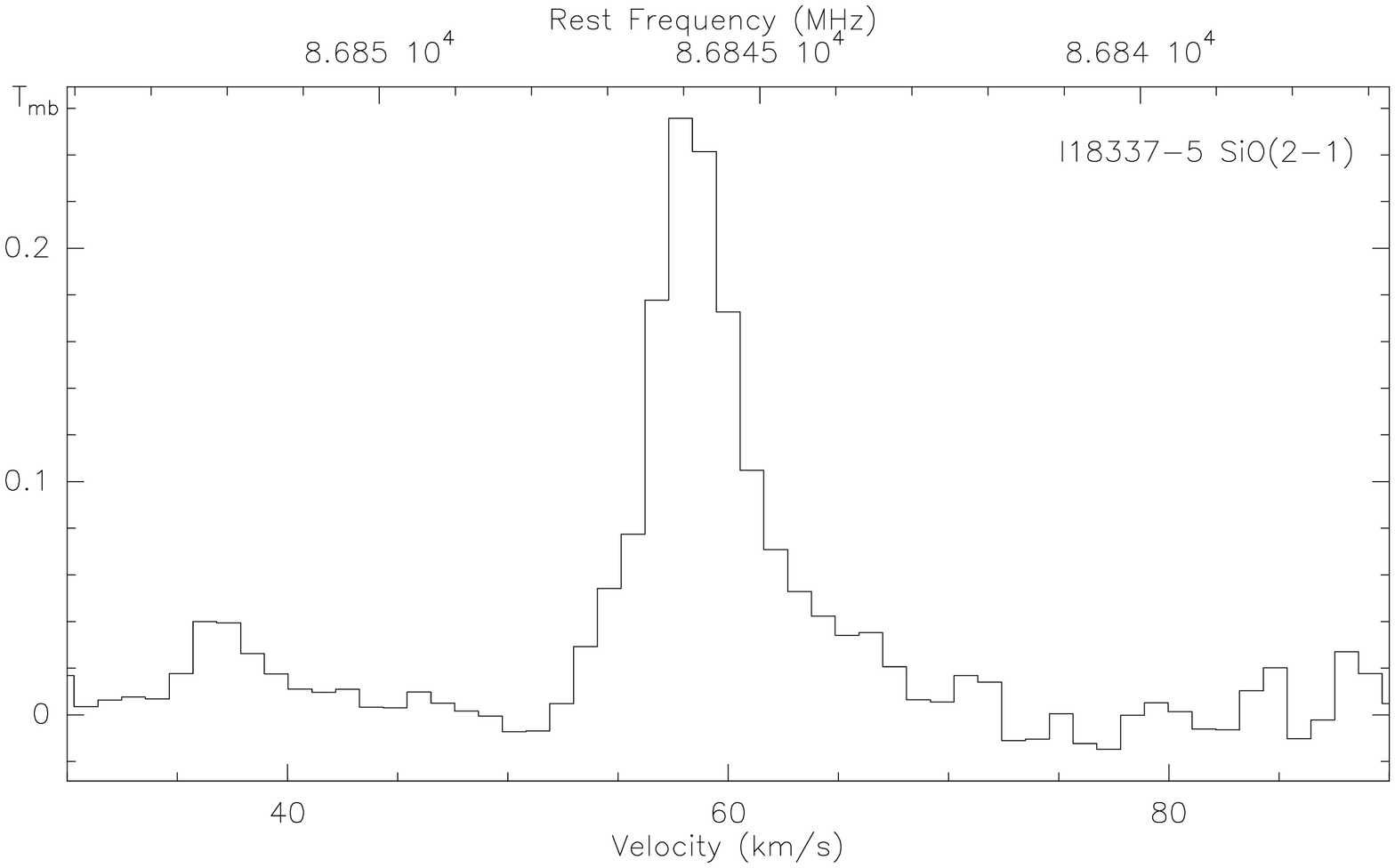}\\
\includegraphics[angle=-90,width=5.4cm]{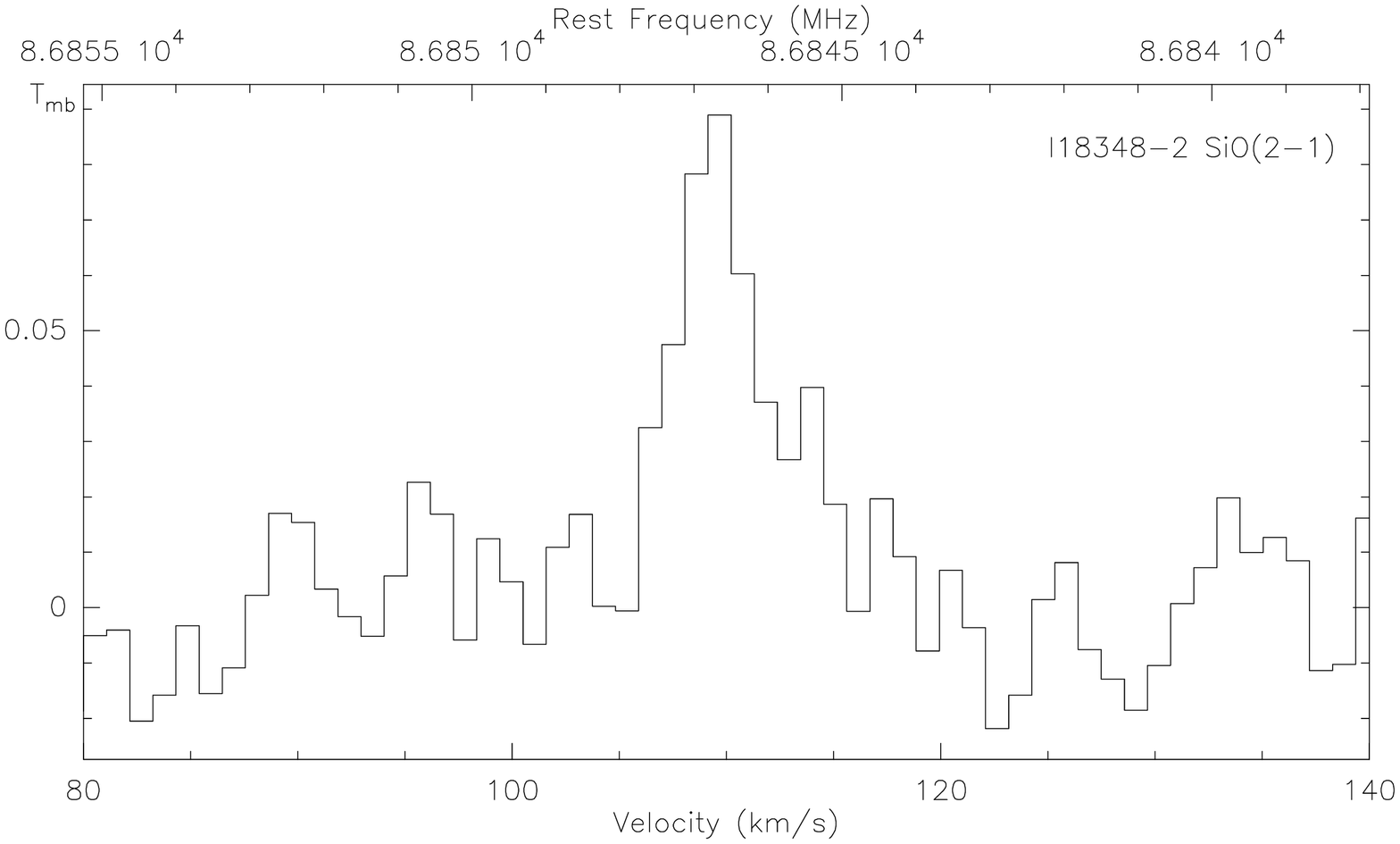}
\includegraphics[angle=-90,width=5.4cm]{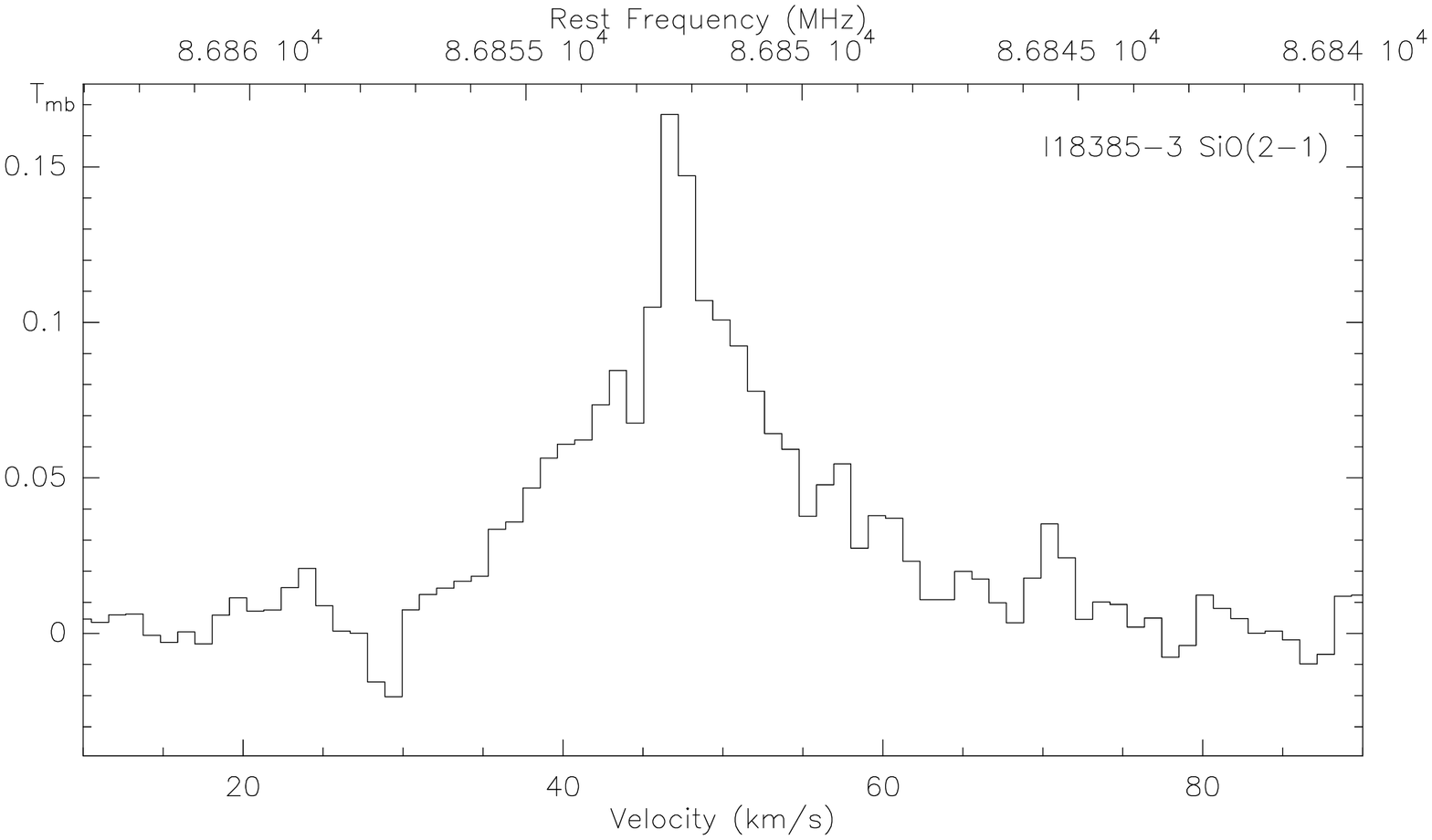}
\includegraphics[angle=-90,width=5.4cm]{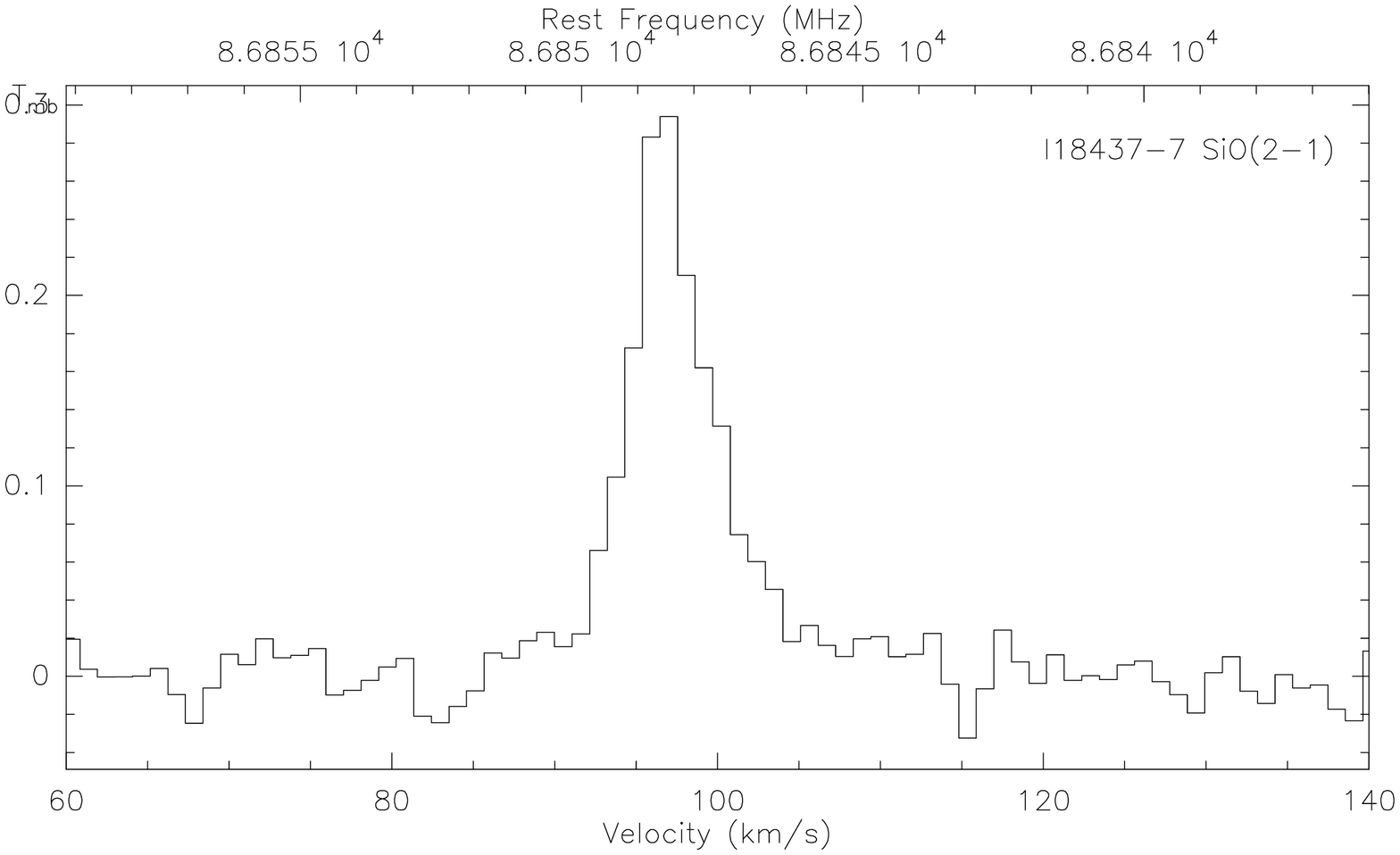}\\
\includegraphics[angle=-90,width=5.4cm]{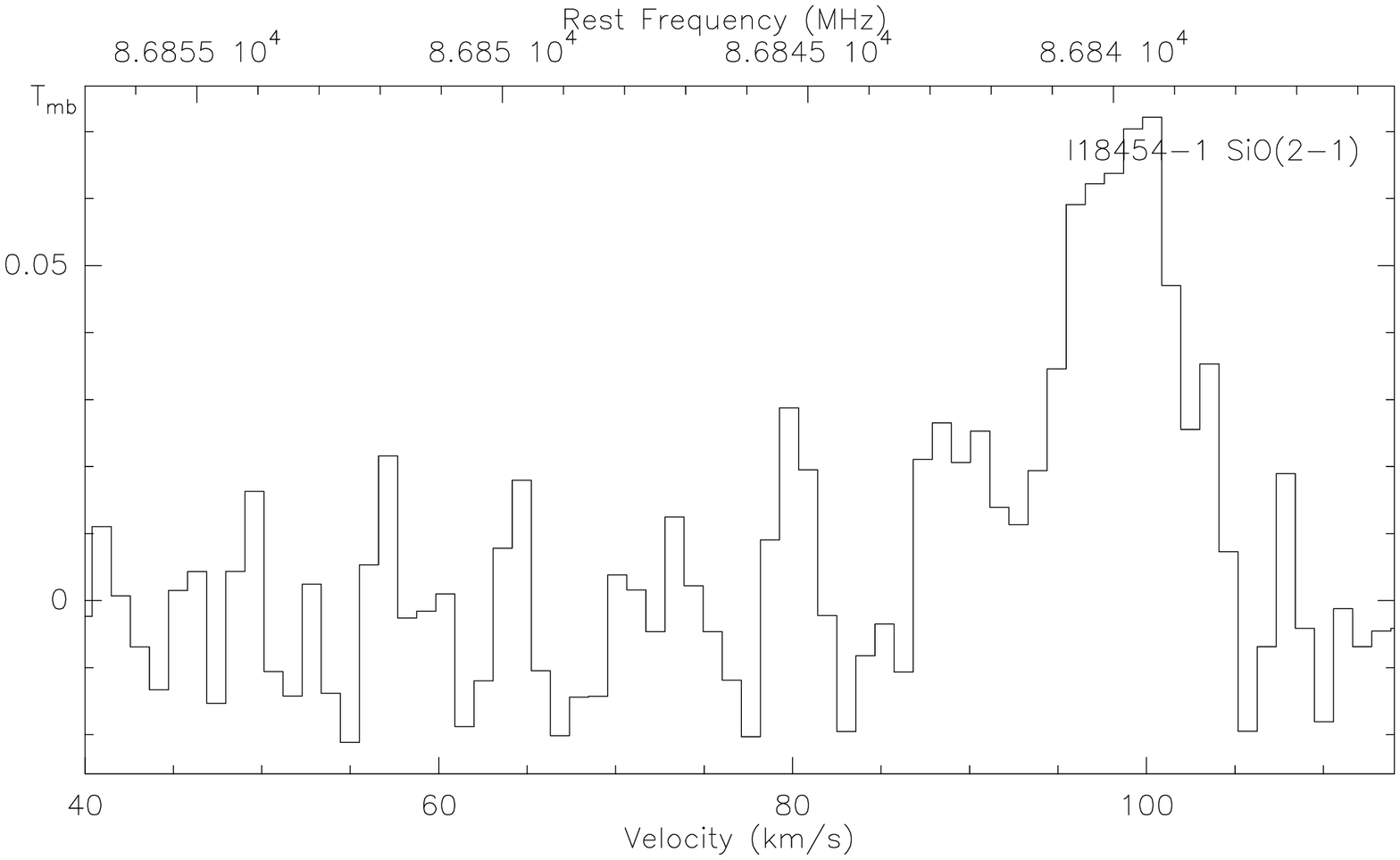}
\includegraphics[angle=-90,width=5.4cm]{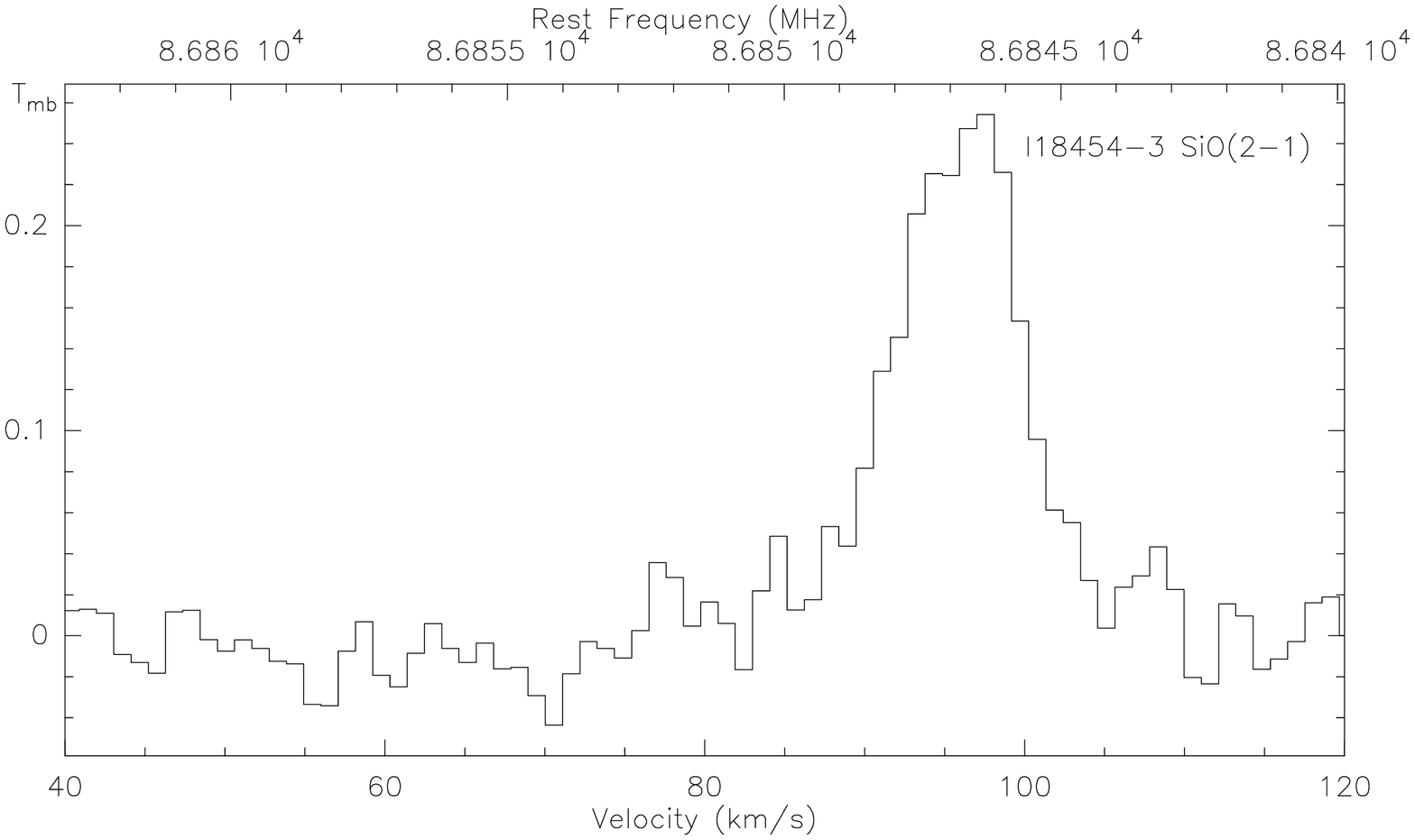}
\includegraphics[angle=-90,width=5.4cm]{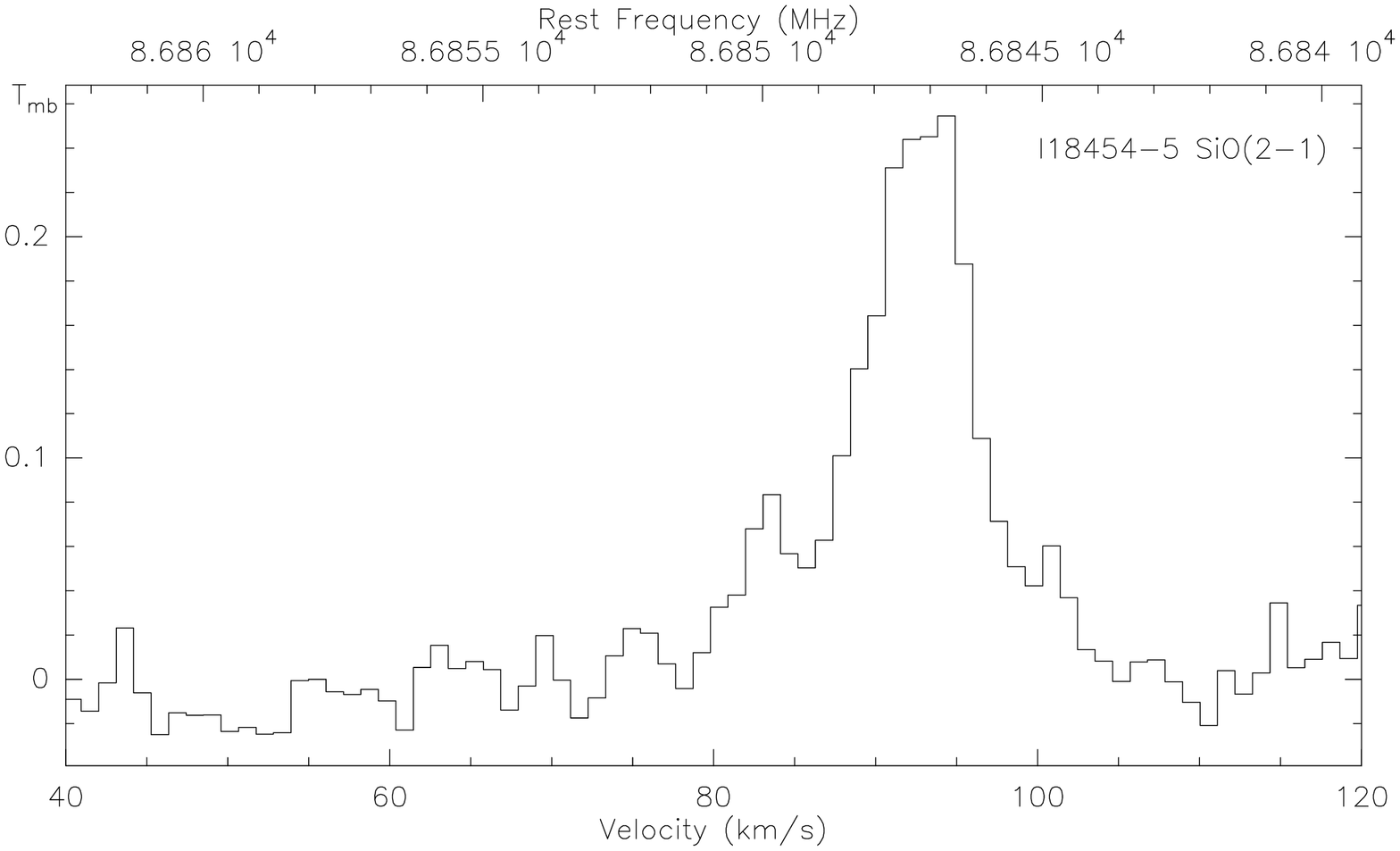}\\
\includegraphics[angle=-90,width=5.4cm]{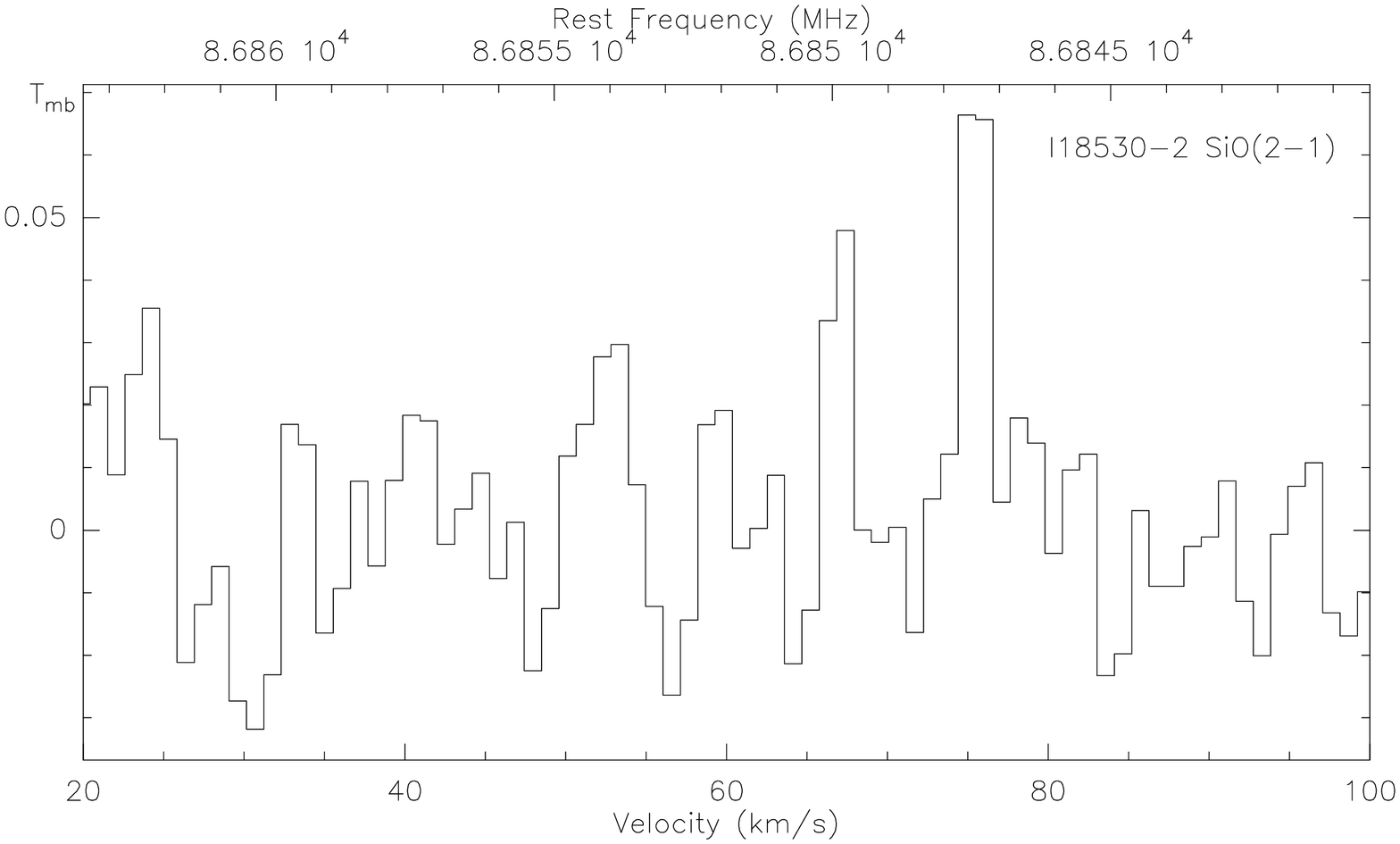}
\includegraphics[angle=-90,width=5.4cm]{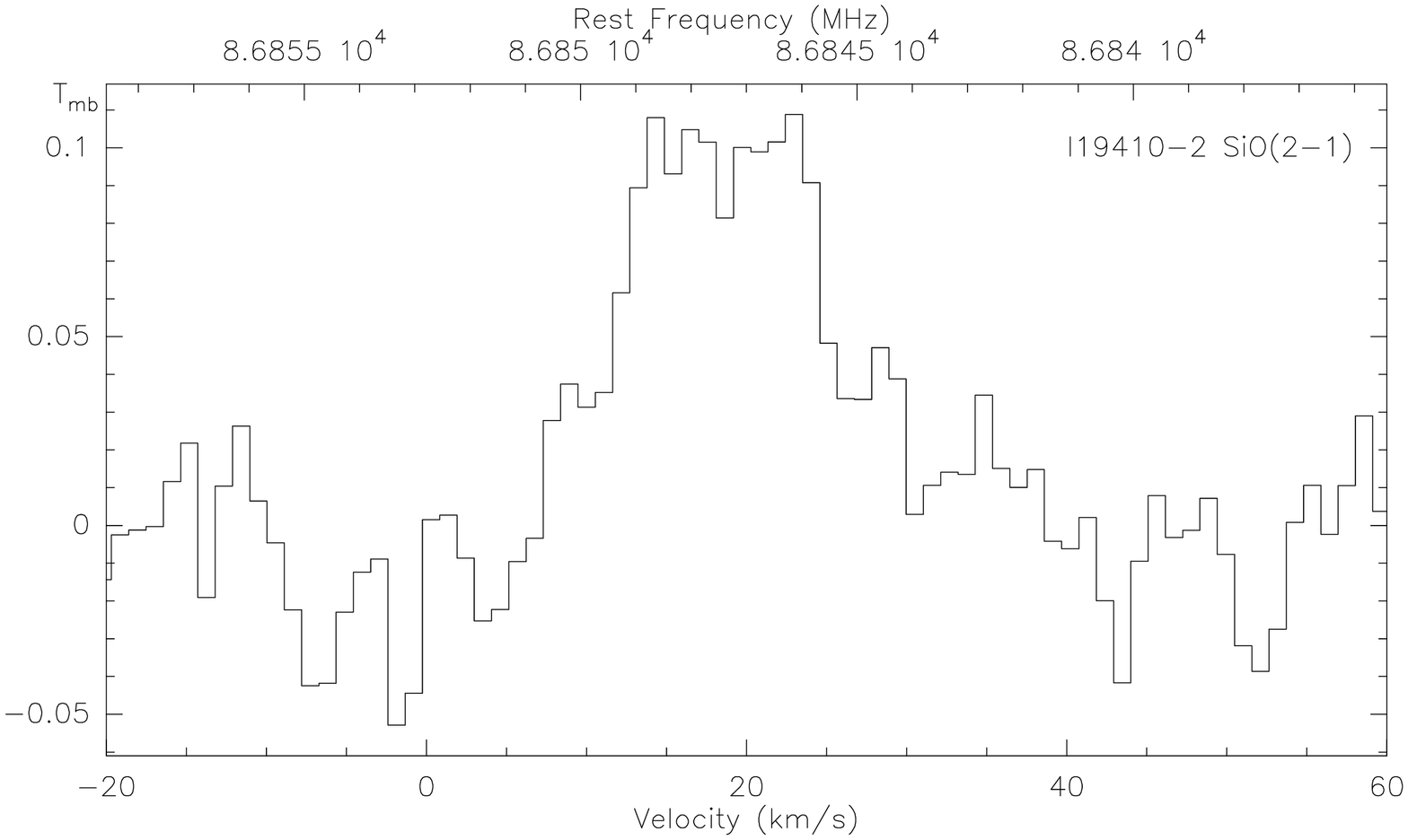}
\includegraphics[angle=-90,width=5.4cm]{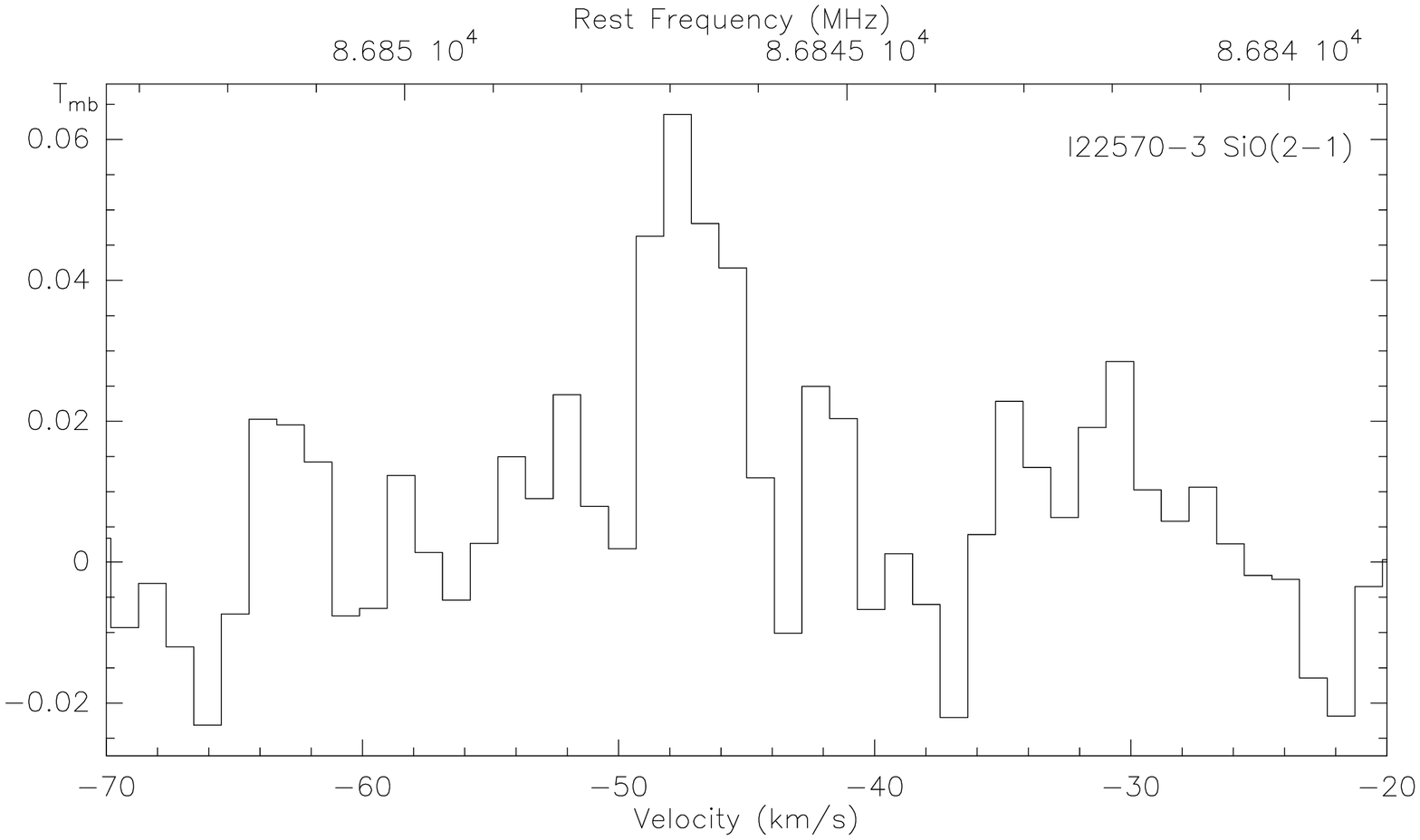}
\label{siospectra}
\caption{SiO(2--1) detections.}
\end{figure*}

One of the aims of this study was to identify how many of the sources
would show clear SiO emission caused by molecular outflow activity.
SiO is believed to be produced in the gas-phase after sputtering
Si-bearing species from the dust grain surfaces when the outflow/jet
impinges on the ambient gas and dust \citep{schilke1997a}. SiO is usually
barely detectable in quiescent clouds. Figure \ref{siospectra}
presents the 18 SiO(2--1) detections toward our source sample of 43
regions. If this source sample is representative for typical Infrared
Dark Clouds, this implies that in at least 40\% of the IRDCs star
formation has already started (absence of SiO emission does not
necessarily imply absence of star formation activity). However, it is
not obvious whether this sample is representative for all IRDCs
because it was selected to be close to sites of already ongoing
massive star formation.

Table \ref{sourceparameters} also lists the width-to-zero-velocity
$\Delta v_0$ of the observed SiO(2--1) lines.  If one divides the
observed SiO spectra by their width-to-zero-velocity $\Delta v_0$ into
3 regimes (a) low velocity $\Delta v_0 < 10$\,km\,s$^{-1}$, (b)
intermediate velocity 10\,km\,s$^{-1}<\Delta v_0<20$\,km\,s$^{-1}$ and
(c) high velocity $\Delta v_0 > 20$\,km\,s$^{-1}$, one finds six
sources in the low-velocity regime, five at intermediate velocities
and seven in the high-velocity regime. While the sources at
intermediate to high velocities obviously show gas at relatively high
velocities with respect to the ambient gas, this is less clear for the
sources in the low-velocity regime: those sources with $\Delta v_0 >
5$\,km\,s$^{-1}$ probably also have outflow contributions, but the
SiO(2--1) spectra toward two sources with $\Delta v_0 <
2.5$\,km\,s$^{-1}$ (IRDC\,18223-6 \& IRDC\,18530-2) hardly look like
typical outflow spectra. A similar spread in SiO velocities has also
previously been observed by, .e.g., \citealt{lefloch1998,codella1999}.
What are possible reasons for the different observed SiO $\Delta v_0$?

The observed velocity spread could partly be produced by the angles
between the outflows and the plane of the sky since we only observe
line-of-sight velocities. However, assuming a random distribution of
outflow inclination angles $\theta$ ($\theta = 0$ is an outflow along
the line of sight) and similar unprojected outflow velocities, one
would expect fewer outflows at high velocities than at low velocities
($\int_0^{45} sin(\theta)d\theta\sim 0.29$ vs.  $\int_{45}^{90}
sin(\theta)d\theta\sim 0.71$). This is in contrast to our relatively
even distribution of $\Delta v_0$ from low to high values. Hence,
inclination effects are unlikley to explain the observations.

Outflows/jets from massive protostars have intrinsically higher
velocities than those of their low-mass counterparts (e.g.,
\citealt{richer2000}). In addition, more luminous sources usually
exhibit stronger SiO emission (e.g., \citealt{codella1999}).
Therefore, the sources with smaller line-width could harbor less
massive protostars that may or may not form a massive star at the end
of their evolution. Furthermore, all objects with low SiO velocity
spread are also the weakest or non-detections in CH$_3$CN and CH$_3$OH
(see \S\ref{ch3cn_ch3oh}) which is also expected in the framework of
embeedded protostars with different masses. Thus, different
masses/luminosities of the embedded outflow-driving sources can
potentially cause the observed SiO line-width difference.

Since IRDCs are believed to represent the earliest evolutionary stages
of massive (and low-mass) star formation, we expect that in some of
them low- to intermediate-mass protostars may be embedded that are
likely destined to grow, accrete and become high-mass stars at the end
of their evolution (e.g., IRDC\,18223-3,
\citealt{beuther2005d,beuther2007a}). However, a priori we cannot
differentiate whether an object will become massive at the end of its
evolution or whether it will remain a low-mass protostar.  In this
scenario, one can also speculate about an evolutionary sequence where
the growing protostars accelerates the gas to higher velocities in the
course of its early evolution, i.e., the smaller line-widths may
correspond to the still less massive and younger objects whereas the
broad line-width sources may have already accreted more mass and could
hence be a little bit more evolved. This picture is consistent with
the above assertion that the line-width distribution can depend on the
mass and luminosity of the central object, just with the additional
assumption that the lower-mass objects may still be able to become
massive stars during their ongoing evolution. This scenario may appear
counter-intuitive compared to typical low-mass outflows where the
strongest emission is observed from class 0 sources and where the SiO
emission gets weaker formore evolved class I and II sources (e.g.,
\citealt{gibb2004}). However, even class 0 sources have typically
already ages of the order $10^4$\,yrs, and IRDCs like those presented
here are likely still significantly younger (e.g.,
\citealt{beuther2007a,krumholz2006b}). Furthermore, this low-mass
classification scheme is unlikely to be well applicable to high-mass
star formation because, among other reasons, massive stars reach the
main sequence while accretion processes can still be ongoing (e.g.,
\citealt{schaller1992,mckee2003}). Therefore, our picture applies only
to the earliest evolutionary stages of strongly accreting protostars
destined to become massive stars at the end of their evolution.  In
principle, the non-detection of SiO line-wings could also be caused by
poorer signal-to-noise ratios or larger distances to the respective
region.  However, both scenarios are unlikely the case here because we
have approximately the same rms for all sources, and the line-width
differences are found also toward sources at the same distances (e.g.,
IRDC\,18223-1243-3 \& IRDC\,18223-1243-4 versus IRDC\,18223-1243-6).
 
Do the $\Delta v_0 < 2.5$\,km\,s$^{-1}$ fit into any of these
pictures? Since more than 50\% of the target sources exhibit no SiO
emission at all down to our sensitivity limit, it is interesting that
two objects show these extremely narrow line-widths.
\citet{lefloch1998} and \citet{codella1999} try to explain similar
observations by outflow/dense clump interactions and in an outflow
evolutionary model. In the latter scenario, the highest velocity
components are found at the earliest evolutionary class 0 stages and
the smaller line-width components could be remnants from earlier
interactions of molecular outflows with the ambient gas
\citep{codella1999}. They estimated the corresponding time-scale to
decrease the SiO emission and the velocities to be of the order
$10^4$\,yrs.  We consider this scenario as unlikely for our sample
because massive star formation proceeds on relatively short
time-scales (of the order $10^5$ years, \citealt{mckee2002}), and
after $10^4$\,yrs the regions are expected to be strong enough to be
detetable at near- and mid-infrared wavelengths (e.g., the samples
presented by \citealt{molinari1996} or \citealt{sridha}) which is in
contrast to our sample selection as Infrared Dark Clouds. Another
possibility is to speculate whether these objects could represent the
earliest observed stage of star formation within our source sample: In
this picture, the Si-bearing species would just be released from the
grains by shock interaction and SiO would only have begun to form in
the gas phase, but the young jet/outflow would not have had the time
yet to accelerate large amounts of gas to detectable higher
velocities. To confirm or discard this picture additional indicators
of star formation activity would need to be observed, e.g., infall of
gas via asymmetric line profiles (e.g., \citealt{myers1996}) or
shocked H$_2$ emission either in the K-band (e.g.,
\citealt{davis2004}) or in the IRAC 4.5\,$\mu$m band (e.g.,
\citealt{rathborne2005}).

\subsection{CH$_3$CN$(6_k-5_k)$ and CH$_3$OH(5--4)}
\label{ch3cn_ch3oh}

In addition to the previously discussed molecules, the spectral setup
allowed to observe also the typical hot-core tracer CH$_3$CN as well
as a series of four CH$_3$OH lines. Although the critical densities
and excitation parameters of both molecules do not differ that much
(see Table \ref{observing}), CH$_3$OH typically has one to two orders
of magnitude larger gas column densities and is hence easier to
observe (e.g., \citealt{vandishoek1998,nomura2004}). Furthermore,
CH$_3$OH is known to be present in massive star formation right from
the beginning, hence it is a so-called parent molecule, whereas
CH$_3$CN is a daughter molecule which needs time to form through
chemical processes during the massive star formation process (e.g.,
\citealt{nomura2004}). Therefore toward a sample of young IRDCs, one
would expect CH$_3$CN to be either undetectable or only weakly
detected in a few sources, whereas CH$_3$OH should be detectable
toward a larger fraction of the sample.

\begin{figure*}[h]
\includegraphics[angle=-90,width=5.4cm]{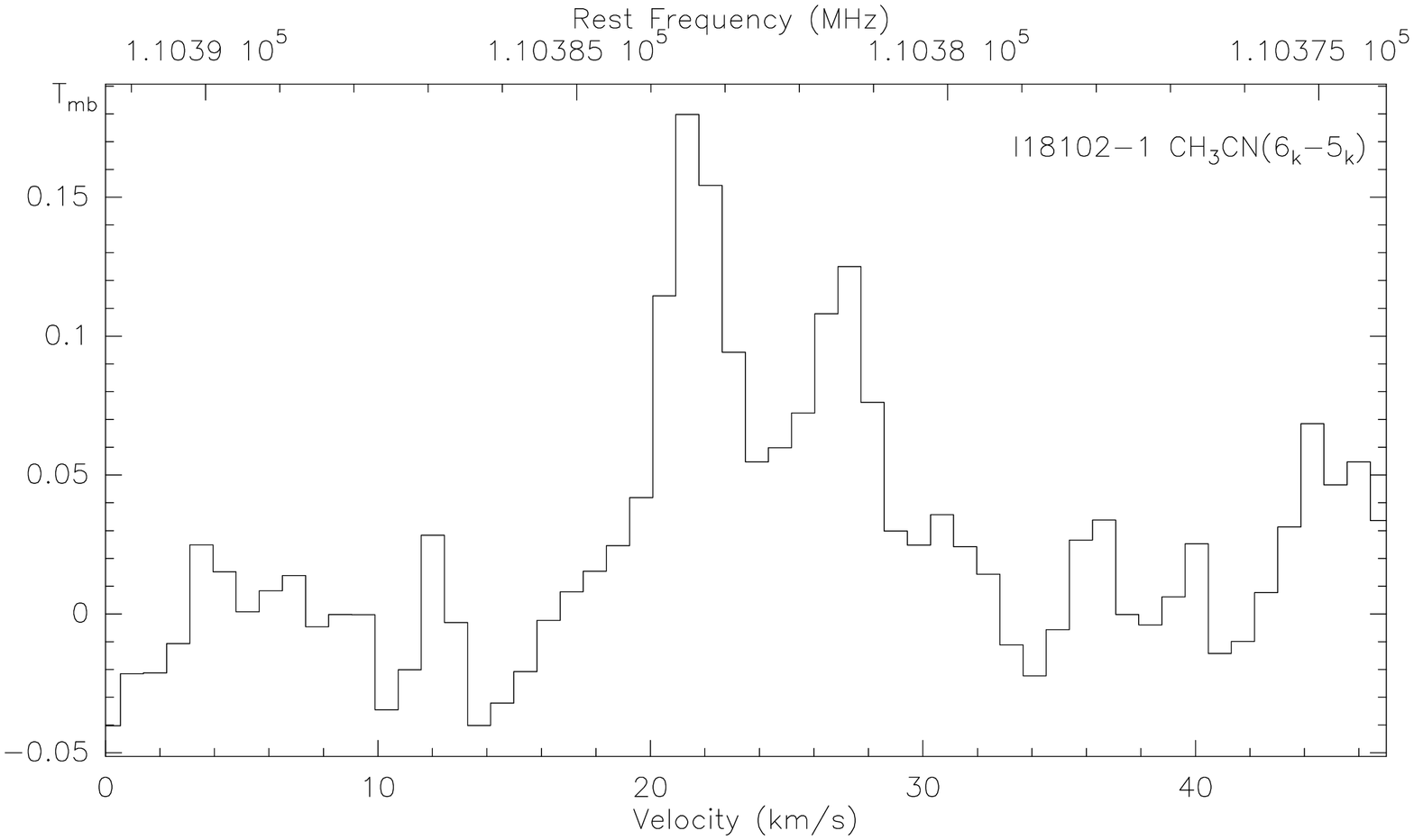}
\includegraphics[angle=-90,width=5.4cm]{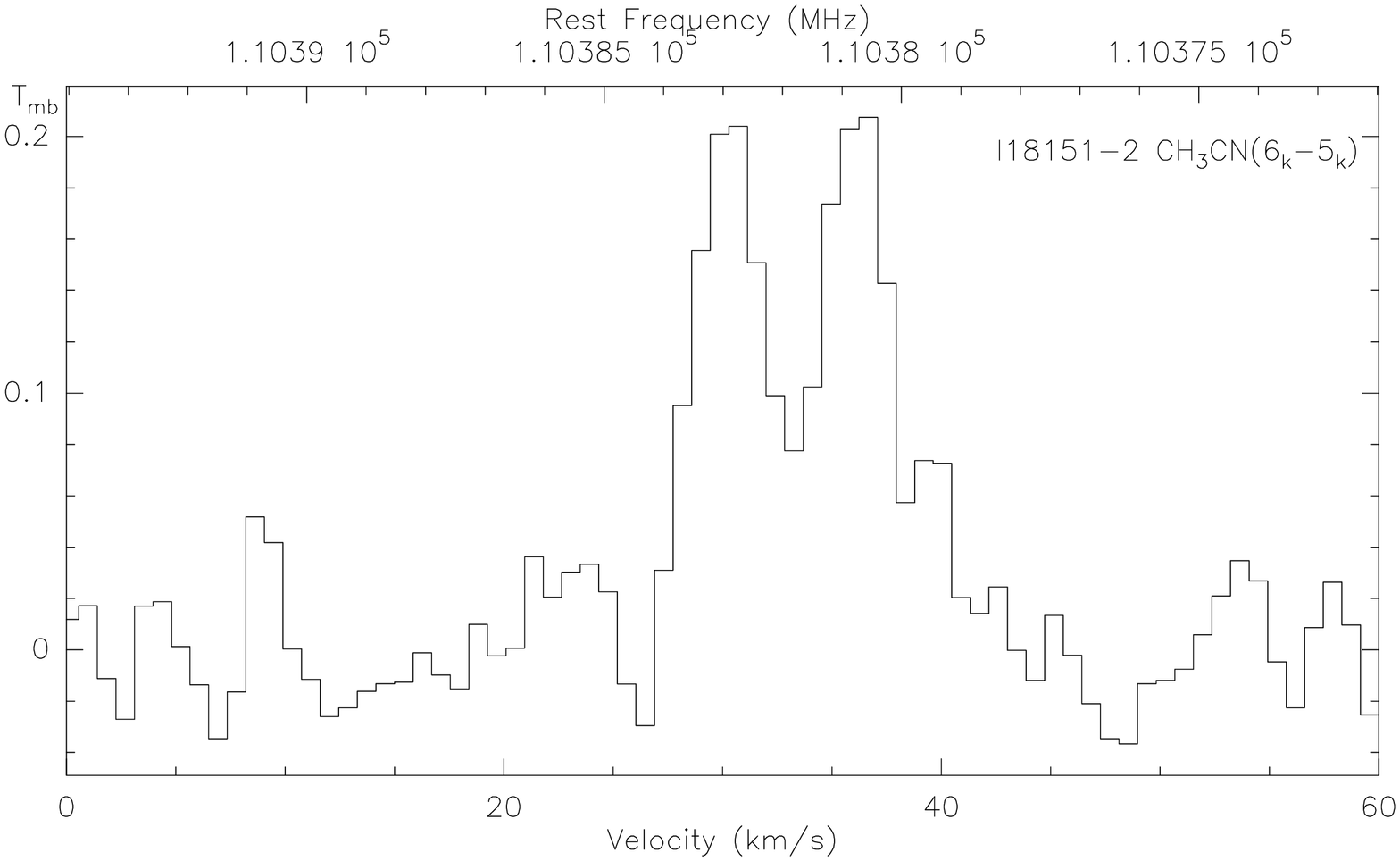}
\includegraphics[angle=-90,width=5.4cm]{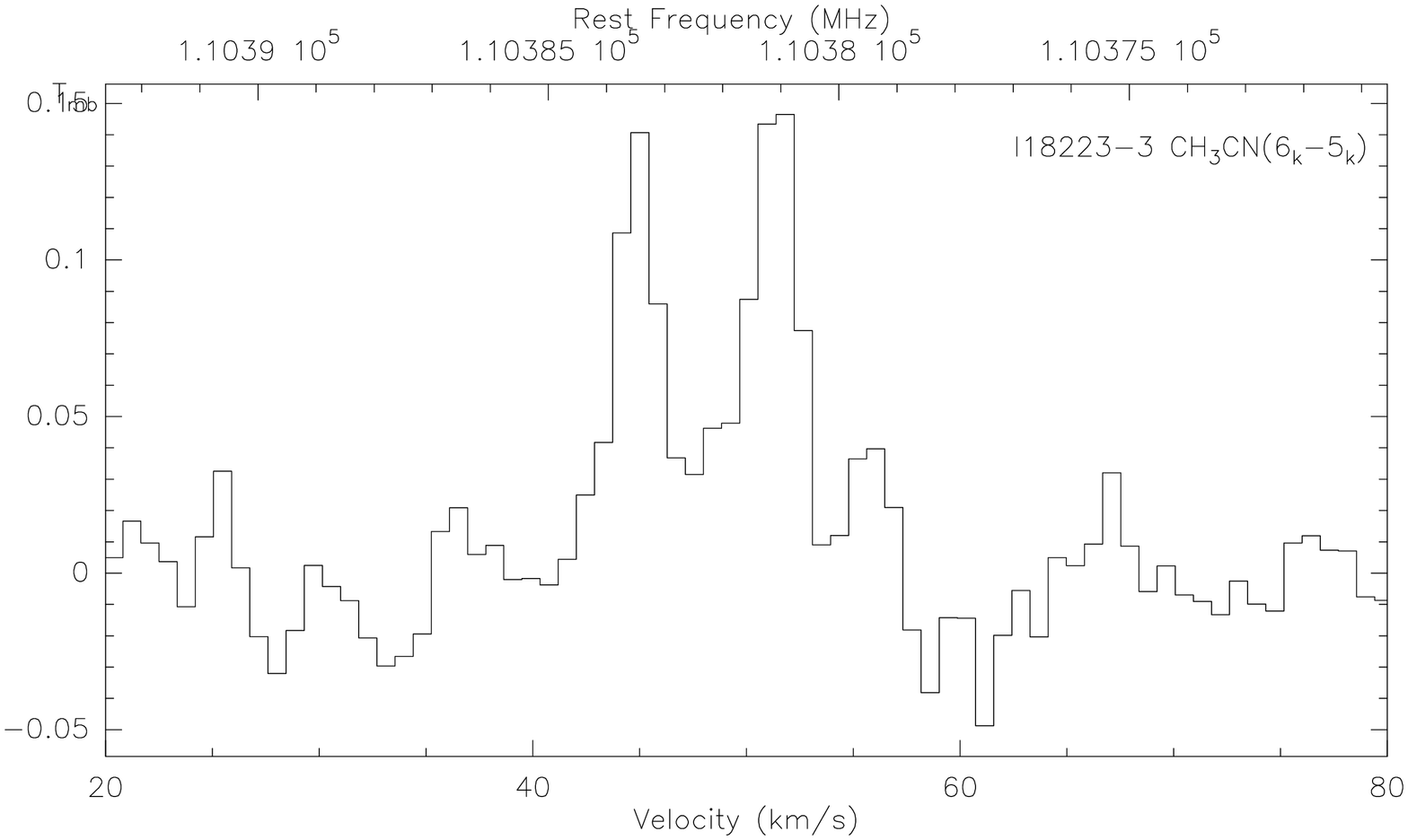}\\
\includegraphics[angle=-90,width=5.4cm]{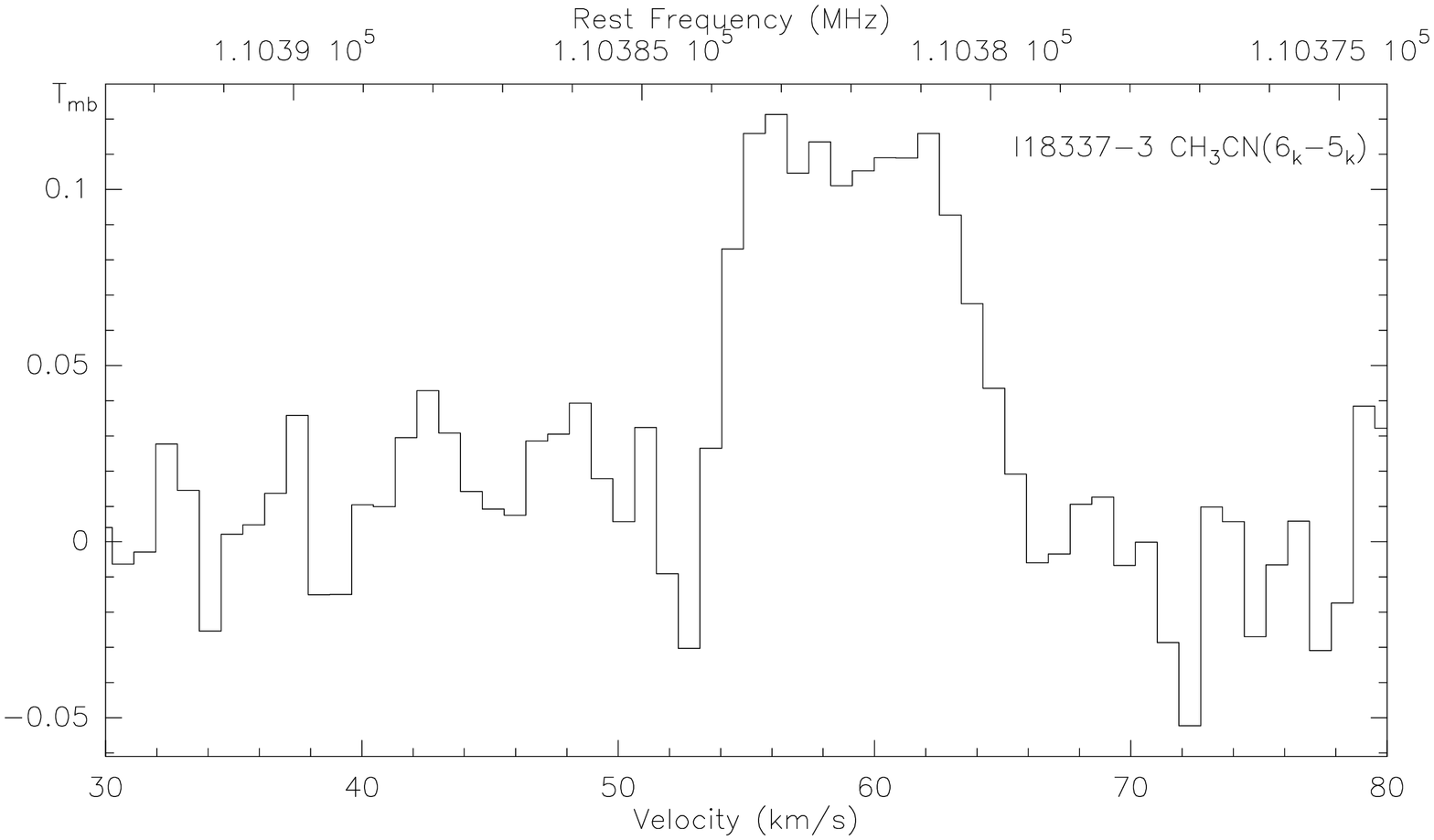}
\includegraphics[angle=-90,width=5.4cm]{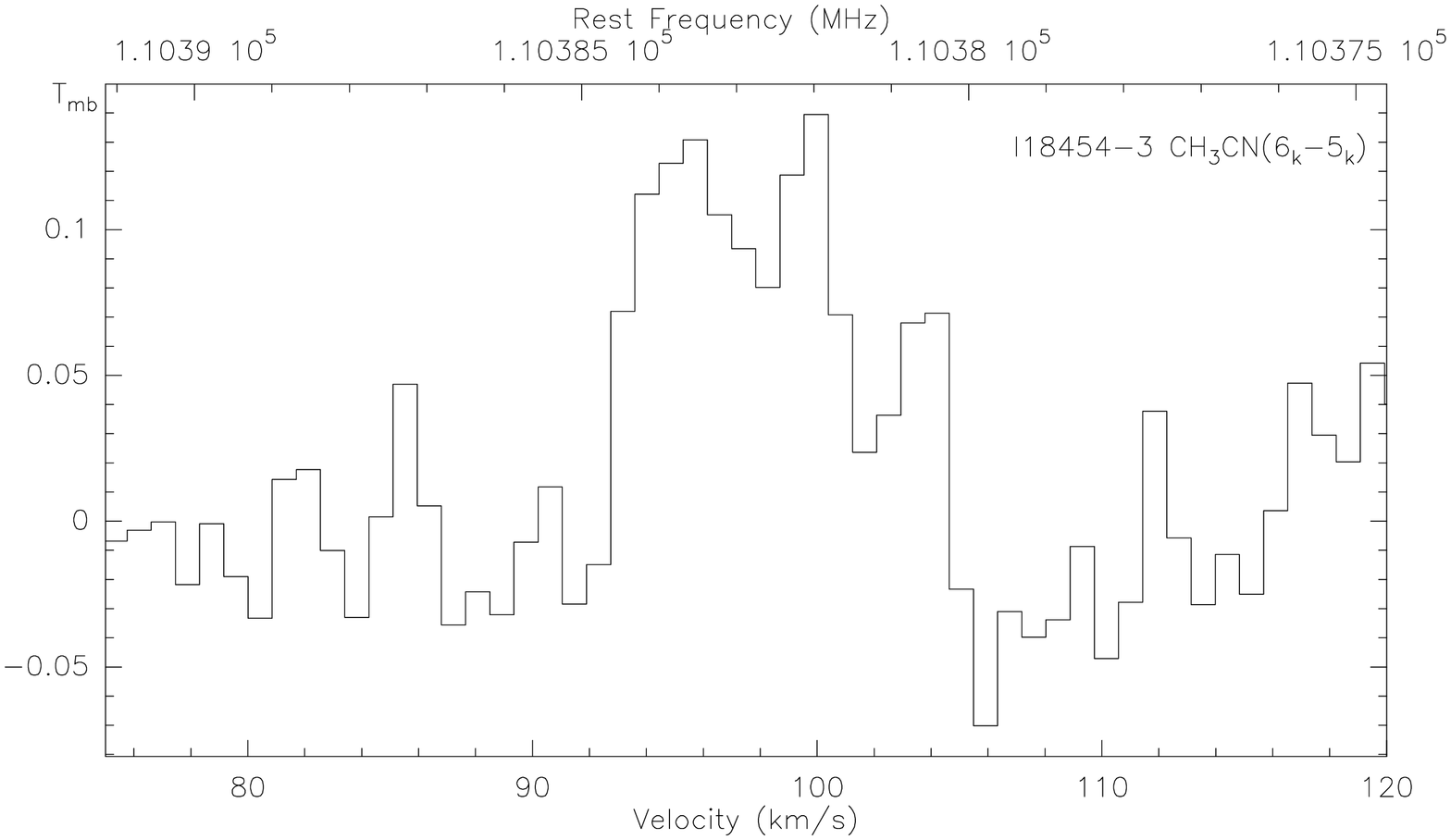}
\includegraphics[angle=-90,width=5.4cm]{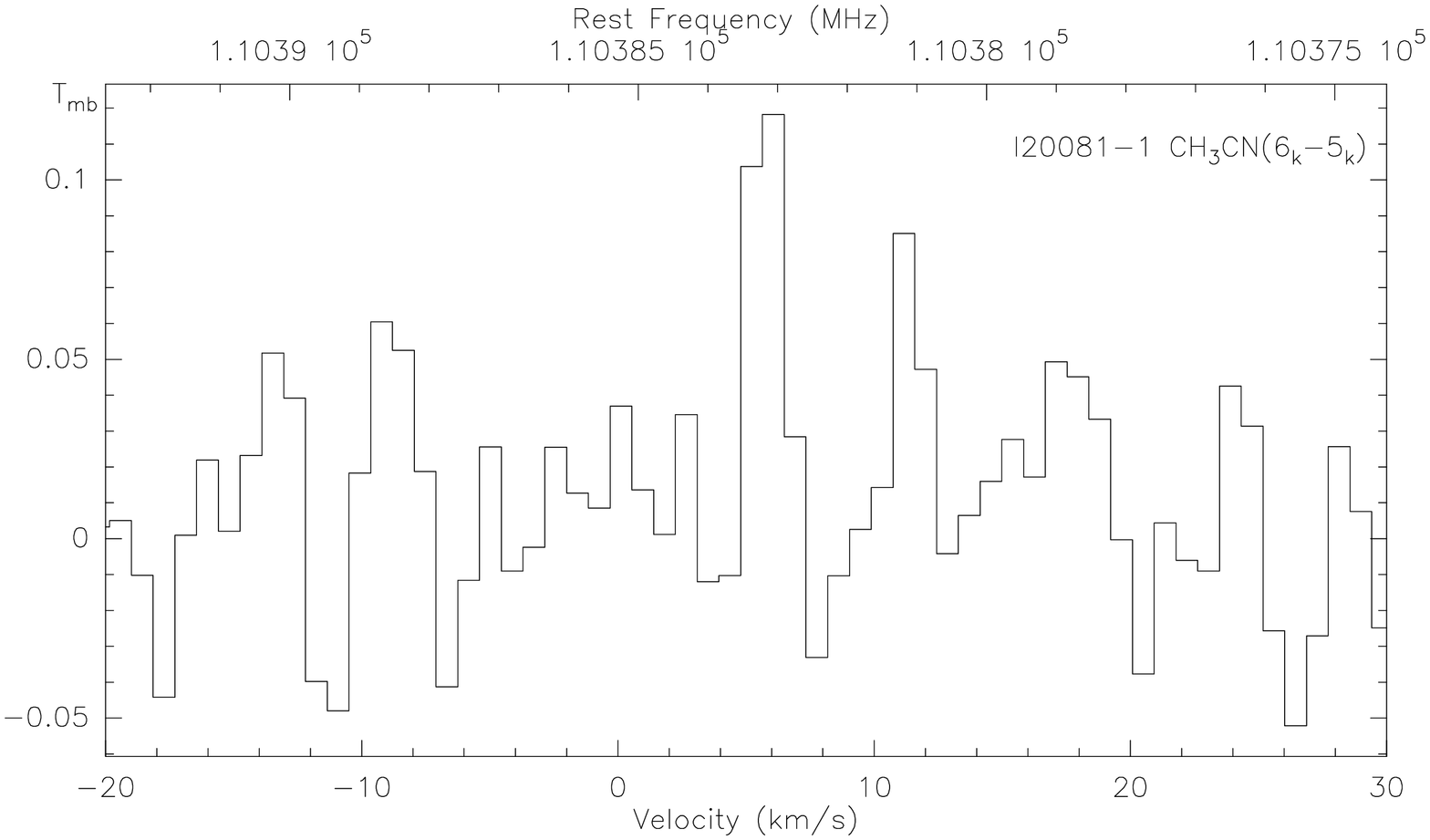}
\caption{CH$_3$CN$(6_k-5_k)$ detections. The rest frequencies of the
  $k=0, 1, 2$ components are at 110383.522, 110381.404 and
  110375.052\,MHz, respectively. Because of the small intrinsic
  line-widths, the $k=0,1$ components are spectrally resolved which is
  usually not the case for more evolved High-Mass Protostellar Objects
  or Hot Molecular Cores.}
\label{ch3cn}
\end{figure*}

\begin{figure*}[h]
\includegraphics[angle=-90,width=5.4cm]{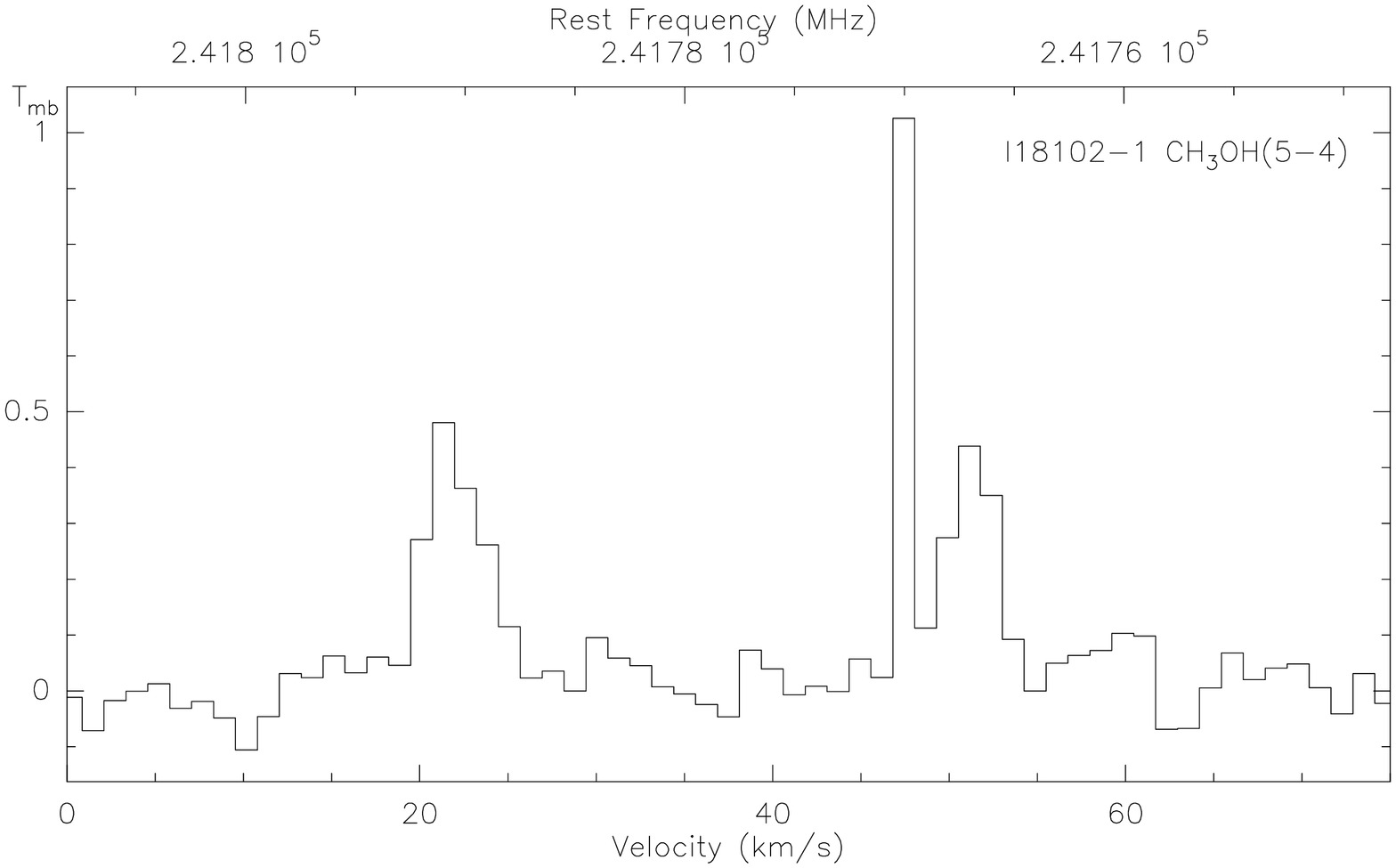}
\includegraphics[angle=-90,width=5.4cm]{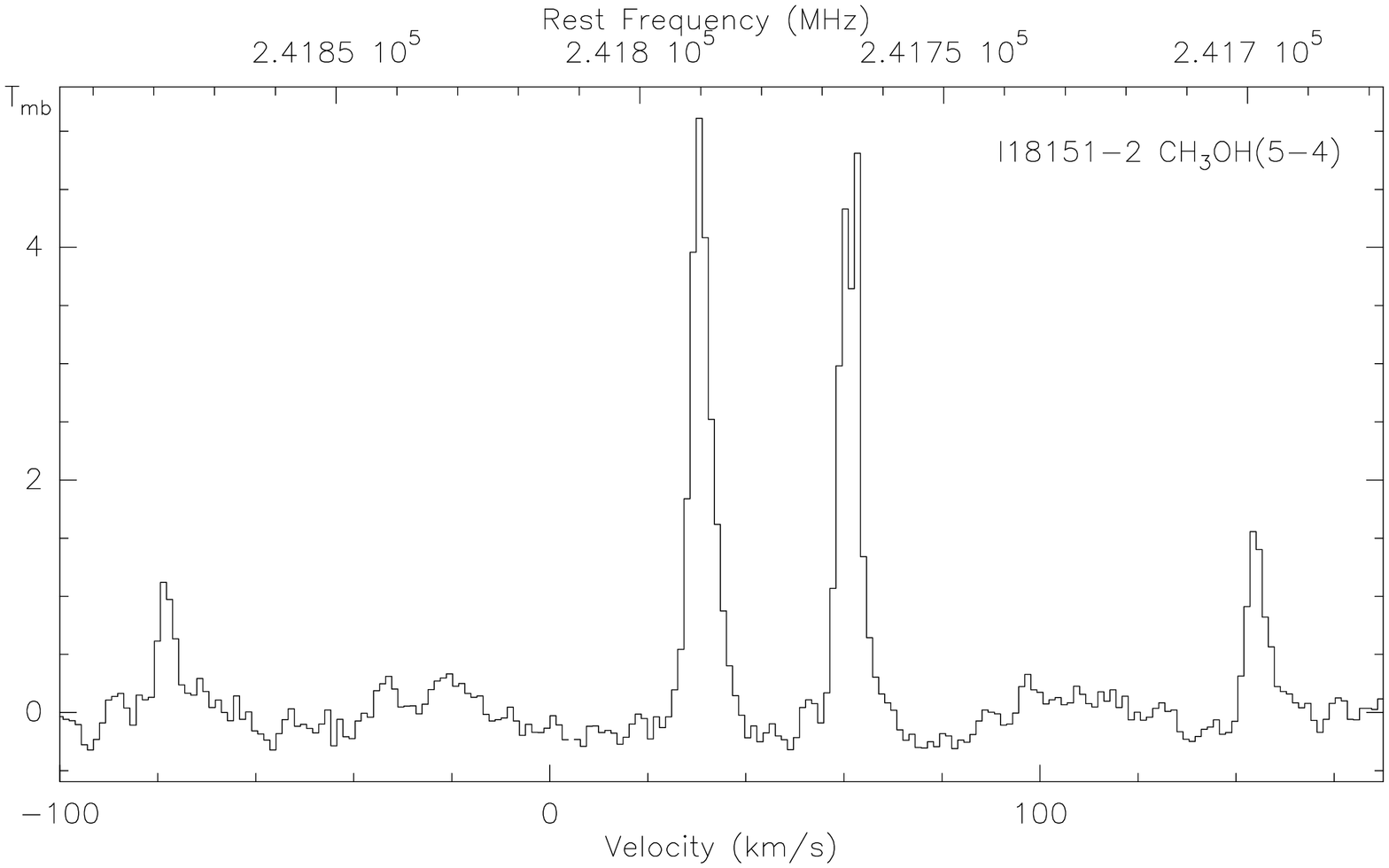}
\includegraphics[angle=-90,width=5.4cm]{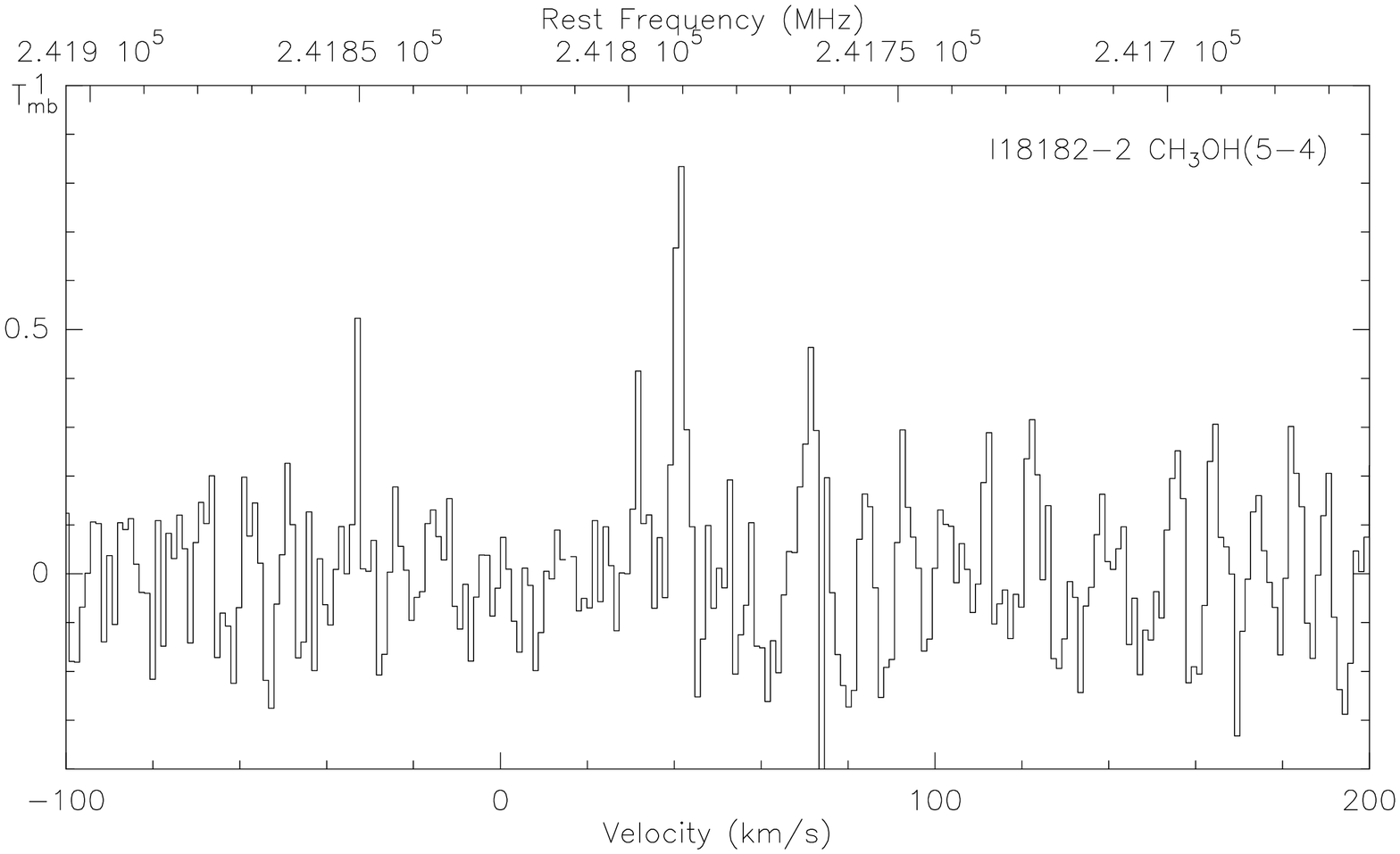}\\
\includegraphics[angle=-90,width=5.4cm]{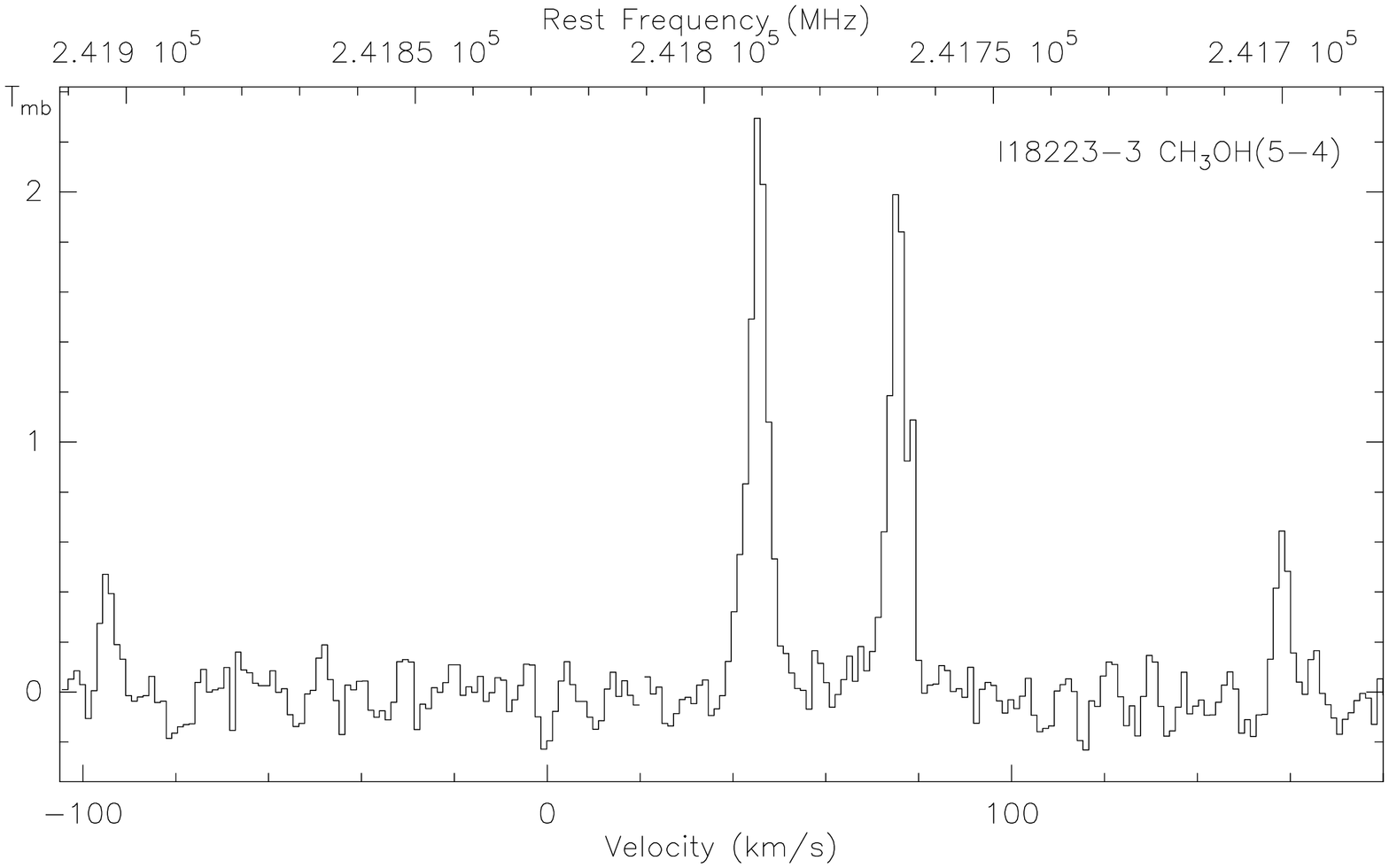}
\includegraphics[angle=-90,width=5.4cm]{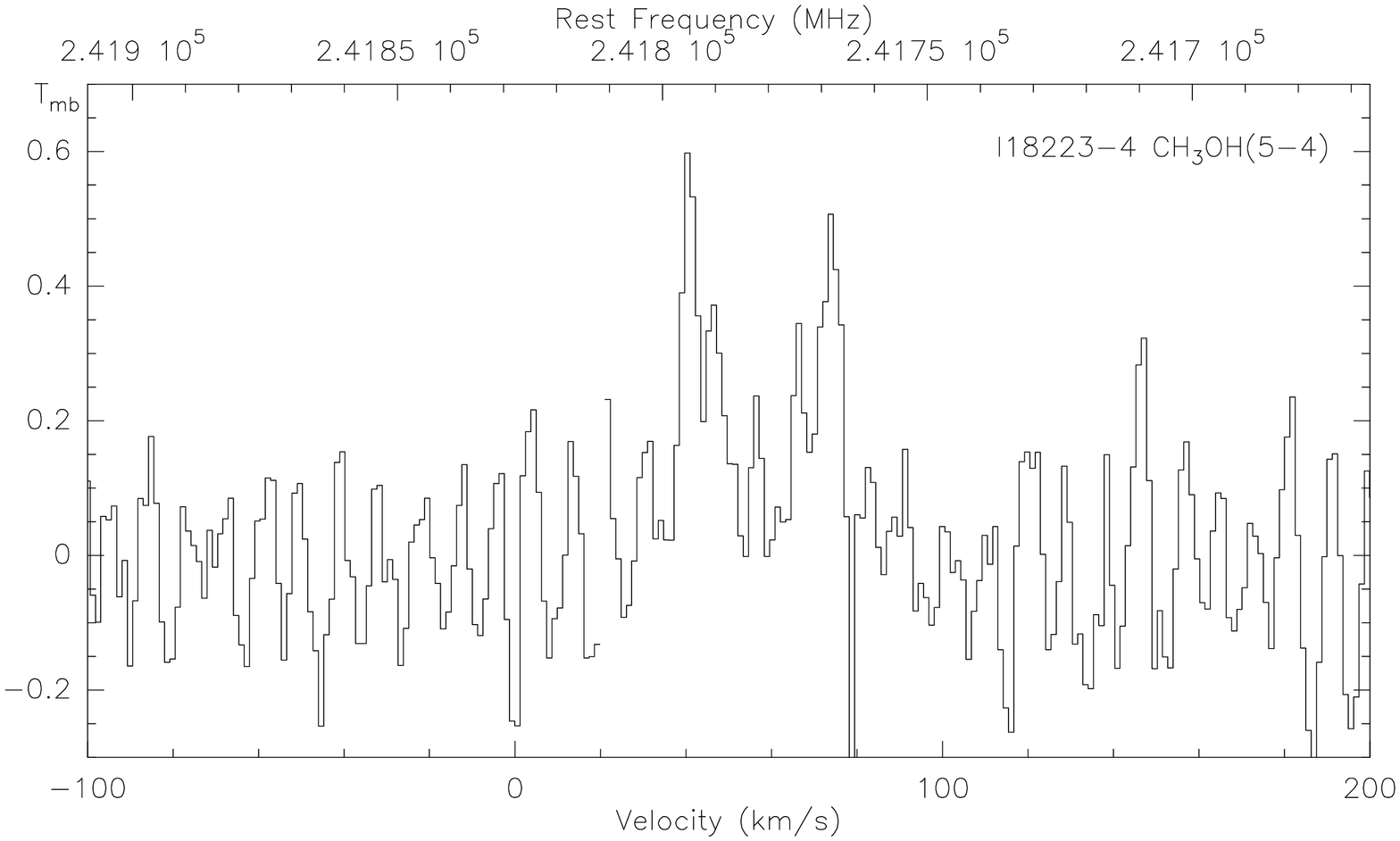}
\includegraphics[angle=-90,width=5.4cm]{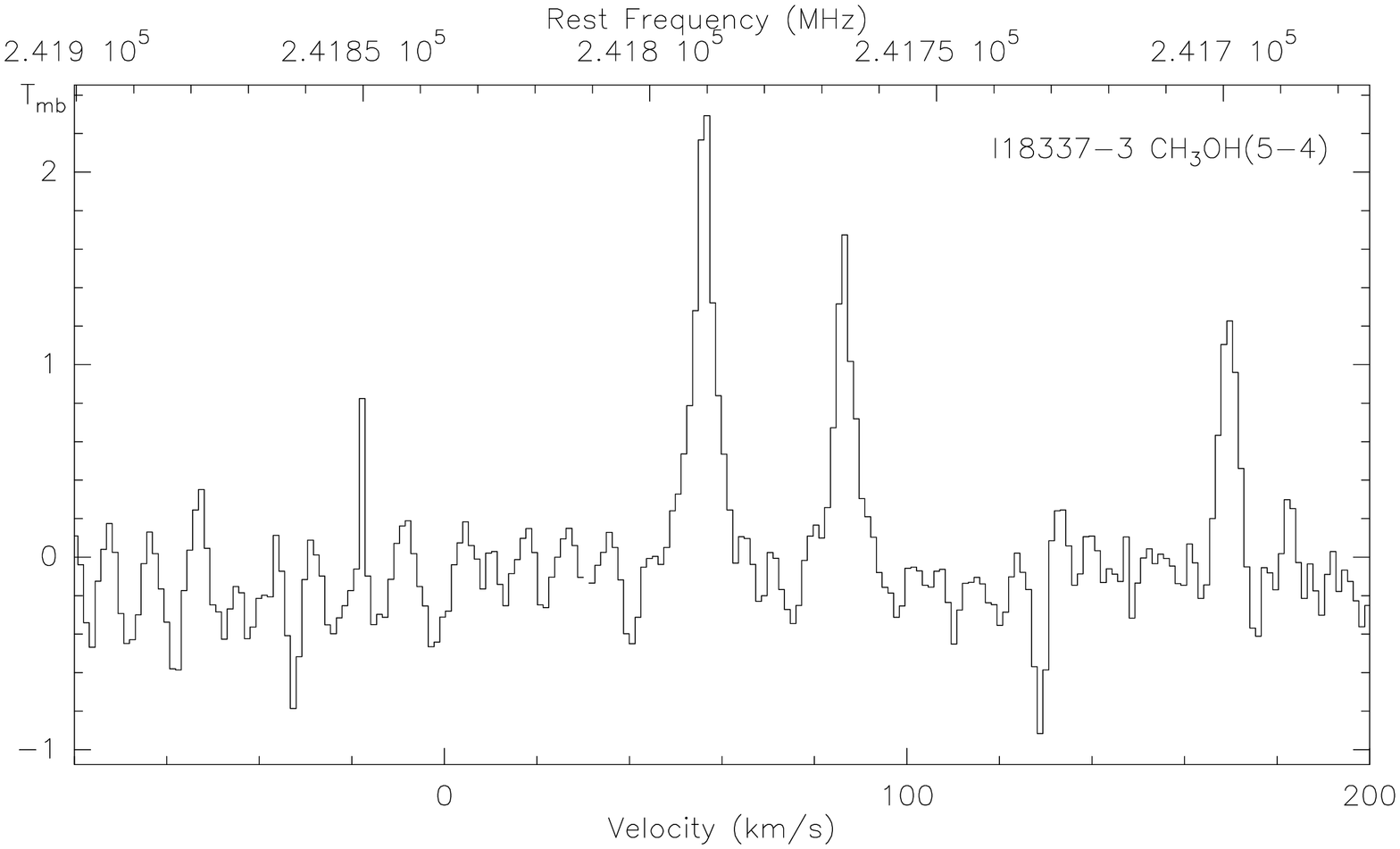}\\
\includegraphics[angle=-90,width=5.4cm]{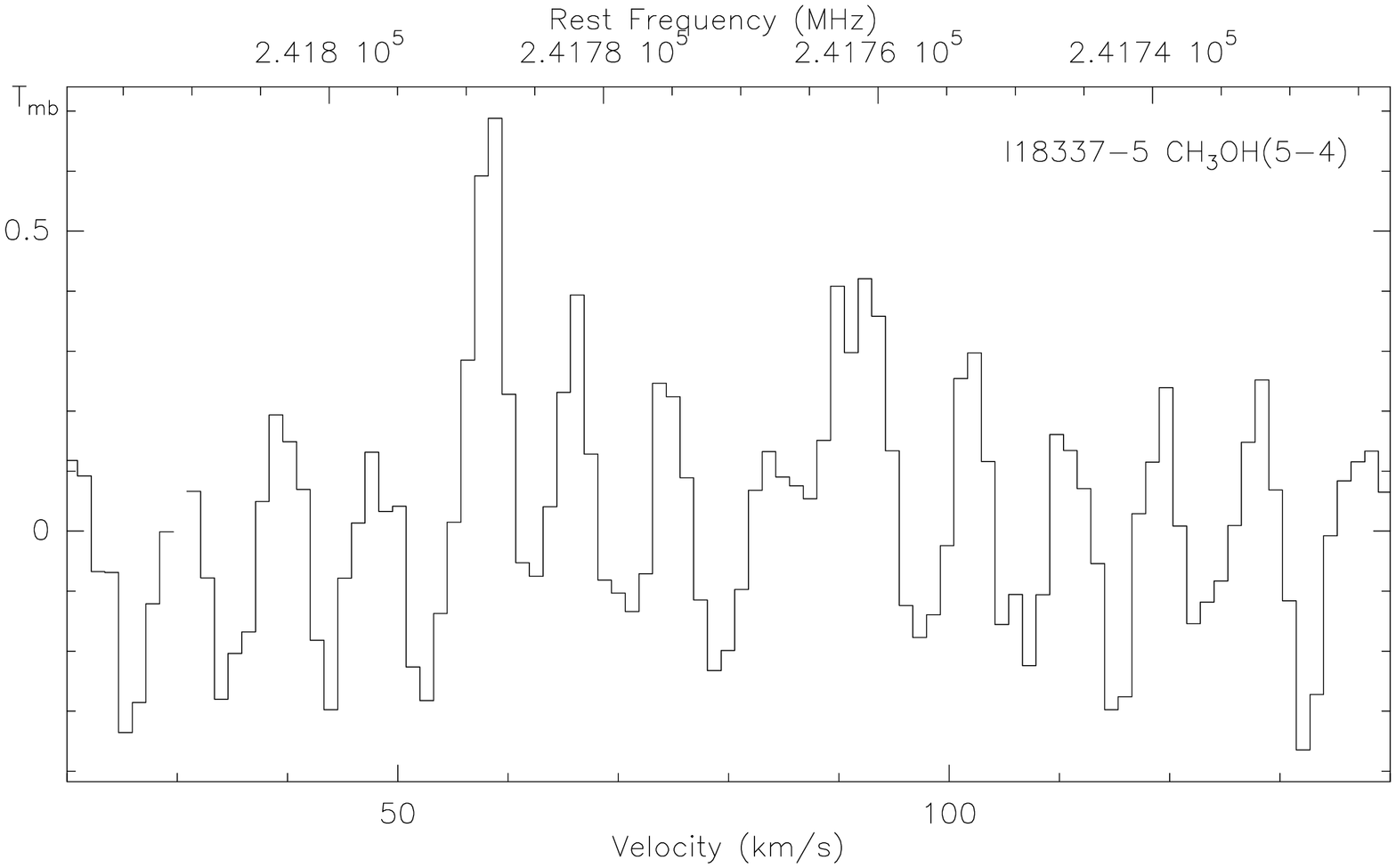}
\includegraphics[angle=-90,width=5.4cm]{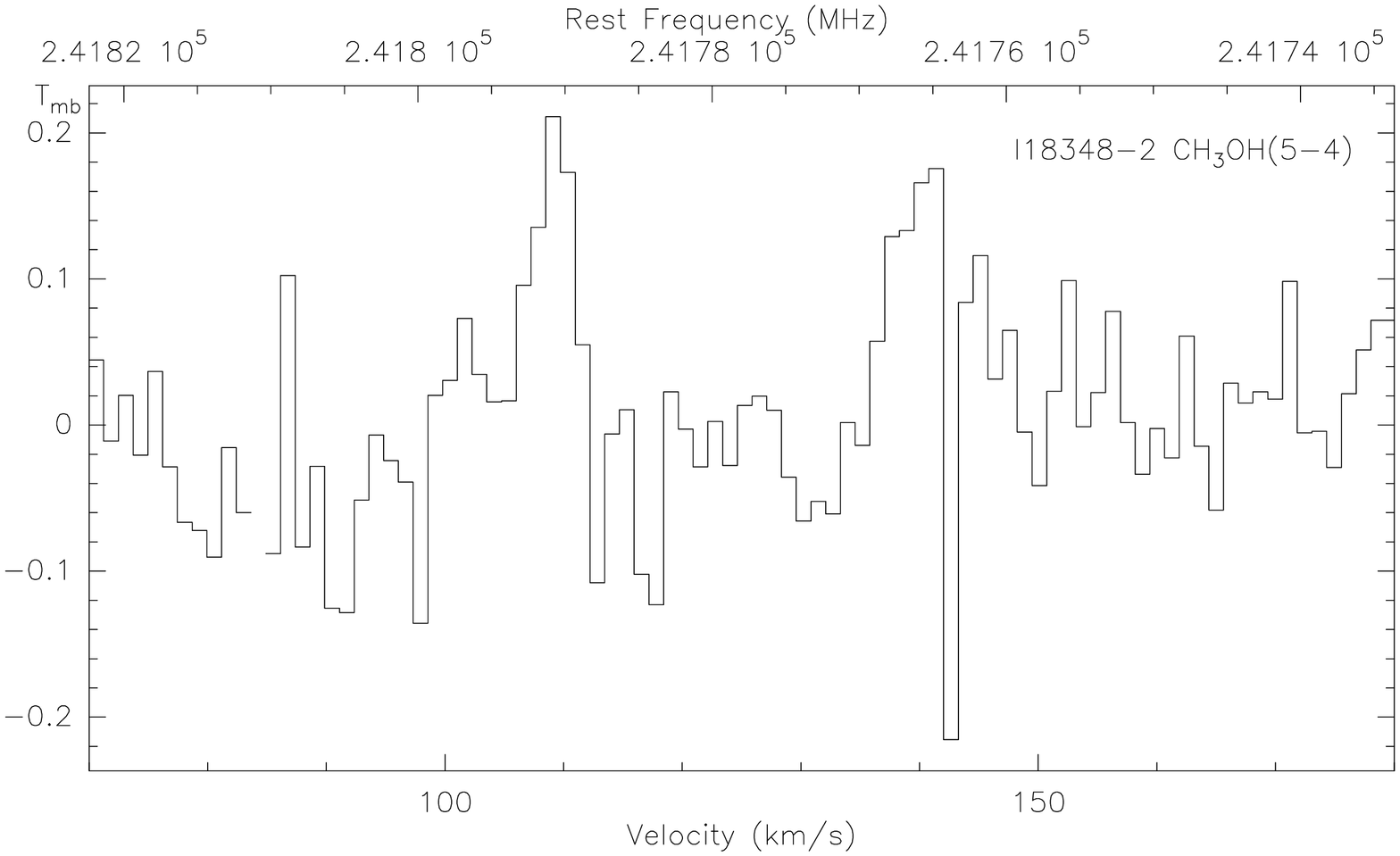}
\includegraphics[angle=-90,width=5.4cm]{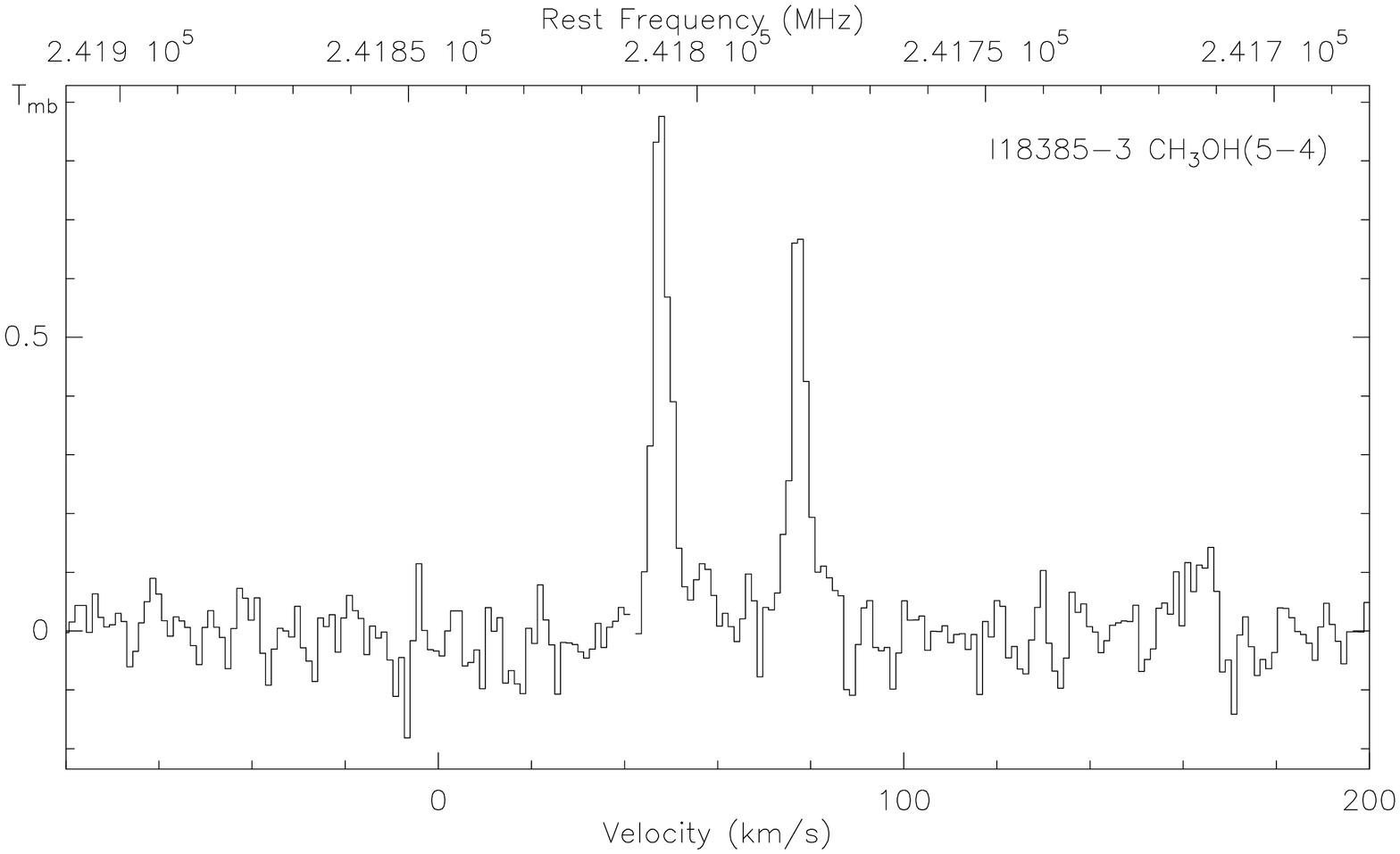}\\
\includegraphics[angle=-90,width=5.4cm]{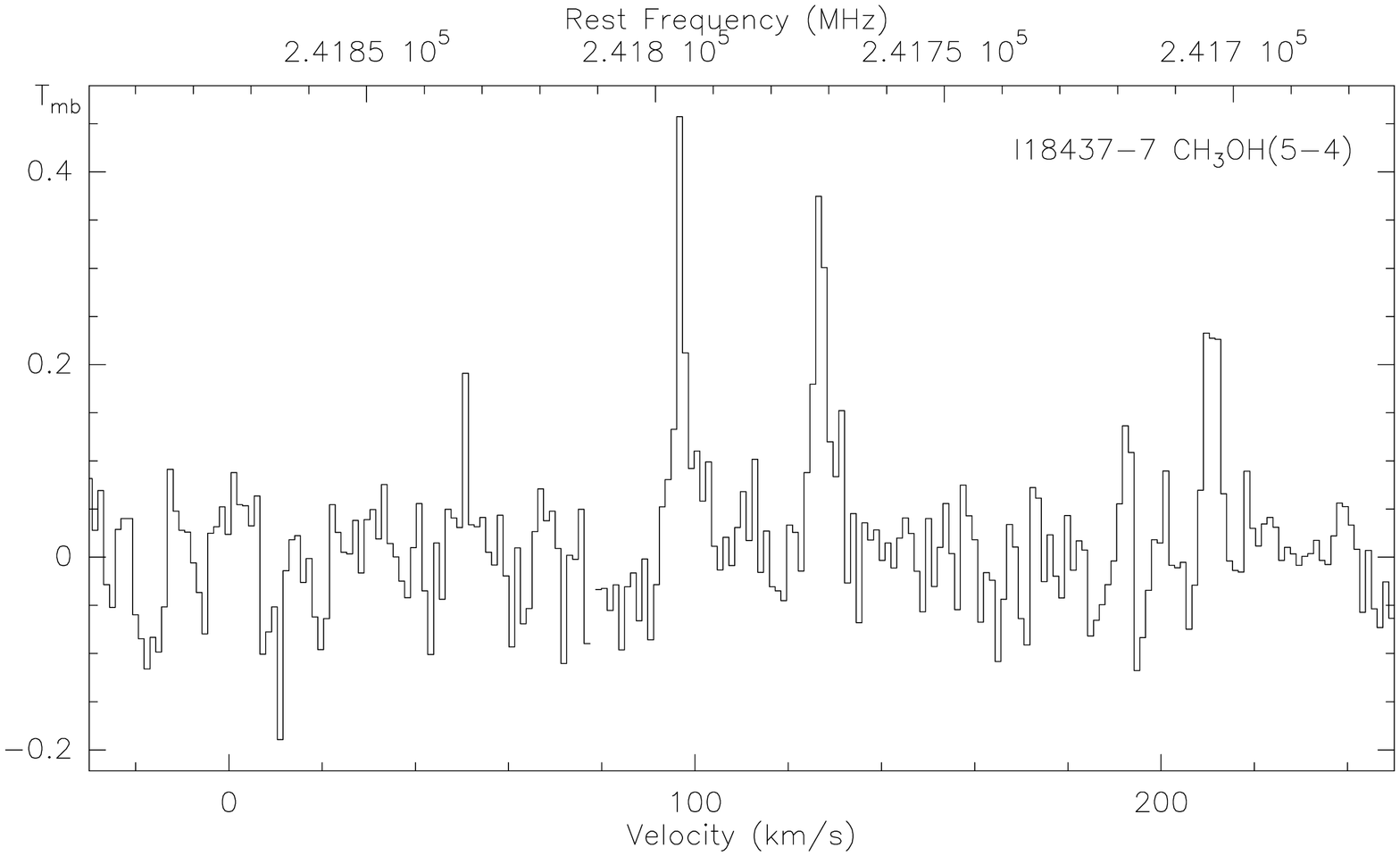}
\includegraphics[angle=-90,width=5.4cm]{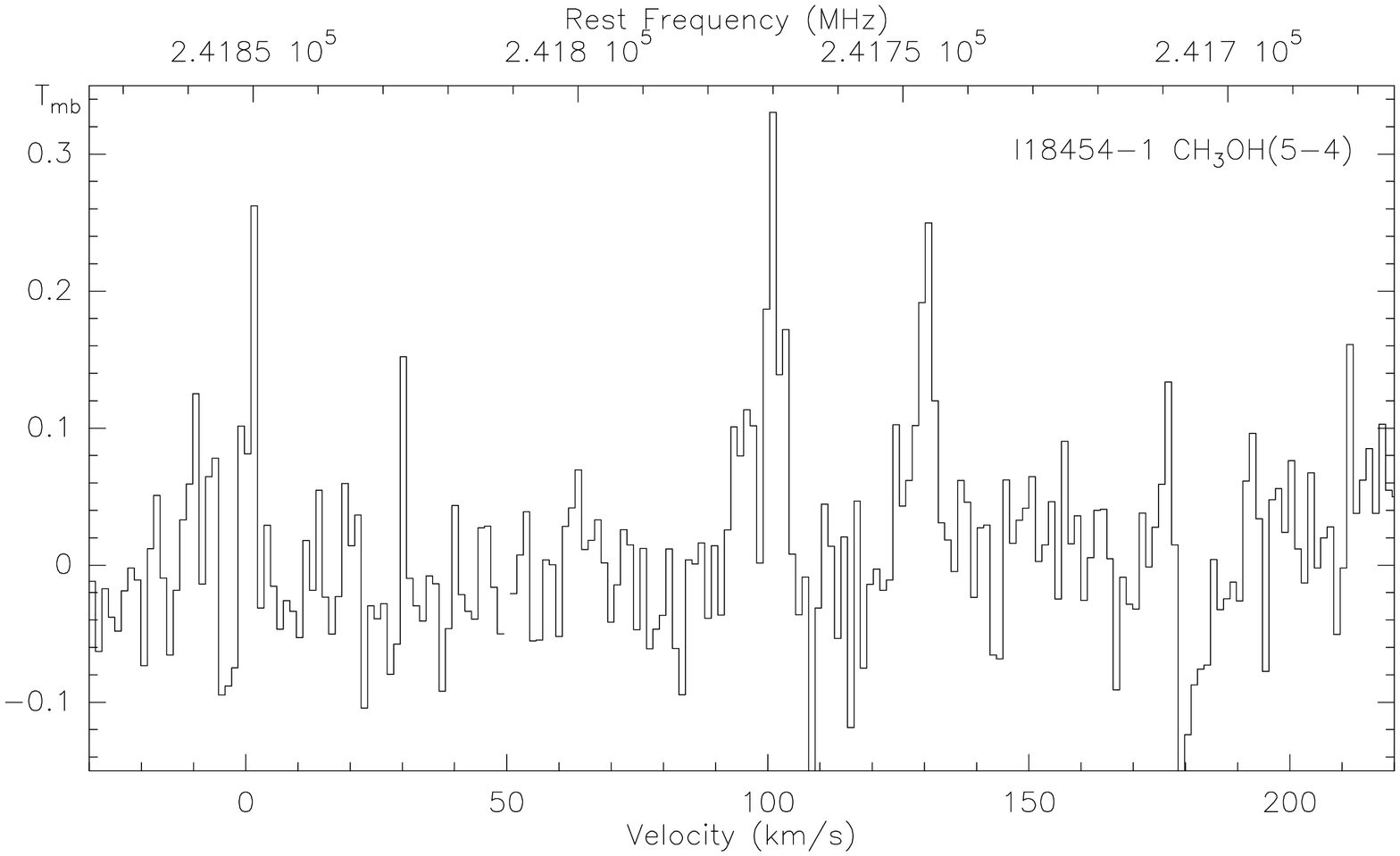}
\includegraphics[angle=-90,width=5.4cm]{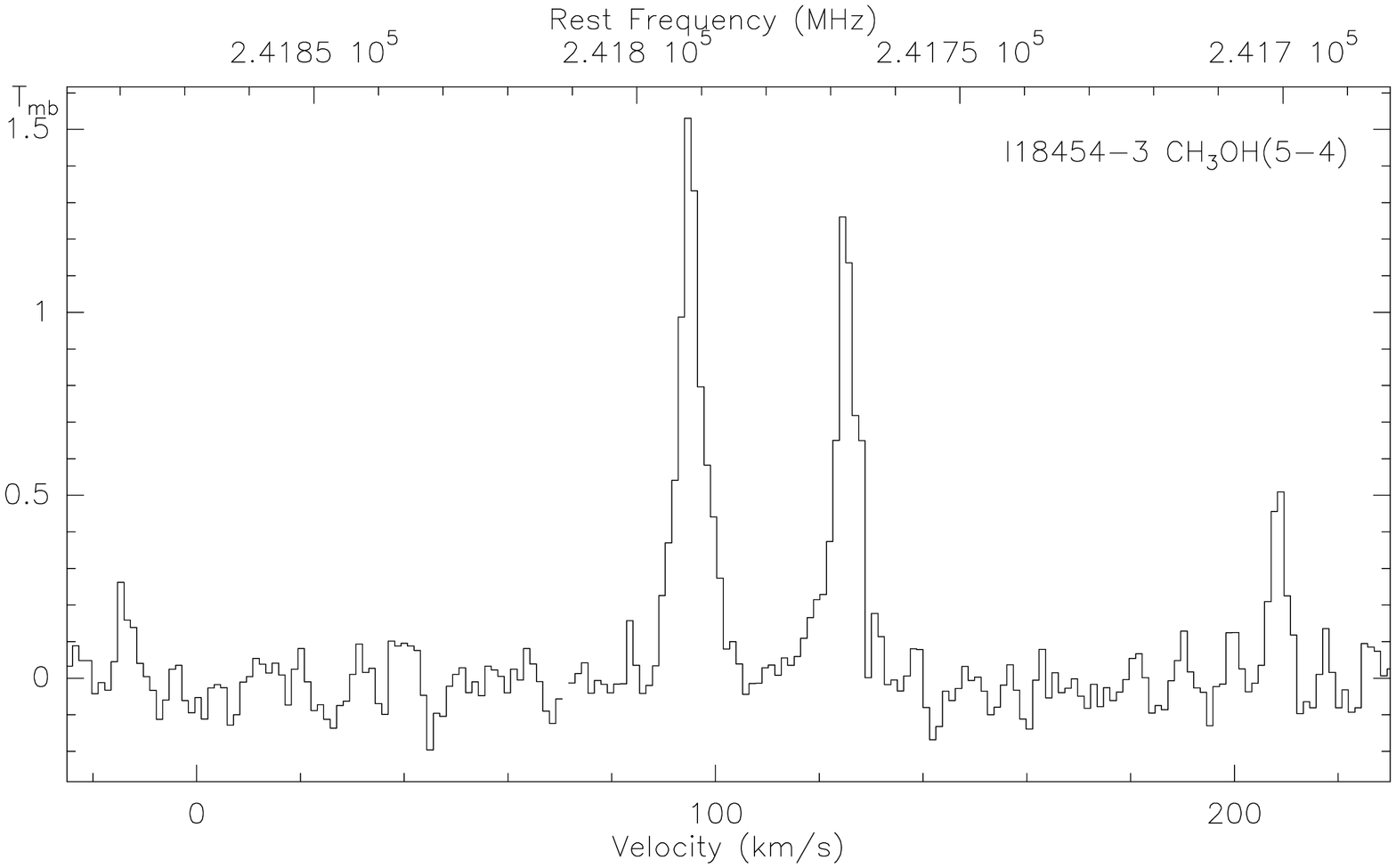}\\
\includegraphics[angle=-90,width=5.4cm]{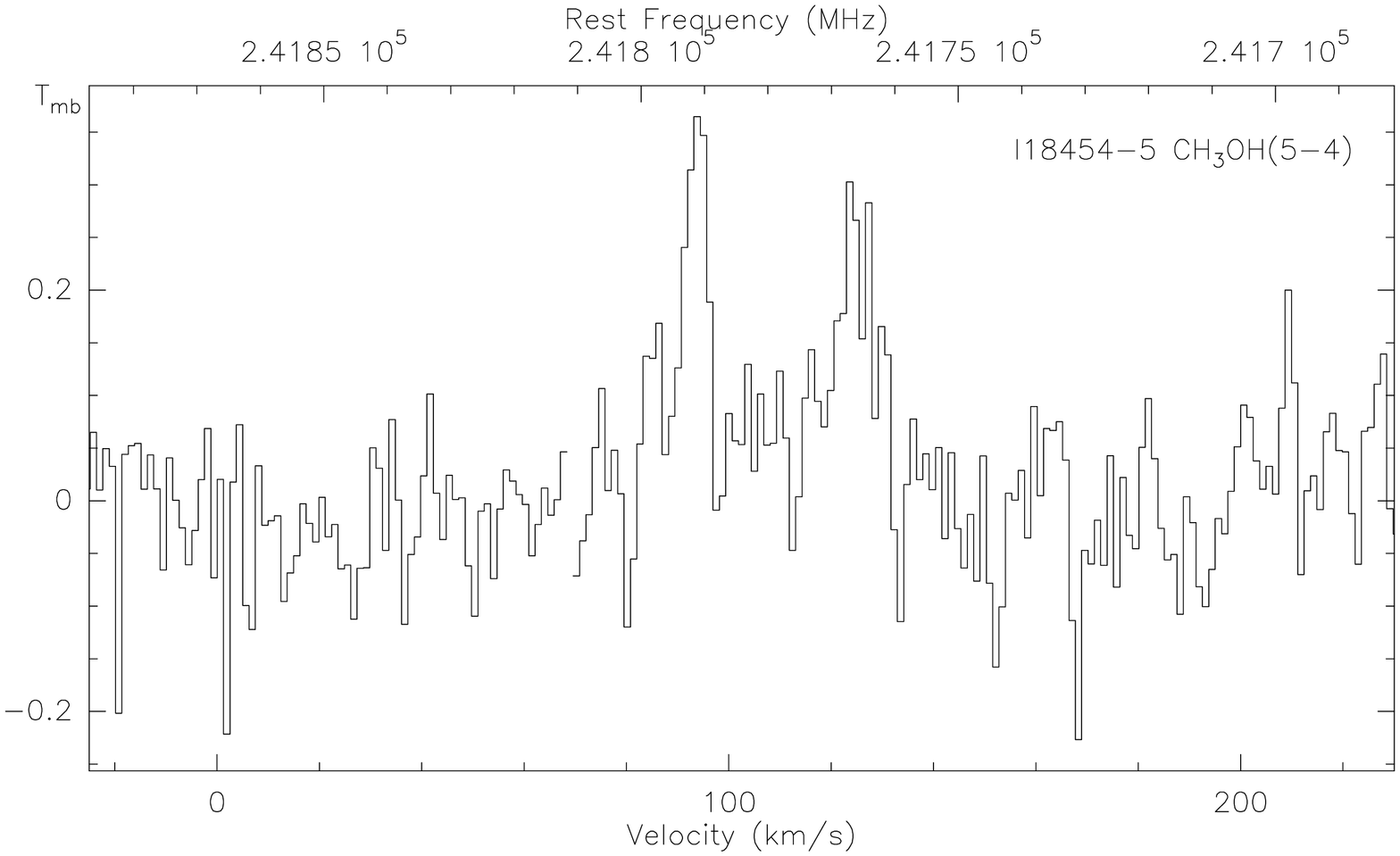}
\includegraphics[angle=-90,width=5.4cm]{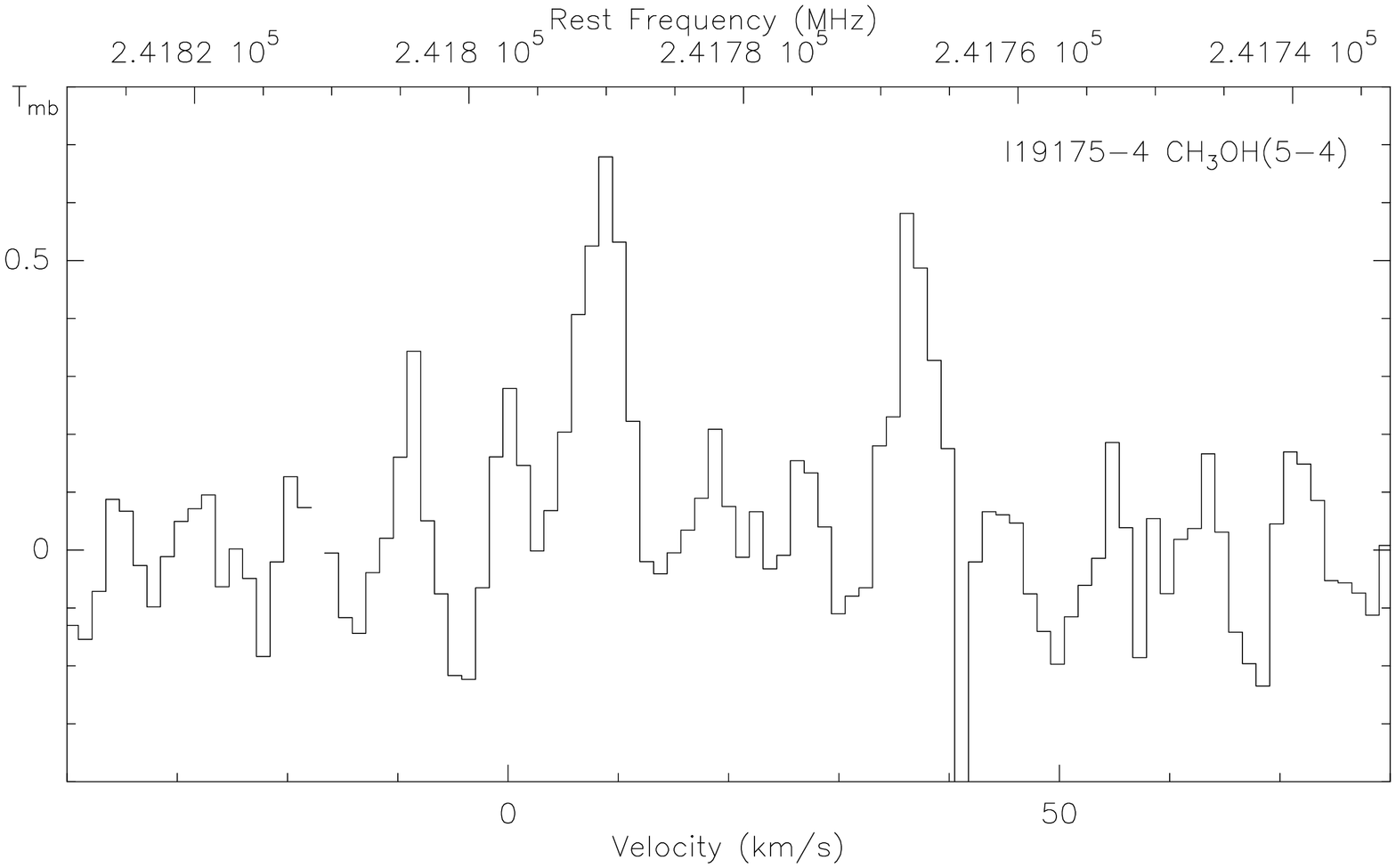}
\includegraphics[angle=-90,width=5.4cm]{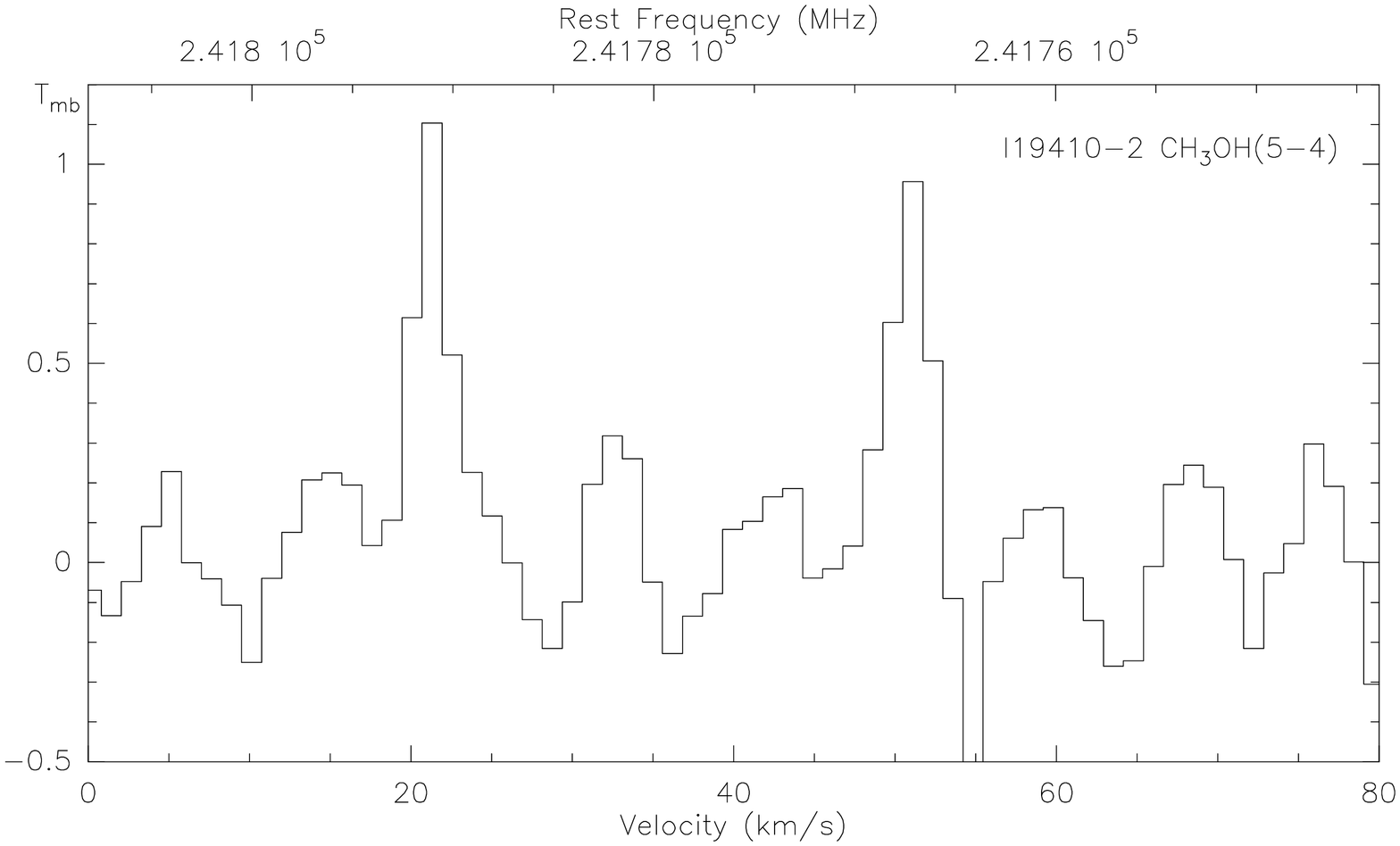}\\
\includegraphics[angle=-90,width=5.4cm]{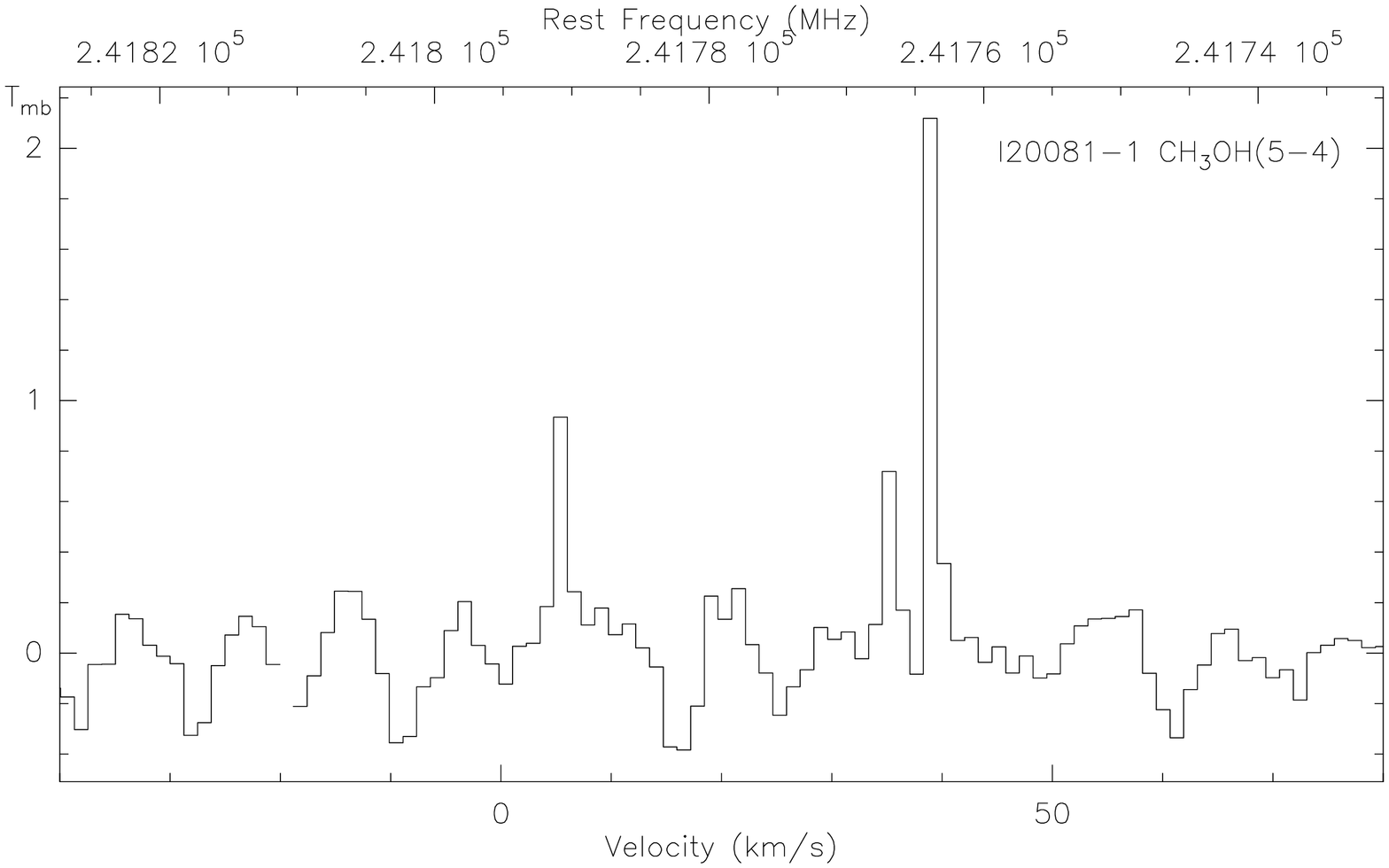}
\includegraphics[angle=-90,width=5.4cm]{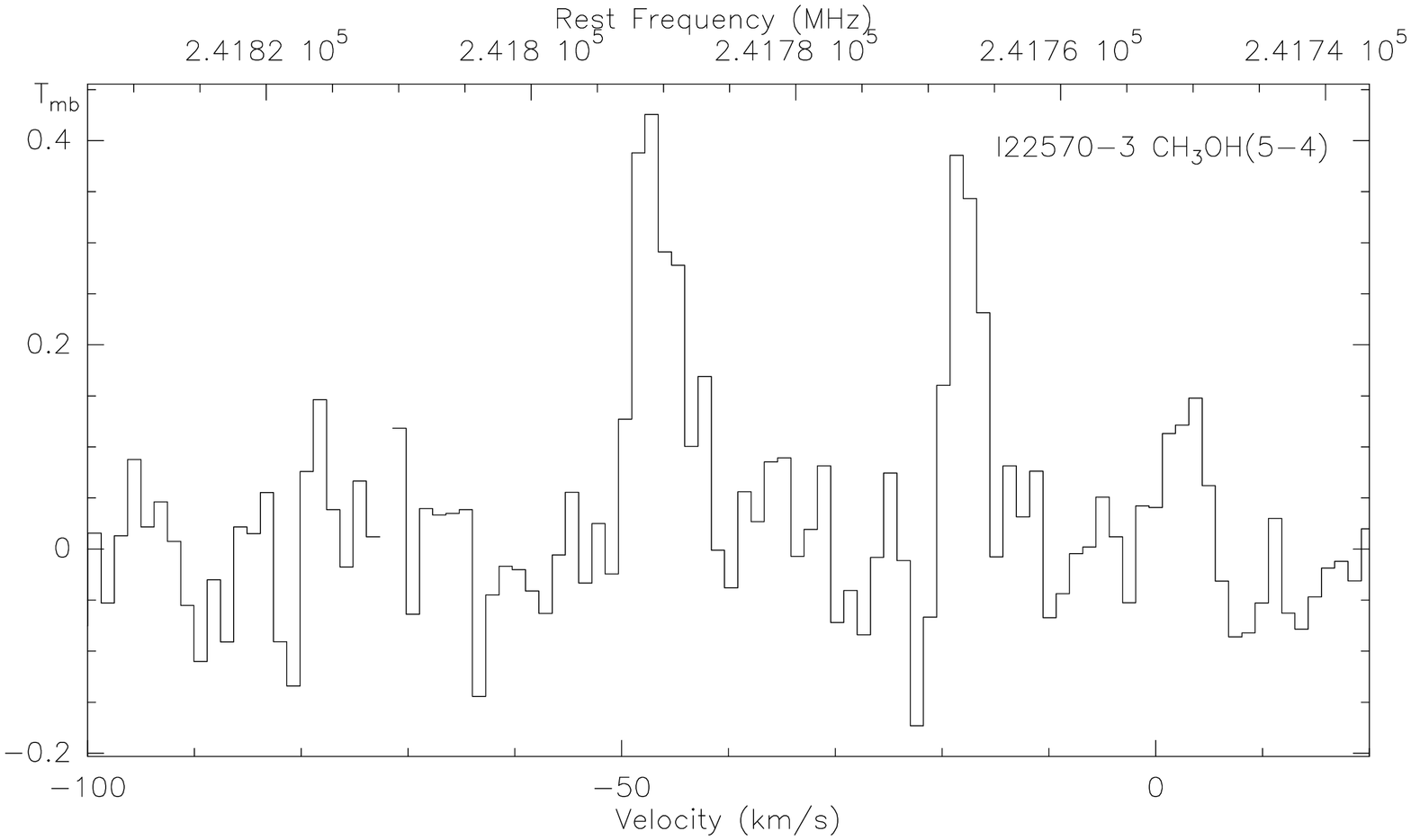}
\caption{CH$_3$O(5--4) detections. The rest frequencies of the four
  detected lines are: CH$_3$OH$(5_1-4_1)$E at 241879.073\,MHz,
  CH$_3$OH$(5_0-4_0)$A+ at 241791.431\,MHz, CH$_3$OH$(5_{-1}-4_{-1})$E
  at 241767.224\,MHz, CH$_3$OH$(5_0-4_0)$E at 241700.219\,MHz. The few
  single-channel spikes are artifacts caused by the correlator.}
\label{ch3oh}
\end{figure*}

These expectations are largely confirmed by our observations (Figs.
\ref{ch3cn} \& \ref{ch3oh}), we detect CH$_3$CN only toward 6 sources
($\sim$14\%) whereas CH$_3$OH is observed toward 17 regions
($\sim$40\%). Furthermore, if detected, CH$_3$CN$(6_k-5_k)$ is
observed only in its two lowest $k$-components with excitation
temperatures $<30$\,K, the $k=2$ transition is detected in no source
at all. Similarly, for CH$_3$OH, we detected toward the 17 regions the
two lower excitation lines, whereas the two higher excitation lines
are detected only toward a small sub-sample of 5 regions. This is
additional confirmation of the youth and low average temperatures of
the selected source sample. Table \ref{sourceparameters2} lists the
line-widths $\Delta\nu_{k=0}$ and $\Delta\nu_{k=1}$ of the CH$_3$CN
$k=0,1$ components from Gaussian fits to each line. While we cannot
fit one source because of multiple components and in one other source
$\Delta\nu_{k=0}$ and $\Delta\nu_{k=1}$ are of approximately equal
width, for the four other sources we find consistently larger
line-widths in the $k=1$ component compared to the $k=0$ component.
This is consistent with already ongoing star formation activity in
these regions, where the warmer more central gas is subject to more
non-thermal motion, either of turbulent nature or due to infall,
rotation and/or outflow processes.

One can try to use the observed lines also to get more quantitative
temperature and column density estimates. Following \citet{loren1984},
assuming optically thin emission and that a single temperature
characterizes the populations within a $k$-ladder and between
different $k$-ladders, the ratios of the intensities of different $k$
lines can be used to estimate this temperature. Specifically, for the
$J=6-5$ line and $k=0$ and $k=1$ components, we have

$$T = 7.2/ln(I_0/I_1)$$

with $I_k$ the corresponding observed peak intensities.  For
IRDC\,18102$-$1800-1 and IRDC\,20081+2720-2, the relative line
intensities are consistent with optically thin emission and we
estimate the temperatures to be $16\pm 16$ and $16\pm 13$\,K,
respectively. For the other sources, while they are unlikely to be
optically thick, the intensities are inconsistent with optically thin
emission, so we do not estimate temperatures. It is interesting that
the estimates, where possible, although not very accurate, turn out to
be low and similar to estimates from NH$_3$ observations
\citep{sridharan2005}. We cannot use the two mainly detected CH$_3$OH
lines for temperature estimates because they are of different symmetry
(A- and E-CH$_3$OH) and can be considered for radiative transport
purposes as distinct molecules.

However, we can calculate the column densities in the detected
CH$_3$OH and CH$_3$CN lines, and then derive the total molecular
column densities following \citet{rohlfs2006}. As partition functions
we used the values listed in the JPL catalog at $T=18.75$\,K
\citep{poynter1985}. Furthermore, we are interested in CH$_3$OH and
CH$_3$CN molecular abundances. To estimate these, we extracted the
1.2\,mm peak fluxes from the images presented in
\citet{beuther2002a}\footnote{\citet{beuther2002a} presented peak
  fluxes from 2D Gaussian fits to the images whereas here we use the
  exact fluxes toward the peak positions. The former are on average a
  bit lower.}. Following \citet{hildebrand1983} and
\citet{beuther2002erratum}, we then calculated the H$_2$ column
densities at $T=19$\,K (corresponding to the partition functions for
the molecular emission) assuming optically thin emission and a dust
opacity index $\beta=2$. The spatial resolution of the 1.2\,mm and
CH$_3$OH data is approximately the same, for the comparison with the
lower-resolution CH$_3$CN data, we smoothed the 1.2\,mm images also to
the corresponding spatial resolution of $23''$. The 1.2\,mm peak
fluxes and H$_2$ column densities (both at $11''$ resolution) are
listed in Table \ref{sourceparameters}. The different H$_2$ column
densities were then used to estimate the CH$_3$OH and CH$_3$CN
abundances. The calculated molecular column densities are estimated to
be correct within a factor 5, the accuracy of the estimated abundances
is correspondingly worse. Table \ref{sourceparameters2} lists the
derived molecular CH$_3$OH and CH$_3$CN column densities and
abundances which are on average relatively low.

The average CH$_3$CN column density and abundance are
$3.1\times10^{12}$\,cm$^{-2}$ and $1.2\times 10^{-10}$, respectively.
This CH$_3$CN abundance compares well to the CH$_3$CN abundance
observed for example in the relatively cold and quiescent Orion
extended ridge whereas it is two orders of magnitude below the values
obtained for the Orion hot core \citep{vandishoek1998}. Furthermore,
it fits well to the chemical modeling of young high-mass star-forming
regions by \citet{nomura2004}: they find very low CH$_3$CN column
densities of the order $10^{11}-10^{12}$\,cm$^{-2}$ after only a few
hundred years which then progressively increase to values
$>10^{15}$\,cm$^{-2}$ after several $10^5$ years. In this picture,
even the sources with detected CH$_3$CN emission are as expected
extremely young massive star-forming regions.

The average CH$_3$OH column density and abundance are
$8.2\times10^{13}$\,cm$^{-2}$ and $4.3\times 10^{-10}$, respectively.
This abundance is orders of magnitude below the values obtained toward
the typical regions in Orion (e.g., $8\times10^{-9}$ for the extended
ridge and $2\times10^{-7}$ for the hot core,
\citealt{vandishoek1998}). However, it is very close to the recently
reported CH$_3$OH abundances of $6\times 10^{-10}$ in the low-mass
starless cores L1498 and L1517B \citep{tafalla2006}. Although the
statistics are still small, this similarity between CH$_3$OH
abundances in low-mass starless cores and high-mass IRDCs indicates
that the initial CH$_3$OH molecular abundances at early evolutionary
stages do not differ significantly between low- and high-mass
star-forming regions.

\section{Conclusions and summary}

These single-dish mm spectral multi-line observations toward a sample
of 43 massive IRDCs revealed a range of interesting results. Since IRDC are
expected to harbor sources in different evolutionary stages from
genuine HMSCs to High-Mass cores with embedded low- to
intermediate-mass protostars potentially destined to become massive stars,
it is not surprising that the observational signatures do vary
significantly over the sample. Because our spectral line selection
covered tracers sensitive to different processes (outflows, dense gas,
lower density gas, hot core chemistry) we are able to constrain various
characteristics. In summary they are:

\begin{itemize}

\item Toward 40\% of the sources SiO(2--1) as a typical outflow/jet
  tracer was detected. Assuming that SiO is exclusively produced by
  gas phase reactions after sputtering Si-bearing species from dust
  grains, this implies that at least 40\% of the IRDCs presented here
  have ongoing star formation activity (non-detection of SiO does not
  imply no star formation activity, hence the lower limit).  The
  observed width-to-zero-velocity of the SiO lines varies between 2.2
  and 65\,km\,s$^{-1}$. While inclination effects are unlikely to
  cause this velocity spread, we discuss other potenatial origins,
  e.g., embedded objects with different masses and/or possible
  evolutionary effects.

\item While we usually detect many velocity components in the
  low-density tracers $^{12}$CO(2--1) and $^{13}$CO(1--0), the
  high-density tracer H$^{13}$CO$^+$(1--0) is also present in all
  except one source, but in general tracing only one velocity
  component. Therefore, it is well suited to derive velocities with
  respect to the standard of rest and hence kinematic distances for
  IRDCs. The observed FWHM line-width of H$^{13}$CO$^+$(1--0) is on
  average a factor 1.5 larger than that of the previously observed
  NH$_3$(1,1) line. Several possibilities for that behavior are
  discussed: While the typical line-width--size relation is unlikely
  the cause for this behavior, the data are consistent with
  gravitational instabilities of the cores. However, this
  interpretation is not unambiguous. Since H$^{13}$CO$^+$ has higher
  critical densities, it traces the denser central cores with likely
  larger internal motions. This motions could be caused by turbulence
  or by active star formation processes like outflows/infall,
  rotation.

\item As expected CH$_3$CN is only rarely detected (toward $\sim$14\%
  of the sources), whereas CH$_3$OH is observed toward a larger
  fraction of the sample ($\sim$40\%). Both molecules are only
  detected in their lower-energy transitions. While temperature
  estimates are hardly feasible with these data, we are able to derive
  the CH$_3$CN and CH$_3$OH abundances with average values of
  $1.2\times 10^{-10}$ and $4.3\times 10^{-10}$, respectively. The low
  CH$_3$CN column densities are consistent with chemical models where
  the CH$_3$CN abundances are very low at the earliest evolutionary
  stages and progressively increase with time. The derived CH$_3$OH
  abundances are close to recently reported values for low-mass
  starless cores, indicating that the initial CH$_3$OH abundances at
  the onset of star formation processes do not vary significantly
  from low- to high-mass cores.

\end{itemize}

\acknowledgments{We like to thank Sven Thorwirth for continuous
  spectroscopic help.  H.B.~acknowledges financial support by the
  Emmy-Noether-Program of the Deutsche Forschungsgemeinschaft (DFG,
  grant BE2578).}


\newpage
\footnotesize
\begin{longtable}[htb]{lrrrrrrrr}
  \caption{Source parameters}\\
  \hline \hline
  Name               & R.A.      & Dec.       & D$^a$& $v(\rm{H}^{13}\rm{CO}^+)$ & $\Delta v(\rm{H}^{13}\rm{CO}^+)$ & $\Delta v_0(\rm{SiO})^b$ & $S_{1.2mm}^c$ & $N_{\rm{H_2}}^d$ \\
  & J2000     & J2000      & kpc& km\,s$^{-1}$ & km\,s$^{-1}$ & km\,s$^{-1}$ & $\frac{\rm{mJy}}{(11'')^2}$ & $\frac{10^{23}}{\rm{cm}^{2}}$ \\
  \hline
  IRDC18089$-$1732-3 &18 11 45.3 &$-$17 30 38 &3.6     &  34.2(0.1) & 3.9(0.2) & 5.5      & 147 & 1.5 \\
  IRDC18090$-$1832-2 &18 12 02.0 &$-$18 31 27 &6.6     & 111.2(0.1) & 2.2(0.2) &          & 89  & 0.9 \\
  IRDC18102$-$1800-1 &18 13 11.0 &$-$18 00 23 &2.7$^e$ &  21.4(0.1) & 3.4(0.1) & 42.2     & 632 & 6.3 \\
  IRDC18151$-$1208-2 &18 17 50.3 &$-$12 07 54 &3.0     &  29.8(0.1) & 3.0(0.1) & 65.0     & 620 & 6.2 \\
  IRDC18182$-$1433-2 &18 21 14.9 &$-$14 33 06 &3.6     &  40.9(0.1) & 2.2(0.1) &          & 144 & 1.4 \\
  IRDC18182$-$1433-3 &18 21 17.5 &$-$14 29 43 &4.6$^e$ &  60.2(0.1) & 1.7(0.3) & 5.5      &     &     \\
  IRDC18182$-$1433-4 &18 21 14.0 &$-$14 34 20 &3.6     &  41.4(0.1) & 2.0(0.2) &          &     &     \\
  IRDC18223$-$1243-2 &18 25 10.0 &$-$12 44 00 &3.7     &  45.5(0.1) & 3.4(0.1) &          & 156 & 1.6 \\
  IRDC18223$-$1243-3 &18 25 08.3 &$-$12 45 27 &3.7     &  45.7(0.1) & 2.7(0.1) & 35.5     & 289 & 2.9 \\
  IRDC18223$-$1243-4 &18 25 06.8 &$-$12 48 00 &3.7     &  45.7(0.1) & 2.4(0.1) & 38.9     & 77  &  0.8 \\
  IRDC18223$-$1243-5 &18 25 12.0 &$-$12 53 23 &3.7     &  45.1(0.2) & 1.7(0.2) &          &     &     \\
  IRDC18223$-$1243-6$^f$&18 25 16.2&$-$12 55 33&3.7    &  45.7(0.1) & 2.0(0.3) & 2.4      &     &     \\
  IRDC18223$-$1243-6$^f$ &18 25 16.2 &$-$12 55 33 &3.7     &  50.9(0.2) & 2.1(0.4) &          &     &     \\
  IRDC18247$-$1147-3 &18 27 31.0 &$-$11 44 46 &6.7     &  59.7(0.2) & 2.2(0.6) &          & 110 &  1.1 \\
  IRDC18306$-$0835-3 &18 33 32.1 &$-$08 32 28 &3.8$^e$ &  54.8(0.1) & 2.4(0.1) &          & 154 & 1.5 \\
  IRDC18306$-$0835-4 &18 33 34.8 &$-$08 31 20 &3.8$^e$ &  55.2(0.1) & 2.4(0.2) &          &     &     \\
  IRDC18337$-$0743-3 &18 36 18.2 &$-$07 41 00 &4.0     &  56.2(0.1) & 2.8(0.1) & 39.8     & 123 & 1.2 \\
  IRDC18337$-$0743-4 &18 36 29.9 &$-$07 42 05 &4.0     &  55.8(0.1) & 4.2(0.3) &          &     &     \\
  IRDC18337$-$0743-5 &18 36 41.0 &$-$07 39 56 &4.0     &  58.4(0.1) & 4.1(0.2) & 16.2     &     &     \\
  IRDC18337$-$0743-6 &18 36 36.0 &$-$07 42 17 &4.0     &  56.6(0.1) & 2.5(0.2) &          &     &     \\
  IRDC18337$-$0743-7 &18 36 19.0 &$-$07 41 48 &4.0     &  55.9(0.1) & 2.7(0.2) &          &     &      \\
  IRDC18348$-$0616-2 &18 37 27.6 &$-$06 14 08 &6.3     & 109.7(0.1) & 2.6(0.1) & 9.9      & 169 & 1.7 \\
  IRDC18348$-$0616-8 &18 37 14.7 &$-$06 17 25 &6.3     & 109.3(0.2) & 4.4(0.4) &          &     &     \\
  IRDC18385$-$0512-3 &18 41 17.4 &$-$05 10 03 &3.3$^e$ &  47.0(0.1) & 2.1(0.1) & 41.1     & 77  &  0.8 \\
  IRDC18431$-$0312-3 &18 45 45.0 &$-$03 08 56 &6.6$^e$ & 105.4(0.3) & 2.2(0.7) &          &     &     \\
  IRDC18431$-$0312-4 &18 45 53.0 &$-$03 09 01 &        &  non det.  & non det. &          &     &     \\
  IRDC18437$-$0216-3$^f$ &18 46 21.9 &$-$02 12 24 &6.2$^e$ &  97.6(0.2) & 2.1(0.4) &          & 96  &  1.0 \\
  IRDC18437$-$0216-3$^f$ &18 46 21.9 &$-$02 12 24 &7.3$^e$ & 111.3(0.1) & 2.3(0.4) &          & 96  &  1.0 \\
  IRDC18437$-$0216-7$^f$ &18 46 22.0 &$-$02 14 10 &6.2$^e$ &  96.1(0.1) & 2.9(0.1) & 13.0     &     &   \\
  IRDC18437$-$0216-7$^f$ &18 46 22.0 &$-$02 14 10 &7.3$^e$ & 111.2(0.7) & 1.9(1.8) &          &     &     \\
  IRDC18440$-$0148-2 &18 46 31.0 &$-$01 47 08 &5.6$^e$ &  97.9(0.2) & 2.9(0.4) &          &     &     \\
  IRDC18447$-$0229-3 &18 47 42.0 &$-$02 25 12 &6.4$^e$ & 102.1(0.2) & 2.6(0.5) &          &     &     \\
  IRDC18447$-$0229-4 &18 47 38.9 &$-$02 28 00 &6.4$^e$ &  99.4(0.1) & 1.9(0.3) &          &     &     \\
  IRDC18454$-$0158-1$^f$ &18 48 02.1 &$-$01 53 56 &3.5$^e$ &  52.8(0.4) & 1.7(0.7) &          & 178 & 1.8 \\
  IRDC18454$-$0158-1$^f$ &18 48 02.1 &$-$01 53 56 &6.4$^e$ & 100.2(0.1) & 2.7(0.2) & 10.9     & 178 & 1.8 \\
  IRDC18454$-$0158-3$^f$ &18 47 55.8 &$-$01 53 34 &6.0$^e$ &  94.3(0.1) & 3.5(0.3) & 17.3$^g$ & 204 & 2.0 \\
  IRDC18454$-$0158-3$^f$ &18 47 55.8 &$-$01 53 34 &6.4$^e$ &  98.4(0.1) & 3.3(0.3) & 17.3$^g$ & 204 & 2.0 \\
  IRDC18454$-$0158-5 &18 47 58.1 &$-$01 56 10 &6.0$^e$ &  94.6(0.1) & 2.7(0.1) & 17.8     & 102 &  1.0\\
  IRDC18454$-$0158-12$^f$&18 48 05.7 &$-$01 53 28 &3.1$^e$ &  46.3(0.4) & 2.5(0.8) &          &     &      \\
  IRDC18454$-$0158-12$^f$&18 48 05.7 &$-$01 53 28 &6.4$^e$ & 101.4(0.1) & 2.7(0.3) &          &     &     \\
  IRDC18460$-$0307-3 &18 48 36.0 &$-$03 03 49 &5.2     &  84.6(0.1) & 1.8(0.2) &          & 84  &  0.8 \\
  IRDC18460$-$0307-4 &18 48 46.0 &$-$03 04 05 &5.2     &  84.4(0.1) & 2.4(0.3) &          & 177 &  1.8 \\
  IRDC18460$-$0307-5 &18 48 47.0 &$-$03 01 29 &5.2     &  84.0(0.1) & 2.1(0.1) &          & 74  &  0.7 \\
  IRDC18530$+$0215-2 &18 55 29.0 &$+$02 17 43 &5.0$^e$ &  75.9(0.1) & 2.1(0.2) & 2.2      &     &     \\
  IRDC19175$+$1357-3 &19 19 52.1 &$+$14 01 52 &1.1     &   7.7(0.2) & 2.9(0.7) &          & 53  &  0.5 \\
  IRDC19175$+$1357-4 &19 19 50.6 &$+$14 01 22 &1.1     &   7.8(0.1) & 2.1(0.2) &          & 87  &  0.9 \\
  IRDC19410$+$2336-2 &19 43 10.2 &$+$23 45 04 &2.1     &  21.5(0.1) & 2.3(0.1) & 36.8$^h$ & 343 & 3.4 \\
  IRDC20081$+$2720-1 &20 10 13.0 &$+$27 28 18 &0.7     &   5.7(0.1) & 1.8(0.1) &          & 244 & 2.4 \\
  IRDC22570$+$5912-3 &22 58 55.1 &$+$59 28 33 &5.1     & -47.5(0.1) & 2.8(0.1) & 5.5      & 88  &  0.9\\
  \hline \hline
  \multicolumn{9}{l}{$^a$ Most distances are taken from \citet{sridharan2005}. The exceptions are marked.}\\
  \multicolumn{9}{l}{$^b$ Width of SiO(2--1) line down to zero intensity.}\\
  \multicolumn{9}{l}{$^c$ Here we present the 1.2\,mm peak fluxes extracted directly from the peak positions. In the original paper,}\\
  \multicolumn{9}{l}{\hspace{0.2cm} \citet{beuther2002a} present peak fluxes from 2D Gaussian fits which are on average a bit lower.}\\
  \multicolumn{9}{l}{$^d$ The column density caluclations follow \citet{beuther2002erratum} at an average temperature of 19\,K to be}\\
  \multicolumn{9}{l}{\hspace{0.2cm} better comparable with the CH$_3$OH and CH$_3$CN column densities calculated at $T=18.75$. Masses and}\\
  \multicolumn{9}{l}{\hspace{0.2cm} integrated fluxes were already presented in \citet{sridharan2005}}.\\
  \multicolumn{9}{l}{$^e$ Newly derived near kinematic distances following \citet{brand1993} because IRDCs at the far}\\
  \multicolumn{9}{l}{\hspace{0.2cm} distances are unlikely.}\\
  \multicolumn{9}{l}{$^f$ Rows with the same IRDC name correspond to different H$^{13}$CO$^+$ velocity components.}\\ 
  \multicolumn{9}{l}{$^g$ This is a line-blend between both velocity components.}\\
  \multicolumn{9}{l}{$^h$ The outflow was studied in detail by \citet{beuther2003a}.}
\label{sourceparameters}
\end{longtable}

\begin{table*}[htb]
\caption{CH$_3$CN and CH$_3$OH parameters.}
\begin{tabular}{lrrrrrrr}
\hline \hline
Name               & $\Delta v_{k=0}^a$ & $\Delta v_{k=1}^a$ & $N_{\rm{CH_3CN}}$ & $X_{\rm{CH_3CN}}$  & $N_{\rm{CH_3OH}}$ & $X_{\rm{CH_3OH}}$ \\
                   & km\,s$^{-1}$ & km\,s$^{-1}$ & $10^{12}$cm$^{-2}$  & $10^{-10}$& $10^{13}$cm$^{-2}$ & $10^{-10}$\\
\hline
IRDC18102$-$1800-1  & 2.9(0.4) & 3.9(0.8) & 2.6  & 0.3 & 2.8 & 0.4 \\    
IRDC18151$-$1208-2  & 3.4(0.3) & 4.5(0.5) & 5.3  & 0.8 & 37.2& 6.0 \\    
IRDC18182$-$1433-2  &           &         &      &     & 3.0 & 2.1 \\    
IRDC18223$-$1243-3  & 2.8(0.5) & 3.1(0.5) & 2.7  & 0.9 & 17.6& 6.1 \\    
IRDC18223$-$1243-4  &          &          &      &     & 4.6 & 6.0 \\    
IRDC18337$-$0743-3  & 3.1(0.9) & 6.2(1.4) & 4.4  & 3.6 & 13.1& 10.7 \\    
IRDC18385$-$0512-3  &          &          &      &     & 5.3 & 6.9 \\    
IRDC18437$-$0216-7  &          &          &      &     & 2.5 & --  \\  
IRDC18454$-$0158-1  &          &          &      &     & 1.7 & 0.9 \\  
IRDC18454$-$0158-1  &          &          &      &     & 1.7 & 0.9 \\  
IRDC18454$-$0158-3  & ?$^b$    & ?$^b$    &      &     & 11.1& 5.5 \\  
IRDC18454$-$0158-3  & ?$^b$    & ?$^b$    &      &     & 11.1& 5.5 \\  
IRDC19175$+$1357-4  &          &          &      &     & 4.3 & 5.0 \\  
IRDC19410$+$2336-2  &          &          &      &     & 5.0 & 1.5 \\  
IRDC20081$+$2720-1  & 1.4(0.3) & 1.3(0.4) & 0.7  & 0.3 &     &     \\  
IRDC22570$+$5912-3  &          &          &      &     & 2.4 & 2.8 \\     
\hline \hline                                                                  
\end{tabular} 
~\\                                     
$^a$ Line-width of the CH$_3$CN $k=0$ and $k=1$ components.\\
$^b$ No reasonable fits because of multiple velocity components.\\
\label{sourceparameters2}
\end{table*}

\end{document}